%% file: main.tex
\documentclass[fleqn,usenatbib]{mnras}

\usepackage{newtxtext,newtxmath}
\usepackage[T1]{fontenc}

\usepackage{graphicx}	
\usepackage{amsmath}	
\usepackage{xcolor}
\usepackage[flushleft]{threeparttable}
\usepackage{verbatim}
\usepackage{array}
\usepackage{tabularx}
\usepackage{colortbl}
\usepackage{hyperref}
\usepackage{float}
\usepackage{subcaption}
\usepackage{booktabs}
\usepackage{adjustbox}
\usepackage{pdflscape}
\usepackage{longtable}
\usepackage{placeins}
\usepackage{threeparttablex}
\usepackage{tabularray}
\usepackage{cuted}
\usepackage{multirow}
\usepackage{xtab}

\DeclareRobustCommand{\VAN}[3]{#2}
\let\VANthebibliography\thebibliography
\def\thebibliography{\DeclareRobustCommand{\VAN}[3]{##3}\VANthebibliography}

\newcommand{\rstar}{\mbox{$R_{\star}$}}

\newcommand{\rsun}{\mbox{$R_{\odot}$}}

\newcommand{\TESS}{\textit{TESS}}

\title[RAVEN]{RAVEN: RAnking and Validation of ExoplaNets}

\author[A. Hadjigeorghiou et al.]{\parbox{\textwidth}{\Large
Andreas~Hadjigeorghiou$^{1,2}$,
David~J.~Armstrong$^{1,2}$, 
Kaiming~Cui$^{1,2}$, 
Marina Lafarga Magro$^{1,2}$, 
Luis Agust\'{i}n Nieto$^{3,4}$,
Rodrigo F. D\'iaz$^{3,5}$,
Lauren Doyle$^{1,2}$,
Vedad Kunovac$^{1,2}$
}
\vspace{0.2cm}
\\
\parbox{\textwidth}{
$^{1}$Department of Physics, University of Warwick, Gibbet Hill Road, Coventry CV4 7AL, UK\\
$^{2}$Centre for Exoplanets and Habitability, University of Warwick, Gibbet Hill Road, Coventry CV4 7AL, UK\\
$^{3}$Instituto de Ciencias F\'isicas (CONICET / ECyT-UNSAM), Campus Miguelete, 25 de Mayo y Francia, (1650) Buenos Aires, Argentina\\
$^{4}$Gerencia de Tecnología de la información y de las Comunicaciones (GTIC), Subgerencia Vinculación y Desarrollo de Nuevas Tecnologías de la Información,\\ DTE-CNEA. Centro Atómico Constituyentes, Av. Gral. Paz 1499, (1650) Buenos Aires, Argentina\\
$^{5}$Instituto Tecnol\'ogico de Buenos Aires (ITBA), Iguaz\'u 341, Buenos Aires, CABA C1437, Argentina
}\vspace{-0.3cm}}
\date{\vspace{-0.5cm}Accepted XXX. Received YYY; in original form ZZZ}

\pubyear{2025}

\begin{document}
\label{firstpage}
\pagerange{\pageref{firstpage}--\pageref{lastpage}}
\maketitle


\begin{abstract}
We present \texttt{RAVEN}, a newly developed vetting and validation pipeline for \textit{TESS} exoplanet candidates. The pipeline employs a Bayesian framework to derive the posterior probability of a candidate being a planet against a set of False Positive (FP) scenarios, through the use of a Gradient Boosted Decision Tree and a Gaussian Process classifier, trained on comprehensive synthetic training sets of simulated planets and 8 astrophysical FP scenarios injected into \textit{TESS} lightcurves. These training sets allow large scale candidate vetting and performance verification against individual FP scenarios. A Non-Simulated FP training set consisting of real \textit{TESS} candidates caused primarily by stellar variability and systematic noise is also included. The machine learning derived probabilities are combined with scenario specific prior probabilities, including the candidates' positional probabilities, to compute the final posterior probabilities. Candidates with a planetary posterior probability greater than 99\% against each FP scenario and whose implied planetary radius is less than 8$R_{\oplus}$ are considered to be statistically validated by the pipeline. In this first version, the pipeline has been developed for candidates with a lightcurve released from the \textit{TESS} Science Processing Operations Centre, an orbital period between 0.5 and 16 days and a transit depth greater than 300ppm. The pipeline obtained area-under-curve (AUC) scores $>97$\% on all FP scenarios and $>99\%$ on all but one. Testing on an independent external sample of 1361 pre-classified TOIs, the pipeline achieved an overall accuracy of 91\%, demonstrating its effectiveness for automated ranking of \textit{TESS} candidates. For a probability threshold of 0.9 the pipeline reached a precision of 97\% with a recall score of 66\% on these TOIs. The \texttt{RAVEN} pipeline is publicly released as a cloud-hosted app, making it easily accessible to the community. 
\end{abstract}




\section{Introduction} \label{sec:intro}
The era of large scale space-based photometric transit surveys, heralded primarily by the \textit{Kepler} mission \citep{Kepler} and advanced by the \textit{TESS} mission \citep{TESS}, has led to an unprecedented number of discovered exoplanet candidates. However, despite the best efforts of the exoplanet research community, confirming or disproving the planetary nature of candidates continues to be a lengthy and challenging process. 
As of the time of this work, the Exoplanet Archive\footnote{https://exoplanetarchive.ipac.caltech.edu/ Accessed:18/08/2025} lists 7658 \textit{TESS} Objects of Interest (TOI), of which 5152 are still listed as candidates. Only 666 have been confirmed as true exoplanets, joining 558 \textit{TESS} detected candidates which were already known planets. 

At the same time, 1185 \textit{TESS} candidates have been identified as False Positives (FP) and 97 as False Alarms (FA). Their relatively large number serves to emphasize the challenge of confirming exoplanet candidates as true planets. Moreover, the diverse nature of the FPs further complicates the matter. Ranging from eclipsing binaries and hierarchical systems, to blended nearby or unresolved stellar binaries and transiting planets, there are a host of different astrophysical FP configurations that can reproduce the detected candidate event. In addition to these, stellar variability and instrumental noise can also generate false planetary candidates, which constitute the FA detections. All these scenarios must be accounted for, investigated and ruled out when trying to confirm a detected event from a transit survey as a true planetary companion. 

The FP identification process is further challenged by the number of detected events retrieved by search algorithms for large scale transit surveys. For example, the \textit{Kepler} mission over 4 years of observations generated more than 32000 significant Threshold Crossing Events (TCEs) as per \textit{Kepler}'s 25th and final Data Release \citep{Robovveter25}. Of those, just 8054 were released as \textit{Kepler} Objects of Interest (KOI), with the rest representing Non-Eclipsing detections. About half of the KOIs were labelled as planet candidates, with the rest identified as FPs. In the case of \textit{TESS}, the Planet Search from the Science Processing Operations Centre (SPOC) \citep{SPOC} resulted in about 160000 TCEs \citep{TOI} from just the \textit{TESS} Primary Mission, while the Quick Look Pipeline's (QLP) \citep{QLP1, QLP2} Box-Least-Squares(BLS) \citep{BLS} search on the almost 15 million stars observed in \textit{TESS}' Full Frame Images (FFIs) produced more than 2.5 million significant detections \citep{TESS-FaintStar}.

To assist the vetting efforts of the TCE Review Team, the \textit{Kepler} mission employed a Data Validation (DV) component \citep{KEPLER_DVP} in its data processing pipeline, which created a diagnostics report based on metrics produced from the lightcurves and the pixel level data. The purpose of the metrics were to highlight results that were inconsistent with planetary transits, especially for non-transit like events. The Diagnostics Report thus allowed the Review team to more easily reject many of the non eclipsing events and also disposition the eclipsing candidates into Planet Candidates (PC) or FPs. \textit{Kepler}'s DV has been adapted for \textit{TESS} and is used by the SPOC pipeline \citep{SPOC}. 

In addition to the DV processing, the \textit{Kepler} team introduced Robovetter, an automated vetting pipeline based on an "expert system" algorithm. This algorithm relied on a series of quantitative tests to characterise candidates as possible Planet Candidates or as one of 4 FP scenarios. The Robovetter was used to create a fully automated candidate catalogue for \textit{Kepler}'s DR24 \citep{Robovetter24} and DR25 \citep{Robovveter25} and to disposition existing and new KOIs. A similar algorithm inspired by the Robovetter, called TESS-ExoClass\footnote{https://github.com/christopherburke/TESS-ExoClass} is currently used for vetting the \textit{TESS} SPOC TCEs \citep{TOI}. Another Robovetter-inspired pipeline for TESS candidates, called LEO-Vetter \citep{LEO-Vetter}, was recently released, providing a fully automated vetting process.

Other forms of vetting pipelines have also been introduced and in particular Machine Learning Implementations. One of the earliest examples is the Autovetter \citep{Autovetter} which used a Random Forest classifier to vet \textit{Kepler} TCEs. The Autovetter, like the Robovetter, was used by the \textit{Kepler} team to provide an automated catalog in DR24 \citep{AutovetterDR24}. Moreover, the AstroNET \citep{AstroNET-Kepler} vetting framework uses a Convolutional Neural Network trained on lightcurves of dispositioned TCEs to classify candidates \textit{Kepler} candidates as FPs or PCs. The framework has been adapted for \textit{TESS} \citep{AstroNET-Triage, AstroNET-TriageV2} and is used by QLP for the initial vetting of its candidates \citep{TESS-FaintStar}.

Advancing a step beyond vetting the candidates are the validation pipelines \citep{BLENDER, vespa, pastis, pastis2, KeplerPipeline, TRICERATOPS_Validated_large1, ExoMiner}, which have been developed to enable the confirmation of candidates as true planets through statistical analysis. The traditional confirmation process for candidates involves a series of follow-up observations, which aim to rule out FP scenarios and characterise the candidate to prove its planetary nature. A prime example of this process is the TESS Follow-up Observation Program (TFOP) \citep{TFOP}. The final confirmation of a candidate as a planet is ideally achieved through Radial Velocity (RV) observations that can determine the companion's mass. In the absence of a detected RV signal, validation pipelines provide a quantitative measure of the candidates' likelihood to be planets, allowing their prioritisation for follow-up observations. This is especially important in light of observation time constraints, as it allows for valuable resources to be allocated to the more promising candidates. Moreover, for the many candidates of whom their RV signal, either due to their mass, orbit, brightness or stellar host, is not yet detectable by the current generation of spectrogaphs, statistical validation is often the only other pathway with which they can be promoted from candidates.  

Although the exact method varies across the different validation implementations, the core methodology relies on the determination of the likelihood that the candidate is a planet and the likelihood that is a FP. This is usually determined through producing numerous modelled signals with different parameter configurations for the planet and FP scenarios and fitting the observed signal. Each likelihood is combined with prior probabilities that are based on observational constraints and the occurrence rate of each hypothesis. The two competing hypotheses are then compared to derive their statistical probability. A confidence threshold value is set over which the planet hypothesis for the candidate is considered sufficient. Validation frameworks rose in prominence following the launch of the \textit{Kepler} mission and the rapid increase in transiting exoplanet candidates that followed. They have been responsible for the statistical confirmation of thousands of exoplanets to date \citep{vespa_validated, vespa_k2_validated}, including valuable terrestrial exoplanets laying in the Habitable Zone of their host stars \citep{BLENDER_5_Hz, BLENDER_12HZ}.

The most prominent validation pipeline for \textit{TESS} candidates is \texttt{TRICERATOPS} \citep{Triceratops}, which has been the only framework developed specifically for the mission. \texttt{TRICERATOPS}, similar to the methods developed for \textit{Kepler}, relies on producing model lightcurves for the different planet and FP scenarios, the parameters of which are sampled from empirical prior distributions. These are then compared to the observed transit signal, with the likelihood of each scenario computed using an arithmetic mean estimation computed over a million sampled lightcurves. \texttt{TRICERATOPS} examines 18 distinct astrophysical configurations for the signal, of which 3 represent positive planet detections. Its standout feature is its integration of the known nearby sources to the target star of the detected event, as listed in the \textit{TESS} Input Catalog (TIC) based on \textit{Gaia} DR2 \citep{GaiaDR2}, which allows it to rigorously examine scenarios where the event is located on one of these nearby sources. \texttt{TRICERATOPS} has been used to validate dozens of \textit{TESS} exoplanet candidates to date \citep{TRICERATOPS_Validated_large1, VaTEST2, VaTEST3, triceratops_validated1}. The pipeline was recently updated to add multi-band photometry support from follow-up observation of \textit{TESS} candidates in its analysis, as presented in \cite{triceratops_plus}.

An alternative validation framework, which utilised 4 varied Machine Learning (ML) models was developed and presented for \textit{Kepler}
in \cite{KeplerPipeline}. The posterior probability score for the planetary hypothesis was derived directly from the ML classification probability score combined with prior probabilities for each FP scenario. With its new methodology, the pipeline essentially replaced the complex and lengthy modelling process, while still allowing for detailed characteristics of the event to shape the probability through its extensive list of metrics that the ML models relied on. The use of ML also provides the pipeline with extreme scalability and efficiency as after the training is completed, new candidates can be dispositioned fast and with ease. The pipeline achieved very positive results based on its testing on pre-dispositioned \textit{Kepler} KOIs and especially those in systems with multiple KOIs. It also demonstrated its capability for validation, producing 50 new validated planets.

Another ML-based validation framework, \texttt{ExoMiner}, was introduced in \cite{ExoMiner} for \textit{Kepler} TCEs, employing a Deep Neural Network (DNN) model. Similar to the other ML based pipelines on \textit{Kepler} data, \texttt{ExoMiner} was successfully tested on pre-dispositioned \textit{Kepler} TCEs and was used to validate 301 undispositioned \textit{Kepler} candidates. The pipeline has since been adapted for \textit{TESS} 2-minute candidates, albeit only in a vetting and ranking capacity, as presented in \cite{ExoMinerTESS}. A unique feature of this newest model, called \texttt{ExoMiner++} is the fact that it is trained on both \textit{Kepler} and \textit{TESS} TCEs, using its superior performance on the former to aid its classification of the latter.

The work presented here builds upon the \textit{Kepler} pipeline of \cite{KeplerPipeline} to present a new vetting and validation pipeline for \textit{TESS} candidates. The new pipeline, called \texttt{RAVEN} (RAnking and Validation of ExoplaNets), is not just an adaptation of the original for the \textit{TESS} mission but rather a significant overhaul and expansion of the framework. The quintessential catalyst of this change is the introduction of synthetic training sets instead of relying only on TCEs produced by the mission, which significantly expands and enhances the parameter space for the planet and FP scenarios explored by the ML models. A training set composed almost exclusively of non eclipsing detections from the \textit{TESS} mission is also used, to retain the pipeline's capability in classifying them. Features are now generated locally rather than relying on the \textit{Kepler} pipeline. The pipeline also now includes its own Positional Probability generation, which is defined as the probability that the event is located on the target star or on a known nearby source. This part of the pipeline has been introduced in \cite{PosProbs} and released as a standalone tool. Moreover, the use of synthetic sets has allowed us to refine the process as the planet scenario is compared individually against each FP scenario, a capability that was previously reserved only for validation frameworks that relied on model fitting. This not only enables a more rigorous derivation of the probability but also allows insight into the nature of FP detections and limitations of validation in general. All of this is achieved while still retaining fast run-times, with a typical candidate taking about a minute to be processed, and scalability due to multiprocessing support. 

The pipeline design and process is presented in Section \ref{sec:Design}.
The performance of the pipeline was tested on both a test subset of synthetic events and on a sample of pre-classified \textit{TESS} Objects of Interest. The results are presented in Sections \ref{sec: Testset_results} and \ref{sec: TOI_results}.

\section{Pipeline Design}\label{sec:Design}
\subsection{Framework}\label{sec:Framework}
\begin{figure*}
    \centering
    \includegraphics[width=\textwidth]{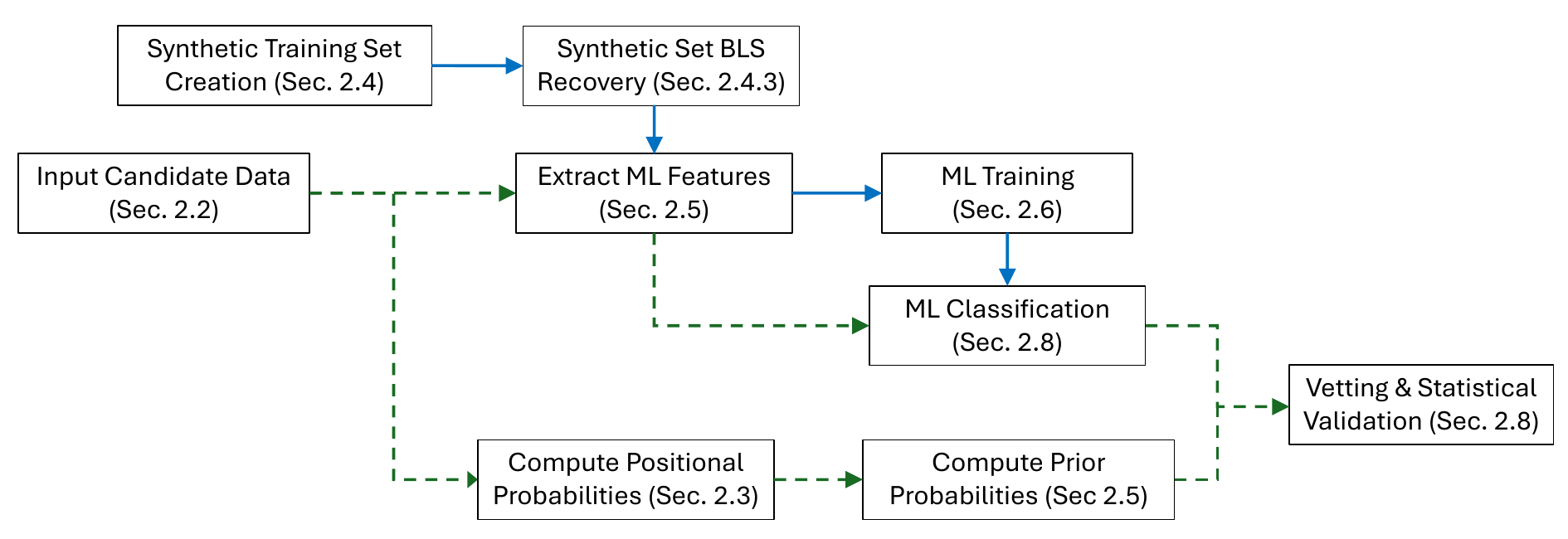}
    \caption{The core process of the pipeline, from the input of the candidate data to the output of the posterior probabilities for vetting and validation. The flow of the candidate data is represented by the dashed green arrows and the flow of the synthetic training data by the blue solid arrows.}
    \label{fig:FlowChart}
\end{figure*}

The pipeline is based on the statistical validation framework introduced in \cite{KeplerPipeline} for \textit{Kepler} candidates (hereafter A21). The framework has been adapted in this work to be used on \textit{TESS} data and at the same time expanded and upgraded. The core process of the framework remains unchanged and will be briefly described here. For a more in depth discussion on the statistical implementation of the framework and its results on the \textit{Kepler} data please refer to the original paper. The process of implementing and running the framework is complex and relies on multiple steps. For ease of reading, we present those steps in a flowchart in Figure \ref{fig:FlowChart}, which points to each relevant subsection.

At its core, the framework employs Machine Learning models to derive the posterior predictive probability of the candidate being a true planetary companion or a FP. We denote this probability as $\mathcal{P}(s|x^*,D,\mathcal{M})$, where $x^*$ is the feature vector for the candidate and $D$ and $\mathcal{M}$ represent the training data and the classification model respectively. We set $s=1$ to signify the planet hypothesis and $s=0$ for the FP. The training datasets, their features and the Machine Learning models used are discussed in detail in sections \ref{sec:training_sets}, \ref{sec:Features}  and \ref{sec:ML_training} respectively.

The classification models use equal-representation training sets, implying a uniform prior probability for each scenario. To account for astrophysical or observational prior knowledge of the likelihood of a given scenario, the final probability is adjusted with empirical prior probabilities, $\mathcal{P}(s|I)$,  that express the overall probability of each scenario to occur and be detected in the \textit{TESS} data. A detailed overview of the prior probabilities is presented in Section \ref{sec:priors}. The final posterior probability for a given candidate thus takes the following form: 
\begin{equation} \label{eq: posterior_prob}
    \mathcal{P}(s=1|x^*,D,\mathcal{M},I) = \frac{\mathcal{P}(s=1|x^*,D,\mathcal{M}) \mathcal{P}(s=1|I)}{\sum\limits_{s}\mathcal{P}(s|x^*,D,\mathcal{M}) \mathcal{P}(s|I)},
\end{equation}
with the denominator acting as a normalizing factor to ensure that the probabilities for the planet and FP hypothesis sum to 1. Candidates whose posterior probability for the true planet scenario is greater than 0.99 can be statistically validated, in line with the field norm \citep[e.g.][]{vespa}. 

In an important deviation from A21, the FPs in the \texttt{RAVEN} pipeline are no longer represented by a single class containing events from all possible FP scenarios. Instead, the FP scenario has been split into 8 individual classes, each one representing a specific FP configuration. This change allows for testing the performance against specific scenarios, as well as bolstering the representation of scenarios which are not well represented in existing datasets (e.g. due to difficulty identifying the nature of the FP). To enable this improved representation of difficult or unusual FP cases we implement synthetic lightcurves for the training of the ML models rather than lightcurves of pre-dispositioned candidates from the mission. These synthetic lightcurves consist of hundreds of thousands of simulated events that we inject into the \textit{TESS} lightcurves, providing the pipeline with expansive training sets for detailed FP identification. The synthetic events and their creation process are presented in Section \ref{sec:training_sets}. 

The splitting of the FPs into individual scenarios necessitated a change to the classification and validation process. As a result, instead of performing one classification of the planet scenario against all FP scenarios, we implement a series of planet-vs-FP classifications, one for each FP scenario. The derived posterior probability from equation \ref{eq: posterior_prob} therefore expresses only the probability of the planet hypothesis in the context of the comparison to the specific FP scenario and we no longer determine the overall posterior probability for the planet hypothesis. Consequently, to statistically validate a candidate, we now conservatively require a posterior probability above 0.99 from each planet-FP classification. This allows us to more rigorously explore the planet posterior probability, as the intricate differences between the planet and each FP scenario, along with their respective prior probabilities, have a more distinctive effect on the final result. 

The only exception to the above validation process is for the class of FP that represent the non-simulated FPs. This FP class contains significant detections from a BLS survey on thousands of \textit{TESS} lightcurvers and essentially represents cases of stellar variability and other sources of noise in the lightcurves that could mimic the transit signal. We use these events to encompass all other Non-Simulated FPs (NSFP) and as a result they form the basis for the pipeline's candidate vetting. Therefore, the NSFP scenario is treated differently to the astrophysical FP scenarios. This is elaborated upon further in Section \ref{sec:nsfp_vetting}. 

Finally, to enable easy ranking of candidates, we collapse the collection of posterior probabilities to one singular final probability estimate that expresses the likelihood of the candidate being a true planetary companion. Since we require a posterior probability of 0.99 across all planet-FP classification pairs to statistically validate a candidate, we take the minimum of the probabilities as the candidate's final probability, which we call the \texttt{RAVEN} probability. This minimum probability essentially represents the lowest likelihood of the candidate being a true planet as determined by our framework. Therefore, we can use this to rank all candidates and validate those with a \texttt{RAVEN} probability above 0.99.

\subsection{Input Data}\label{sec: Data}
The \texttt{RAVEN} pipeline currently works with lightcurves produced from the \textit{TESS} FFIs and released by SPOC \citep{FFI}. These lightcurves are obtained from the FFI observations of each sector using aperture photometry, with a sampling rate of 30 minutes for sectors 1-27 and 10 minutes for sectors 28-55. For the \textit{TESS}' Second Extended Mission, which commenced with sector 56, the FFIs are released with a 200 second sampling rate.
In this work, we use lightcurves up to sector 55.

Due to processing time constraints, the SPOC pipeline provides a limited number of up to 160000 FFI lightcurves per sector. The targets are selected automatically from the pipeline based on the selection criteria outlined in \cite{FFI}. A more complete number of lightcurves for each sector is provided by the QLP. Due to the differences between the lightcurve products from the two pipelines, especially in regards to their aperture implementation and post-processing, the pipeline is currently not able to be run on the QLP lightcurves.
Furthermore, the pipeline is currently constrained to work with candidates that have a period up to 16 days and a minimum depth of 300ppm. This choice means that the focus is on candidates with multiple detected transits in the \textit{TESS} lightcurves, reducing the possibilities for error and limiting the pipeline to well-defined, significant, periodic candidates. Considering that a single observation sector for \textit{TESS} covers a time length of 27.4 days and that some targets have only been observed in one sector, the period limit is well justified. The depth limit ensures that we focus on significant events where analysis is not dominated by noise in the lightcurve.

\subsection{Data Processing}\label{sec:Processing}
The processing of the candidates for the \texttt{RAVEN} pipeline, up to the stage of producing the Positional Probabilities, is almost identical to the one we described in our work in \cite{PosProbs}, hereafter referred as HA24. Readers should refer to HA24 for a detailed description. We provide an overview of the process and minor changes from HA24 here.

\subsubsection{Retrieve Stellar Characteristics and Identify Nearby Sources}\label{sec:Sources}
The input to the pipeline is a list of candidates, along with the TIC ID of their associated target stars and the epoch, period, duration and depth estimate of the transit. From there, the pipeline first identifies the \textit{TESS} sectors on which each target star was observed and then locates their corresponding lightcurve files. If specified, the pipeline can additionally retrieve any missing files from MAST\footnote{https://archive.stsci.edu/missions-and-data/tess}. Following that, the stellar characteristics of the target stars are retrieved from the TIC, namely the stellar radius, effective temperature, \textit{TESS} magnitude, proper motion and their coordinates. The same characteristics are retrieved for all nearby sources to the target within a radius of 168''. We subsequently exclude nearby sources fainter than 10 \textit{TESS} magnitude difference to the target, which, as discussed in HA24, cannot be the hosts of the event. The \textit{TESS} magnitudes for the target and the nearby sources are calibrated into the expected flux counts that would be detected by \textit{TESS} in each sector of observation, as per equation 3 of HA24. Additionally to the above, we further obtain the \textit{Gaia} RUWE score for the target star from Data Release 3 \citep{GaiaDR3}, alongside the radius and effective temperature for those that did not have one listed in the TIC. 

\subsubsection{Compute Per Sector Characteristics and Positional Probabilities}\label{sec:per_sector}
For each candidate we then determine their per sector transit characteristics, namely their depth, transit duration and full eclipse duration, along with their observed centroid offset, by fitting trapezoid transit models to each sector's flux and centroid timeseries data respectively. The flux contribution of both the target star and its known nearby sources is then determined by modelling the pixel-level observation using the TESS Pixel Response Function \citep[PRF,][]{TESS_PRF_software}, which allows us to determine the implied eclipse depth for the event if it would occur on each source. The centroid offset is subsequently modelled by iteratively adjusting the flux of each source by the implied depth and producing a new simulated pixel-level observation that mimics the effect of the eclipse. The modelled centroid offset is compared to the observed to obtain the likelihood of each source being the host of the event. These probabilities are normalised across the known sources to obtain the final Positional Probabilities for the candidate. The entire process is described in detail in Section 2 of HA24.

\subsection{Synthetic Training Set}\label{sec:training_sets}
The \texttt{RAVEN} pipeline has been developed from the ground up to use synthetic lightcurves for the training of its ML models. This offers a distinct advantage, as the training data are not tied to the efforts of the exoplanet community to verify candidates as planets or FPs. While already characterised candidates have their own advantages, their small numbers, owing to the difficulty of verification efforts, limit the construction of expansive training sets. Moreover, for some of the FP scenarios we consider, no clearly identified candidates exist, while the exact nature of the FP candidates can often be uncertain. 

The synthetic training sets allow us to instead include examples of the different scenarios we examine in our training sets, with no uncertainty as to their disposition. The training sets therefore incorporate a wide variety of different configurations for the scenarios, allowing the models to explore in depth the distributions of their characteristics, free from the biases introduced by the constraints of the confirmation methods. 
We discuss the creation of the synthetic training sets and the efforts and considerations taken to ensure that they faithfully represent the expected candidates of their respective scenarios in the rest of this section.

\subsubsection{Simulated Planets and FPs}\label{sec:sims}
The initial collection of synthetic events uses simulated eclipses, which were injected in the SPOC lightcurves. The simulated events were generated using our modified version of the \texttt{PASTIS} \footnote{\url{https://github.com/ckm3/pastis-dev}} \citep{pastis} software and initially include the following scenarios: 
\begin{enumerate}
    \item a Transiting Planet (Planet), where a planet transits the target star,
    \item an Eclipsing Binary (EB), where a secondary star eclipses the target star,
    \item a Hierarchical Eclipsing Binary (HEB), where an unresolved eclipsing binary is bound in a wide orbit to the primary target star,
    \item a Hierarchical Transiting Planet (HTP), where a planet transits an unresolved stellar companion to the target star,
    \item a Background Eclipsing Binary (BEB), where an unresolved eclipsing binary pair is blended with the target star (but not bound to it in a multiple system, unlike the HEB scenario),
    \item a Background Transiting Planet (BTP), where a planet transits an unresolved background star.
\end{enumerate}

Note that these are physical scenarios, which may represent as candidates in different ways such as half or double orbital periods. Further supplemental scenarios are implemented in Section \ref{sec: NFP}. 

In an effort to ensure that the synthetic data closely represent the observed \textit{TESS} population, the primary star in each of the above configurations was randomly selected from a well characterised sample of the \textit{TESS} Input Catalog (TIC) presented in \citet{LaurenSample}. The sample consists of main-sequence sources with SPOC FFI lightcurves observed in sectors 1 to 55 crossmatched with \textit{Gaia} DR2 and DR3 to define their properties. For the purposes of our simulations, we removed from this sample any sources with a \textit{Gaia} RUWE score higher than 1.05 to constrain our selection to single star systems. The RUWE score has been previously shown to correlate with the presence of additional stellar companions, even for values lower than the currently accepted threshold of 1.4 for multiplicity \citep{Gaia_Belokurov, Gaia_Binaries_Stassun}. Furthermore, targets with TOI, community candidates (CTOI), KOI, and K2 candidates listed in ExoFOP\footnote{\url{https://exofop.ipac.caltech.edu/tess/}} were removed from the sample, to ensure that the simulated events would not be injected in lightcurves with known candidates. 
Finally, the sample was also limited to a stellar effective temperature range of 3000\,K to 10000\,K as provided in the TIC. It should be noted that the sample from \citet{LaurenSample} is already limited to a \textit{TESS} Magnitude brighter than 13.5 and a \textit{Gaia} Magnitude brighter than 14.
In total, our target sample consisted of 1,200,520 SPOC FFI stars.

The stellar parameters of the target sources, as provided or derived from \cite{LaurenSample}, were used to inform and constrain our simulated systems, thus making sure that the scenarios were realistic and plausible 
The properties and orbital parameters of the simulated objects were derived based on a combination of established distributions in the literature, isochrone fitting or otherwise uniformly drawn. The specific parameters used for modelling the six primary astrophysical scenarios and their sources are listed in Table \ref{tb:astro_param_distributions}. 
For the isochrone fitting, the MESA Isochrones \& Stellar Tracks \citep{dotterMESAIsochronesStellar2016,choiMesaIsochronesStellar2016,paxtonModulesExperimentsStellar2011,paxtonModulesExperimentsStellar2013,paxtonModulesExperimentsStellar2015} were used through the \texttt{isochrones} software \citep{mortonIsochronesStellarModel2015}. Limb darkening coefficients for all simulated stellar sources were based on \cite{claretGravityLimbdarkeningCoefficients2011}, updated for the TESS bandpass. For the hierarchical systems, the wider stellar binary or star-planet system was set on a fixed 10000 day non-eclipsing orbit around the target, with the parameters of the host star constrained based on the parameters of the target star. 

The orbital parameters of the planetary and stellar close in companions were allowed to fully explore the parameter space, with the only limitation that the period had to be between 0.5 and 16 days as per our pipeline's design. Both the orbital period and planetary radius were drawn from their respective relative occurrence distributions as presented in \cite{Hsu2019}. For stellar companions, periods were assigned according to the log-normal distribution given in \cite{Raghavan2010}. Moreover, the inclinations are drawn from an isotropic distribution, \(p(i) \propto \sin i\) for \(i \in [0,\uppi/2]\), while the eccentricity distribution was obtained based on the statistical results in \cite{MoeDiStefano2017}, where $p_e \propto e^\eta$, and the power law is truncated at $e_\mathrm{max}$. For the simulations, we chose $\eta = -0.3$, with the value of $e_\mathrm{max}$ depending on the orbital period as per Equation (3) of \cite{MoeDiStefano2017}. For periods less than 2 days, we set their eccentricities to zero.

The freedom allowed to the parameters meant that not all simulated scenarios had a detectable transiting event. More specifically, the simulations were considered to result in a non-detectable event if the inclination angle prohibited any transits or eclipses, or if the magnitude of the diluting host star was too faint to cause a drop in the detected flux of the target above the pipeline's 300\,ppm limit. While these non-detectable simulated events were ultimately not used in the training of the pipeline, their numbers and reason of rejection were recorded. This is a core output of our modelling process and is essential to our methodology, as the number of successful and rejected events allows us to compute the likelihood probability of each scenario to be detected by \textit{TESS} which forms an integral part of the pipeline's prior probabilities. The number of successful and rejected events for each scenario are presented in Table \ref{tb:Detection}, with the derived detection probability provided in the final column. The stark differences between the probabilities for each scenario highlights the likelihood of detecting each scenario in the \textit{TESS} data, with EBs being relatively highly detectable at 19.8\% and BTPs being extremely unlikely to be detected at only 0.05\%. It also serves to underscore the difficulty in detecting planetary companions, with a probability of just 2.5\%. As the detectability and hence prior probability of the BTP scenario is extremely low, we decided to not incorporate the scenario in the framework as the number of simulations that would have needed to be produced to create a large enough sample was computationally prohibitive.

The simulated events were injected in the corresponding \textit{TESS} FFI lightcurves of the chosen primary target stars. By choosing real observed \textit{TESS} targets and then injecting the simulated events in their lightcurves, we allowed for their noise properties and the peculiarities related to the \textit{TESS} sector-based observation strategy to be encapsulated in the synthetic training data. As a result, these synthetic lightcurves represent our best attempt at creating a realistic sample of the different scenarios for the training of the ML models. The epoch time for the injection events was randomly selected as a point between the start of the first sector of observation and up to the length of the simulation's orbital period. In rare cases, the events were subsequently injected into gaps in the lightcurve and as such have no transits present. We account for the detectability of these events, along with others that despite successfully passing our simulation process cannot actually be recovered in the \textit{TESS} data by running a retrieval survey as discussed in the following sections.

\input{sim_table}

\subsubsection{Synthetic Nearby False Positives}\label{sec: NFP}
We expand upon the initial set of astrophysical scenarios with the addition of 3 simulated FP scenarios for which the eclipsing event is located on a known nearby source instead of the target. These are known as Nearby False Positives (NFPs) and refer to cases where the nearby host contributes light in the target's aperture, which is then observed as originating from the target. The result is a blended lightcurve, with the eclipsing event appearing diluted and shallower. For the \texttt{RAVEN} pipeline, we take into consideration the following NFP scenarios:
\begin{enumerate}
    \item a Nearby Transiting Planet (NTP), where a planet transits the diluting host,
    \item a Nearby Eclipsing Binary (NEB), where the nearby diluting source is an eclipsing binary.
    \item a Nearby Hierarchical Eclipsing Binary (NHEB), where the nearby diluting source is a hierarchical eclipsing binary.
\end{enumerate}

The NFP events are distinctly different from the 6 primary scenarios described in the previous section in that the event is located on a resolved source, the photometric contribution of which to the target's aperture can be quantified and corrected for. A correction is already applied in the \textit{Kepler} and \textit{TESS} SPOC pipelines \citep{Correction}. This correction does not remove the photometric signature of the blended transit or eclipse but merely alters its apparent depth on the target. Moreover, the extent of the contribution to the aperture does not depend only on the magnitude difference between the target and the host star, as is the case with the Background and Hierarchical scenarios, but also their positions on the CCD plane. As such, modelling these scenarios is inherently more complex and would require a pixel-level injection. 

Due to the complexity and heavy processing workload that would be required, the synthetic events for the NFP scenarios were not directly simulated. Instead, we chose to leverage our existing simulations, the dilution information for the targets from SPOC and the understanding of the NFP scenarios from AH24 to adopt a simpler approach that avoids pixel-level injection. The simulated NFP events are created by altering the simulations to account for different degrees of dilution in the observed target flux. As discussed in AH24, the detected depth of the diluted event, $D_{dl}$, on the target is determined by:
\begin{equation}
    D_{dl} = D_{ns} \frac{f_{ns}}{f_{t}},
\end{equation}
where $D_{ns}$ signifies the depth of the event on the nearby host, with $f_{ns}$ and $f_{t}$ the fraction of flux contribution of the nearby and target star in the aperture respectively. The two fractions effectively encompass all the relevant information for the scenario, namely the magnitude difference and distance of the target and host source, along with the additional contribution of other nearby sources in the aperture. Therefore, the process for creating our simulated NFP scenarios depends only on sampling the two fractions for randomly selected simulated Planet, EB and HEB events. Under this framework, the host sources of our simulated events become the diluting sources of the selected target stars for the NFP events. As such, the three NFP scenarios inherit the detectability of the base scenarios. To then determine their scenario specific detectability, we can simply examine if the depth of the simulated NFP event is above our 300ppm limit. In extremely rare cases, as the flux correction is done by subtracting the contributing fluxes, the depth can cause the flux to artificially drop below a count of 0. These cases are also rejected.

Similar to the 6 initial scenarios, the target stars for the simulated NFP events were drawn from the same Main Sequence sample of the SPOC FFI stars. As such, the target fraction for our simulations was set to be the one already determined for the actual observations by the SPOC pipeline, which is provided in the lightcurve data labelled as "CROWDSAP". This has the distinct advantage that the simulated event is directly tied to the actual observed flux that has been corrected for the specific fraction. It should be noted that the target fraction differs between sectors, usually minimally, due to a combination of aperture changes in different sectors, the telescope pointing and the proper motion of stars. As a result, the depth of the diluted event is accordingly affected, with extreme cases leading to the event effectively disappearing in some sectors. By using the observation's target fraction in each sector, the effects caused by the changing fractions are directly baked into our simulations. 

Selecting a suitable nearby fraction for the simulated events is not as straightforward as the target fraction. The simplest approach would be to use the remainder fraction as the nearby on the assumption that only the host source is diluting the target's aperture. However, such an assumption would be extremely unrealistic as in the vast majority of cases there is more than one contributing source. A more suitable option would be to use the specific fractions of the existing nearby sources for the targets to inform our choice. While these fractions are not readily available, the \texttt{RAVEN} pipeline already has a method to determine the flux contributions of each nearby source as part of its Positional Probability component, described in detail in AH24. We therefore used our pipeline to directly determine the nearby flux fractions of the diluting sources for all target stars. 

\begin{figure}
    \centering
    \includegraphics[width=\columnwidth]{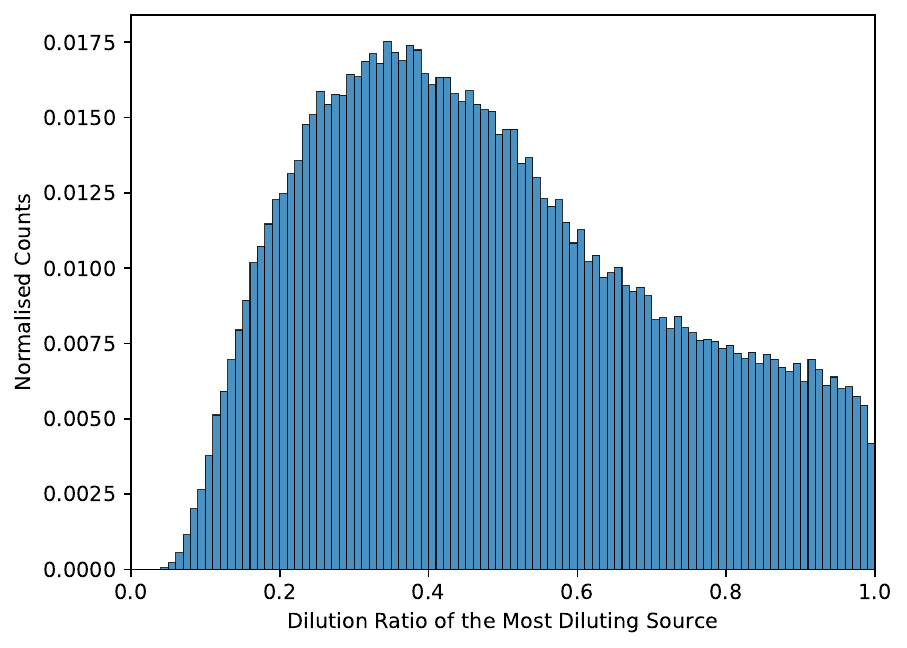}
    \caption{The distribution of the ratio of dilution contribution from the most diluting source in the target's aperture. A ratio of 1.0 means that there is only one diluting source.}
    \label{fig:NearbyFractions}
\end{figure}

However, we did not use our derived nearby fractions directly for our simulations as they correspond to the specific nearby stars of the target, while our framework effectively replaces them with a host of our choice. Instead, we quantified a distribution that expresses how possible diluting hosts contribute to the target's aperture in general. To do this, we selected the most diluting source for each target and then obtained the fraction they contribute to the total diluting flux in each sector. The mean of this fraction across all sectors was then used to establish our distribution, which is displayed in Figure \ref{fig:NearbyFractions}. As can be seen in the plot, the contribution of the most diluting sources ranges from about 0.1 to 1.0, with the latter end designating a single diluting source. The distribution peaks at a value of 0.35, suggesting that often the most diluting sources are only one of several weaker contributors. Effectively, sampling the distribution allows us to include the degree of crowding in the target's neighbourhood as part of our simulations, without being directly tied to the target's actual environment, while at the same time providing a realistically possible nearby fraction for a diluting host. Deriving this distribution also has the benefit that we can provide suitable nearby fractions for the on-target FP events, which becomes important for our ML training where the fraction is used as a feature. It should be noted that to determine the actual nearby fraction, the sampled fraction, denoted as $r_{ns}$, is multiplied with the fractional dilution for each target as per the following equation:
\begin{equation} \label{eq: nearby_fraction}
f_{ns} = (1-f_{t})*r_{ns}
\end{equation}

\begin{figure}
    \centering
    \includegraphics[width=\columnwidth]{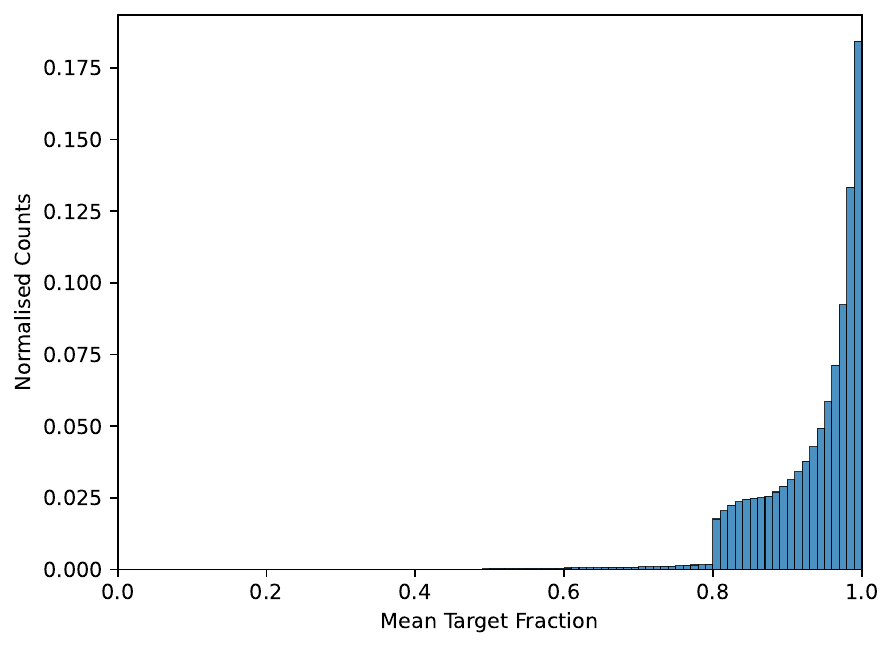}
    \caption{Distribution of the mean fraction of flux from the target star in the aperture across all sectors of observations, as determined by the SPOC pipeline.}
    \label{fig:TargetFractions}
\end{figure}

To ensure a sufficiently large sample of simulated NFPs, all targets in our sample were used, with each one assigned a randomly drawn base scenario simulation and a corresponding nearby fraction. To determine the detectability of the events during the sampling process, the mean target fraction across all sectors was used. The distribution of the mean target fraction for all target stars is shown in Figure \ref{fig:TargetFractions}. A characteristic feature of the distribution is the sharp drop seen at 0.8, which is due to the SPOC selection strategy for the FFI lightcurves that prioritises the least diluted stars. The total number of successful and rejected NFP simulations are provided in Table \ref{tb:Detection}. The detectability likelihood for each scenario is listed in the final column and it is important to highlight that this probability is the product of the success rate for the simulations and the likelihood of the corresponding base scenarios.

The depth of the diluted event $D_{dl}$ is calculated from the depth on the nearby source $D_{ns}$ by:
\begin{equation}
    D_{dl} = D_{ns} \frac{f_{ns}}{f_{t}},
\end{equation}

The injection process for the NFPs is identical to that of the on-target events, with the only exception the transformation of the simulated event to account for the effect of dilution. This is achieved through the following:
\begin{equation}
    Sim_{dl} = (Sim - 1) \frac{f_{ns}}{f_{t}} + 1,
\end{equation}
where $Sim_{dl}$ and $Sim$ denote the diluted and original simulated flux arrays. It should be noted that the creation process for the synthetic NFP lightcurves is aided by the fact that the SPOC lightcurves are already corrected for dilution from nearby sources, which allows us to ignore their contribution. Moreover, the above transformation process takes this dilution correction into account, with the resulting synthetic event replicating what would have been observed after the correction was applied. For further details on the correction, please refer to the \textit{Kepler} Mission Data Processing Documentation \citep{KeplerPDC}. 

\begin{table}
    \centering
    \caption{The total number of simulated systems using our modified \texttt{PASTIS} framework for each scenario and the number that resulted in a detectable signal. The ratio of simulations that succeeded to the total number of simulations provides the detection likelihood of each scenario, with the exception of the NTP, NEB and NHEB events for which the ratio is multiplied by the detectability likelihood of their base scenarios as they inherit it.}
    \label{tb:Detection}
    \begin{tabular}{c|cccc}
    \hline
        \textbf{Scenario} & \textbf{Total} & \textbf{Succeeded} & \textbf{Rejected} & \textbf{Detectability} \\
        \hline
        Planet & 4,645,720 & 115,511 & 4,530,209 & 0.0249 \\
        EB     &   677,740 & 134,080 &   543,660 & 0.1978 \\
        HEB    & 1,043,636 & 115,766 &   927,870 & 0.1109 \\
        BEB    & 1,207,059 & 116,538 & 1,090,521 & 0.0965 \\
        HTP    & 7,847,914 &  36,578 & 7,811,336 & 0.0047 \\
        BTP    &   865,110 &     405 &   864,705 & 0.0005 \\
        NEB    & 1,200,520 & 916,200 &   284,320 & 0.1510 \\
        NHEB   & 1,200,520 & 558,074 &   642,446 & 0.0516 \\
        NTP    & 1,200,520 &  68,049 & 1,132,471 & 0.0014 \\
        \hline
    \end{tabular}
\end{table}

\subsubsection{Box-Least-Squares Survey}\label{sec:BLS}

To match our collection of simulated events to the true population of \textit{TESS} candidates, we applied a BLS \citep{BLS} survey on each of the synthetic scenarios. Only lightcurves recovered through this BLS survey were then used for the training of the ML methods, thus ensuring that the training sets represent the underlying population of detected candidates. This recovery process additionally allowed us to quantify the recovery rate for each of our scenarios when using a BLS or similar period-search based detection methods. The recovery rate, along with the detectability of the scenarios discussed before, form an important part of our priors and are essential to the outcome of our pipeline. The priors and their role are discussed in more detail in Section \ref{sec:priors}. 

It should be noted that the exact recovery rate is dependent on the method and implementation of each survey. However, we consider that the results from our BLS survey act as a proxy for the general recovery rates of the different scenarios. Any differences between comparable, period-search based methods is expected to be minor, but this could be tested by future users. Therefore, the results from our pipeline should be valid for \textit{TESS} candidates retrieved with similar search methods.

\begin{figure}
    \centering
    \includegraphics[width=\columnwidth]{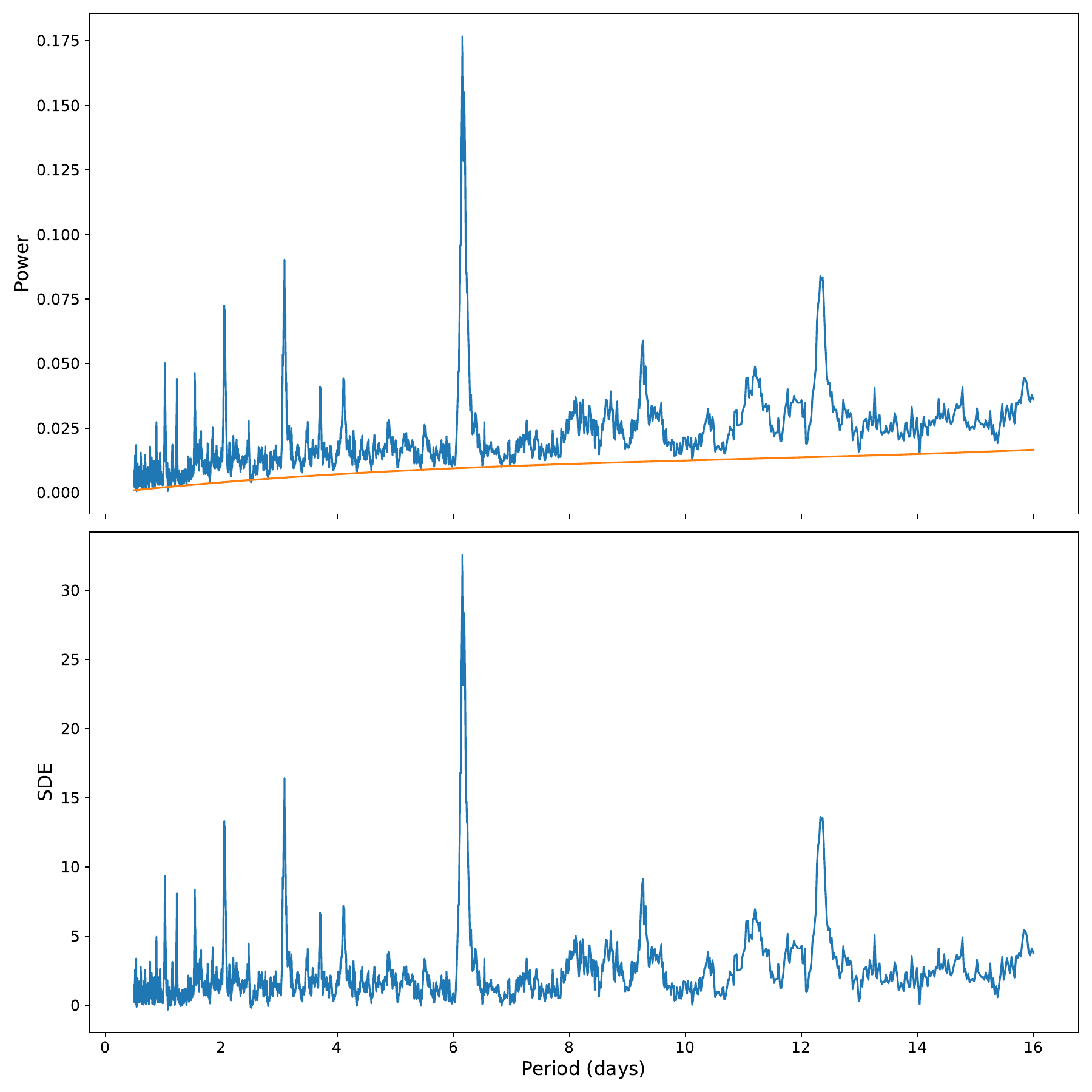}
    \caption{Top: BLS Power Spectrum for a synthetic planet event. The highest peak in the spectrum corresponds to the injected transit. The detrending cuve is overlaid in orange, showcasing its ability to capture the increasing trend of the spectrum. Bottom: The resulting SDE spectrum, following the detrending and normalisation of the power spectrum by its MAD. }
    \label{fig:bls}
\end{figure}

The BLS survey was implemented using the \texttt{cuvarbase} \citep{cuvarbase} python package running on a GPU. The synthetic lightcurves were detrended prior to running the BLS using a Savitzky-Golay filter \citep{SavitzkyGolay}, with a moving $3^{rd}$ degree polynomial and a 4 day window length. As the resulting power spectra from the BLS feature a gradual increasing trend towards longer periods, a minimum filter was used to detrend them. The filter was set up with a window length equal to 1/15th of the total data points, with the derived trend further smoothed using a $3^{rd}$ degree polynomial before detrending. The detrended power spectrum was then divided by its Median Absolute Deviation (MAD) to obtain the Signal Detection Efficiency (SDE) \citep{BLS}. An example BLS power spectrum, detrending curve and SDE spectrum is shown in Figure \ref{fig:bls}.

For the recovery process, we selected the 5 most significant peaks from each BLS spectrum and recorded their periods, transit duration and epoch. In an effort to reduce the number of alias peaks selected, we ignored any peak for which its double or half period is already selected. For each selected peak, we further calculated the depth of the detected events by phase-folding the lightcurve on the corresponding period and computing the difference of the mean in-transit flux with the mean out-of-transit flux spanning 1.5 transit durations on either side of the transit. The term transit is used here liberally to refer to any detected event regardless of its nature. 

In order to check if the injected events were recovered among the 5 selected peaks, we used the following 3 criteria:
\begin{enumerate}
    \item The peak period was within 2\% of the injected.
    \item The difference between the peak and injected epoch was less than 0.5 days. 
    \item The ratio of the injected and peak transit depth and vice-versa was less than 3.
\end{enumerate}
For the synthetic FP events involving binary configurations the transit epoch was checked against both the primary and the secondary eclipse, with the event counted as recovered if any of the two conditions was satisfied. Furthermore, if the peak period was found to be double or half of the injected period while still passing the other 2 criteria, the peak was labelled as a recovered alias of the injected event. 

\begin{figure}
    \centering
    \includegraphics[width=\columnwidth]{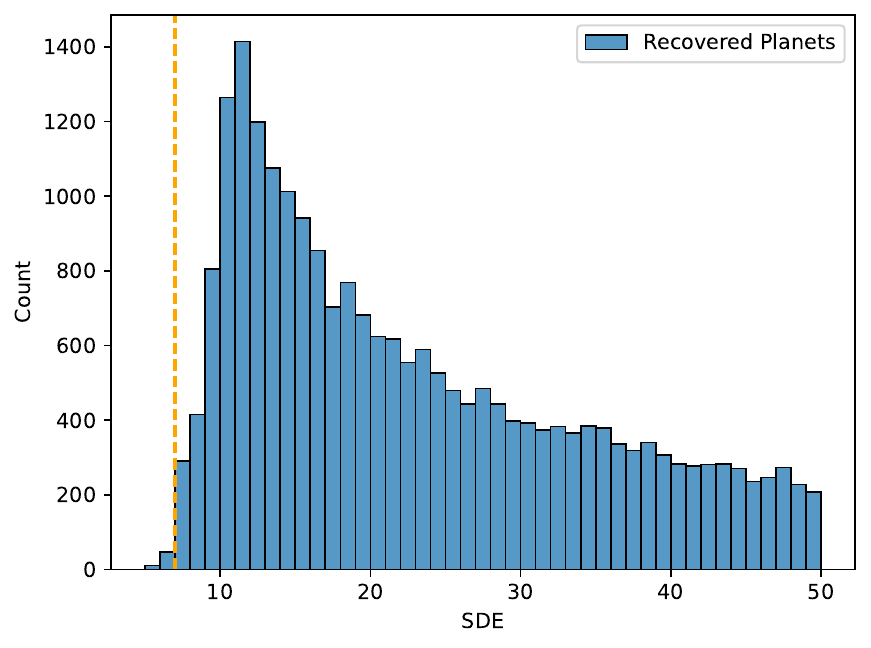}
    \caption{Distribution of the SDE for recovered injected planets from a BLS search without an SDE threshold. The distribution was limited for this plot up to an SDE value of 50. The dashed vertical line highlights the sharp drop of the number of recovered planets below an SDE value of 7.}
    \label{fig:sde}
\end{figure}

To determine the recovery rate and construct our training sets, the results for each synthetic lightcurve were reduced down to 1 peak, which was labelled as either 'Recovered' or 'Not Recovered'. Peaks in which the injected event was correctly detected at the true period were prioritised regardless of whether the peak had the strongest SDE. Period alias peaks were only selected if the true period was not recovered, with the strongest alias chosen if multiple had been detected. For the non-recovered events, the peak with the highest SDE was selected. Examining the SDE distribution of the recovered planets revealed a sharp drop in the number of recovered planets at an SDE value of 7, as shown in Figure \ref{fig:sde}. This threshold coincides with the value at which both the BLS and the Transit Least Squares (TLS) methods were found to cross the 1\% false positive recovery from white noise signals \cite{TLS}. As such, we adopted this limit for our recovery process and discarded any peaks that laid below it, in order to ensure that our training sets and recovery rates are representative of the expected results from such surveys. 

\begin{figure}
    \centering
    \includegraphics[width=\columnwidth]{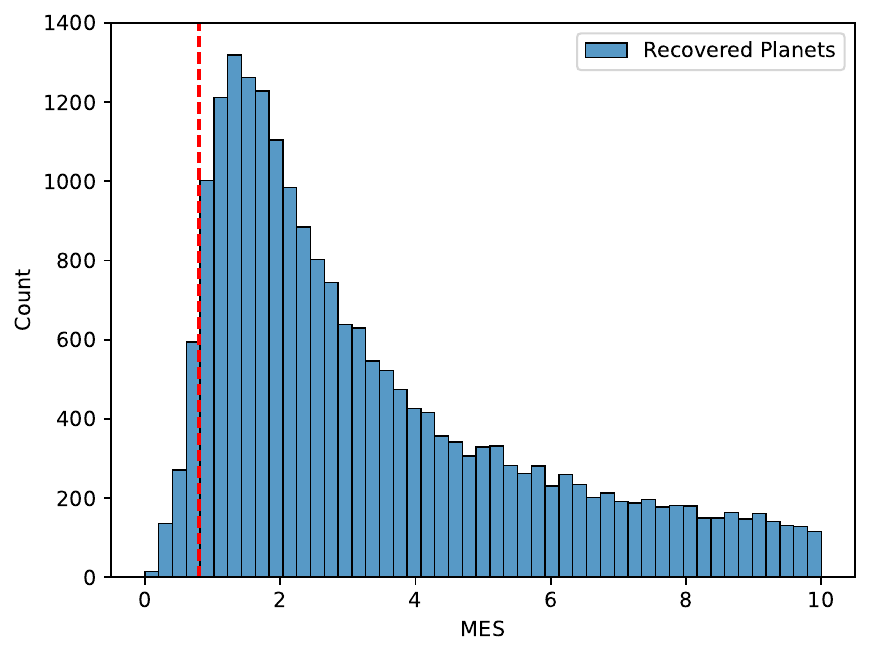}
    \caption{Distribution of the MES for the BLS recovered injected planets, highlighting the drop to the number of recovered planets below a value of 0.8. The value was selected as a threshold for which events, synthetic or real, are considered as recovered by the pipeline and is shown on the plot with a dashed vertical line. The distribution was limited for this plot up to a MES value of 10.}
    \label{fig:MES_cut}
\end{figure}

The \textit{TESS} SPOC pipeline uses a similar threshold value of 7.1 for its detection, albeit using a different significance statistic called the Multiple Event Statistic (MES). We implemented our own version of the MES, inspired by the one used by the SPOC pipeline, for our training features, the details of which are discussed in Section \ref{sec:Features}. We further incorporated this MES in our recovery process as an additional threshold for events passing the SDE limit. The benefit of this is that it allows us to clean up the training sample from low-significance, noise dominated and potentially incorrectly retrieved events. To do this we examined the MES distribution of the recovered planets, which is displayed in Figure \ref{fig:MES_cut} with a cut-off at a value of 10 for clarity. As can be seen in the plot, the distribution peaks at an MES value of about 1.8. This immediately highlights the difference in implementation between \texttt{RAVEN}'S MES and SPOC's. A significant drop at a MES value of 0.8 can be identified, which was chosen as the pipeline's recovery threshold. Events below the threshold were discarded, which for the planet scenario represented 3.7\% of the True recovered events and 40\% of the events in which the injected planet was not detected.

\begin{table}
    \centering
    \caption{The number of injected and recovered synthetic scenarios by the BLS survey. The resulting recovery rate encapsulates the probability to recover each scenario in the \textit{TESS} data and is used to inform our prior probabilities.}
    \label{tb:Recovery}
    \begin{tabular}{c|ccc}
    \hline
        \textbf{Scenario} & \textbf{Injected} & \textbf{Recovered} & \textbf{Recovery Rate} \\ \hline
        Planet & 115,511 & 44,907  & 0.389 \\
        EB     & 134,079 & 128,222 & 0.956 \\
        HEB    & 115,765 & 103,293 & 0.892 \\
        BEB    & 116,538 & 81,532  & 0.700 \\
        HTP    & 36,578  & 16,990  & 0.464 \\
        BTP    &    405  &    107  & 0.264 \\
        NEB    & 135,000 & 103,393 & 0.766 \\
        NHEB   & 135,000 & 75,074  & 0.556 \\
        NTP    & 68,049  & 26,241  & 0.386 \\
        \hline
    \end{tabular}
\end{table}

\begin{figure*}
    \centering
    \includegraphics[width=\textwidth]{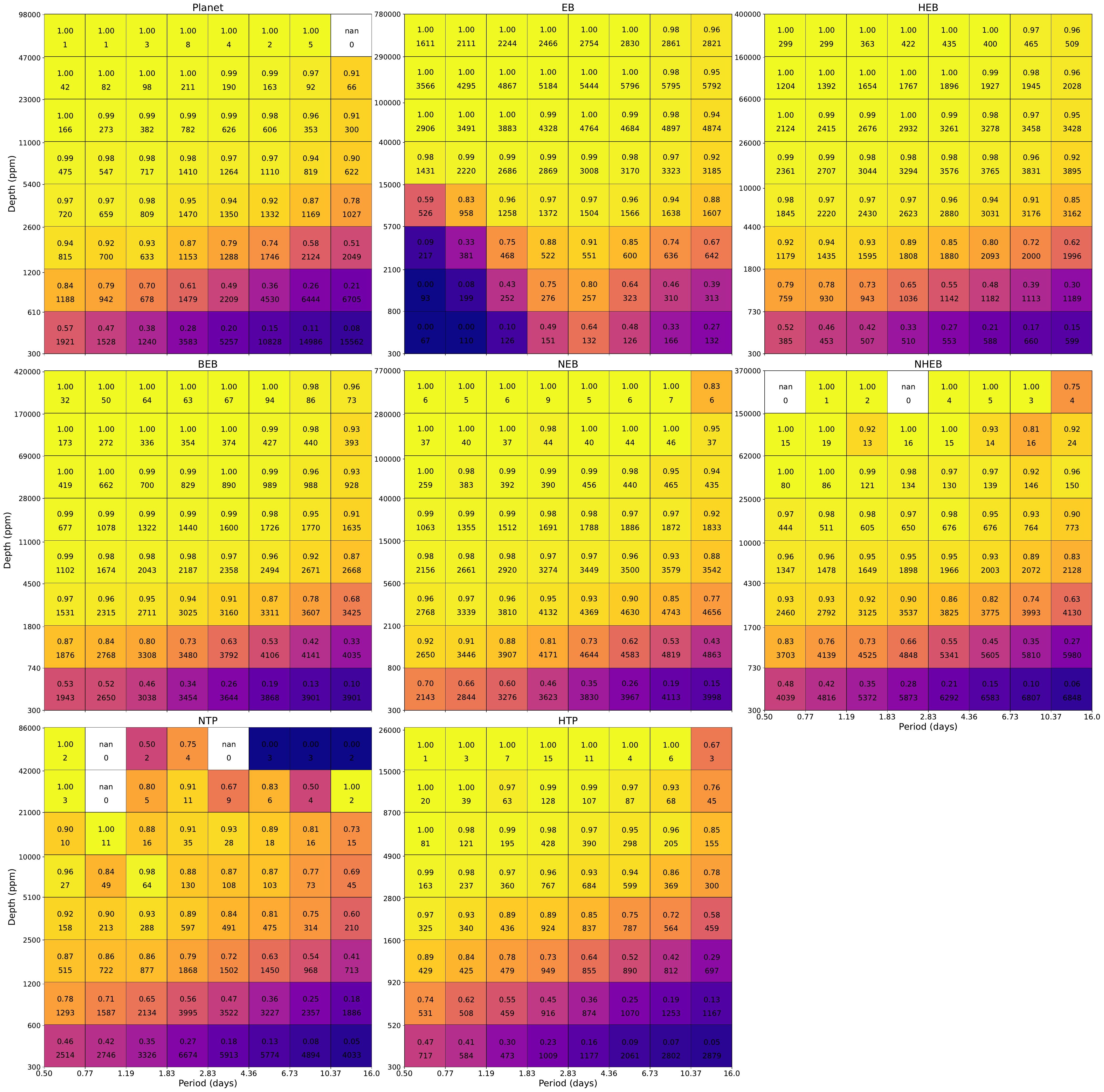}
    \caption{Heatmap of the BLS survey recovery rate for each synthetic scenario as a function of the period and depth of the injected events. Note the changing depth axis for each scenario.}
    \label{fig:recovery_plots}
\end{figure*}

Following the reduction of the detected events based on the above two restrictions, the recovery rates for the different synthetic scenarios were computed and are listed in Table \ref{tb:Recovery}. A finer breakdown of the recovery rates based on the period and depth of the events is presented in the form of heatmaps in Figure \ref{fig:recovery_plots}. As can be seen in the figures, the recovery rate for all scenarios follows a similar trend, with the BLS recovering the majority of objects with deep eclipses and at shorter periods. On the other hand, shallow events and events at longer periods were mostly not recovered, especially for scenarios with planetary transits.

\subsubsection{Non Simulated FP Training Set}\label{sec:NSFP}
Many of the BLS peaks discussed above do not correspond to an injected event. Instead these peaks are caused by the detection of stellar variability and of non-astrophysical noise features in the lightcurve. 
While the SDE and MES thresholds implemented for our survey reduce the occurrence of non-astrophysical detections, they do not completely eliminate them. As a result, we introduce a training set that allows for FP events not represented by one of the simulated scenarios.

For the creation of the training set, we run a BLS survey on a sample of 100000 \textit{TESS} targets with a SPOC FFI lightcurve and no injected simulations. We excluded all of the current TOIs to avoid recovering known events. Overall, the survey resulted in 218073 BLS peaks from 74615 targets which satisfied both the SDE and MES thresholds defined above. From the selected peaks, we removed any with a MES value greater than 100, as we found them to consistently correspond to eclipsing events, along with peaks for which the event depth was below 300ppm. This left a total of 177609 peaks distributed among 65907 target stars. We further reduced the number of detected events by randomly selecting one peak per target to form the final NSFP sample. It should be noted that despite our efforts to minimise them, this sample includes a small number of events represented by our simulated scenarios.

\subsubsection{Training Sets Disposition}\label{sec:train_disp}

\begin{table}
    \centering
    \caption{The sizes of the training, validation and test datasets used to train and optimise \texttt{RAVEN}'s ML classifiers for each planet-FP pair. Each dataset contains an equal number of Planet and FP events.
    }
    \label{tb:TrainingSets}
    \begin{tabular}{c|ccc}
    \hline
    \textbf{Planet-} & \textbf{Training} & \textbf{Validation} & \textbf{Test} \\
    \hline
    EB        & 69,572 & 8,698 & 8,696 \\
    HEB       & 69,572 & 8,698 & 8,696 \\
    BEB       & 69,572 & 8,698 & 8,696 \\
    NEB       & 69,572 & 8,698 & 8,696 \\
    NHEB      & 69,572 & 8,698 & 8,696 \\
    NTP       & 39,960 & 4,996 & 4,994 \\
    HTP       & 26,364 & 3,296 & 3,296 \\
    NSFP      & 69,572 & 8,698 & 8,696 \\
    \hline
    \end{tabular}
\end{table}

In the context of ML classification, balanced training sets are often preferred and found to perform better than imbalanced ones, as seen in \citet{QuasarClassification}. As such, for each planet-FP pair the number of simulated events per scenario were evenly matched by down-sampling the largest of the two. This meant that any preference of one scenario over the other stemming from its combined higher occurrence, detection and recovery probabilities is omitted from the models. Instead, this information is incorporated at a later stage through the scenario prior probabilities.

Training of the ML models was performed on 80\% of the available events for each planet-FP pair for the training of the models, with the remaining 20\% split into equal-size test and validation sets. The validation sets were used for refining the parameters of their corresponding models while the test sets were used only to evaluate their final performance. The class representation remained balanced in all sets of data. A breakdown of the number of events in the training, validation and test sets is provided in Table \ref{tb:TrainingSets} for all planet-FP classification pairs. The selection of events from those recovered by the BLS survey described in Section \ref{sec:BLS} is explained below.

There were 44,907 recovered synthetic planets from the BLS survey. We excluded all events where only the alias was recovered to ensure that the features properly represented the lightcurves of transiting planets. This led to the removal of 1424 alias events. Not removing them would have had unintended consequences due to skewing the features, especially those that focus on changes to the transit. 
 With their removal, it is important to note that any true planet candidates run through the pipeline with a period alias will need to be corrected to receive accurate scores. This left a total of 43,483 synthetic planet events available. These were then split into the training, validation and test sets. 

The alias events were also excluded for the HTP and NTP scenarios, to ensure that all scenarios involving transiting planets were consistent and that the classification would focus on the differences of the transits rather than any incorrectly retrieved period. 
We included recovered aliases for the FP scenarios involving simulated EBs, as they represented a far greater proportion of detections, ranging from 8.5\% to 15\%, and were much more significant compared to the planets. 

\subsection{Machine Learning Features}\label{sec:Features}
With the training sets produced, we generate the features on which the models are trained on. These features include both stellar specific characteristics, such as the radius and \textit{TESS} magnitude, and event specific characteristics, such as the period, duration and depth. More importantly, they encompass metrics extracted from the observed lightcurves. These range from the Signal-to-Noise (SNR) ratio, to odd-even transit comparisons and to more specialised features such as the Single Event Statistic (SES) and the Multiple Event Statistic (MES). 

A full list of the features used in the \texttt{RAVEN} pipeline, along with a brief description, is provided in Table \ref{tb:Features}. The majority of the features are similar in function to those use in A21. It should be noted that some of the features have been dropped for the \texttt{RAVEN} pipeline, while new ones have been added. These changes stem from the use of drastically different training datasets and from changes to our training methodology, which is discussed in Section \ref{sec:ML_training}.  The rest of this section offers a detailed breakdown on how the features are generated for both the synthetic training sets and \textit{TESS} candidates. Prior to computing the features, the lightcurves are detrended with the same Savitzky-Golay based process described in Section \ref{sec:BLS}, with the difference that the events are masked during the detrending.

\begin{table*}
    \caption{List of features used by the pipeline for the training of the ML classifiers. The features are listed with their internal names as provided by the pipeline's feature generation process, along with a brief description. Some of the features are not used at all for training, but only for the generation of other related features.\label{tb:Features}} 
\begin{tabular}{llr}
        \hline
        \textbf{Features} & \textbf{Description} & \textbf{Scenario} \\ \hline
        Tmag & \textit{TESS} Magnitude of target star. & All \\
        Gmag & \textit{Gaia} Magnitude of target star. & All \\
        Tmag-Gmag & Tmag-Gmag colour of target star. & All \\
        BP-RP & BP-RP colour of target star. & All \\
        dist & Distance to target star ($kpc$). & All \\ 
        rstar & Target star radius ($\rsun$). & All \\ 
        per & Orbital period of eclipsing event (days) from Pipeline input. & All \\ 
        t0 & Mid-eclipse epoch of the event (BTJD) from Pipeline input. & None \\ 
        tdur & Duration of the eclipsing event (days) from Pipeline input. & None \\  
        depth & Eclipse depth (ppm) from Pipeline input. & None \\ 
        fit\_aovrstar & Ratio of the orbital semi-major axis to host stellar radius from transit model fitting. & All \\ 
        fit\_rprstar & Ratio of companion radius to host stellar radius from transit model fitting. & All \\ 
        prad & Implied planet radius ($R_{\oplus}$). & All \\ 
        fit\_tdur & Duration of eclipsing event from trapezoid fitting (days). & All \\ 
        fit\_tdur23 & Duration of eclipsing event excluding the ingress and ingress from trapezoid fitting (days). & All \\ 
        fit\_depth & Eclipse depth of the event from trapezoid fitting (ppm). & All \\ 
        fit\_tdur\_per & Event duration from the trapezoid fitting divided by period. & All \\
        ingressdur & Ingress duration (days). & All \\ 
        grazestat & Ratio of full-eclipse duration to the eclipse duration. & All \\ 
        evenodd\_durratio & Ratio of eclipse duration for the odd and even numbered eclipses from trapezoid fitting. & All \\ 
        evenodd\_depthratio & Ratio of eclipse depth for the odd and even numbered eclipses from trapezoid fitting. & All \\ 
        robstat & A measurement of depth variation across all observations. & All \\ 
        SNR & Signal to Noise Ratio for the eclipsing event. & All \\ 
        max\_SES & Maximum value of the Single Event Statistic (SES) across all eclipses. & All \\ 
        min\_SES & Minimum SES across all eclipses. & None \\ 
        med\_SES & Median value of the SES across all eclipses. & None \\
        rSES\_med & Ratio of Maximum to Median SES. & All \\ 
        rSES\_min & Ratio of Maximum to Minimum SES. & All \\ 
        MES & Multiple-Event Statistic (MES) corresponding to the centre of the eclipse. & All \\
        med\_MES & Median value of the MES curve in the absence of the primary eclipse. & None \\
        min\_MES & Minimum value of the MES curve in the absence of the primary eclipse. & None \\
        mad\_MES & Median Absolute Deviation (MAD) of the MES in the absence of the primary eclipse. & None \\
        rmesmed & Ratio of the MES to the median MES. & NSFP-only \\ 
        rmesmad & Ratio of the MES to the MAD of the MES. & NSFP-only \\ 
        rminmes & Ratio of the minimum MES to the MES of the primary eclipse. & NSFP-only \\ 
        max\_secmes & MES of the most significant secondary event, in the absence of the primary eclipse. & All \\ 
        max\_secmesp & Phase of the potential secondary eclipse. & All \\ 
        rsecmesmad & Ratio of the MES for the potential secondary eclipse to the MAD of the MES. & All \\ 
        sec\_depth & Depth corresponding to the potential secondary eclipse (ppm). & All \\ 
        sec\_robstat & A measurement of depth variation for the secondary eclipse across all observations. & All \\ 
        albedo\_stat & A measurement of the significance of the geometric albedo as derived from the secondary eclipse depth. & All \\ 
        target\_fraction & The mean flux fraction from the target star in the aperture across all sectors of observation. & NFP-only \\ 
        nearby\_fraction & The mean flux fraction from the most contributing nearby source in the aperture across all sectors of observation. & NFP-only \\ 
        SOM\_stat & Statistic derived from the application of a SOM classification algorithm. & All \\
        SOM\_dist & Distance statistic related to the compatibility of the SOM classification. & All \\
            \hline
\end{tabular}
\end{table*}

\subsubsection{Stellar Metrics}\label{sec:stellar_metrics}
The features incorporate a variety of stellar characteristics in an effort to provide an insight into the nature of the event based on its source. Namely, these include the \textit{TESS} and \textit{Gaia} magnitudes, denoted as \textbf{Tmag} and \textbf{Gmag}, along with their derived colour (\textbf{Tmag-Gmag}) and their \textbf{BP-RP} colour. In addition, the distance (\textbf{dist}) to the target star and its radius (\textbf{rstar}) are also included. The \textbf{Tmag} and stellar radius are retrieved from the TIC, while the \textbf{Gmag} and \textbf{BP-RP} from \textit{Gaia} DR3. The distance feature is generated from its measured parallax, \textit{Plx}, in DR3, as per:
\begin{equation}
    \textbf{dist} = \frac{1000}{Plx},
\end{equation}

It is important to highlight that for the diluted "on-target" simulations, namely the HTP, HEB, BEB and BTP, the magnitude based features are computed by combining the target's magnitude with those of the additional stellar sources, as determined by the isochrone fitting. This is done to emulate the magnitudes that would be observed for such configurations, as the unresolved sources would appear as a single star. To that effect, the \textbf{BP} and \textbf{RP} magnitudes for the target are retrieved separately from DR3, so that they can be combined with those of the diluting sources before the \textbf{BP-RP} colour is computed. These adjustments ensure that the simulation values match the TIC and DR3 values for similar real stellar configurations. 

Overall, the stellar characteristics were confined to those that could be readily retrieved by photometry, as they could be more easily replicated for the diluted simulations. As such, we intentionally excluded potentially useful features such as the star's surface gravity and effective temperature, which require a more involved derivation. The only exception is the stellar radius, which remains unchanged from that of the target's for all simulations as it could not be trivially adjusted for the contribution of the unresolved stellar sources. This assumes that the flux from the target star in these configurations will dominate the observed photometry and thus the derived radius even for the diluted scenarios will be closely matching that of the target's. This is indeed the case for the majority of our simulations.
 
\subsubsection{Eclipse Parameters and Model Derived Metrics}\label{sec:model_metrics}
The pipeline makes use of the transit and eclipse parameters for initial estimates, namely the period (\textbf{per}), epoch (\textbf{t0}), transit duration (\textbf{tdur}) and the transit depth (\textbf{depth}). For the simulated events, these parameters are taken directly from their creation and used as is. For candidates, the parameters are provided as an input to the pipeline, in the case of this work from the TOI catalogue, and are then further refined using a fine-grained BLS along a small range around the input period. As the NSFP events are true FP \textit{TESS} candidates they undergo the same process, with their input event parameters taken directly from the BLS results.

Since the event parameters differ in their derivation, only the period is included in the ML features, as any small differences in period caused by the derivation are irrelevant to FP classification. The rest of the parameters are used to perform a least-squares fit of the data with a transit model using the \texttt{batman} package \citep{batman}. The transit model assumes a circular orbit with a quadratic model for the limb darkening. Even though the events, both simulated and real, will often be non-circular, the consistency of the model fit between the two sets ensures that the results are relevant for the purpose of ML classification. The transit model is fit to the data to obtain a value for the orbital semi-major axis to stellar radius ratio, \textbf{fit\_aovrstar}, and the ratio of the companion's radius to that of the host star, \textbf{fit\_rprstar}. It is important to clarify that the values obtained from the fit should not be taken as the actual physical values for the system, but rather metric features to be used for the classification. The input event depth, period and transit duration are used to provide estimated initial values for the two parameters using the following relations:
\begin{equation}
    \frac{R_{p}}{\rstar} = \sqrt{D},
\end{equation}
\begin{equation}
    \frac{\alpha}{\rstar} = \frac{1 + \frac{R_{p}}{\rstar}}{\sin(\pi \frac{d}{p})},
\end{equation}
where $R_{p}$ denotes the companion radius, $\rstar$ the host star's radius, $D$ the event depth (\textbf{depth}), $\alpha$ the semi-major axis and $d$ and $p$ the transit duration (\textbf{tdur}) and period (\textbf{per}) respectively. In case the model fit fails, we fall back to the estimated values for the features. It should be noted that during the fit, we let both the period and epoch fit with bounds of 1\% of the input value, but we do not make use of the results. Finally, the computed \textbf{fit\_aovrstar} metric is used to obtain the implied planet radius, \textbf{prad}, based on the target's stellar radius. 

The events are further fit using a trapezoid model. The model uses a fixed period and epoch, fitting only for the transit duration (\textbf{fit\_tdur}), the full-eclipse duration (\textbf{fit\_tdur23}) and the depth (\textbf{fit\_depth}). The full-eclipse duration refers to the portion of the eclipse between the ingress and egress. 
The fit is performed on the phase-folded lightcurve, with the duration parameters expressed in terms of phase. We use the input eclipse parameters as the initial values for the fit. The fit results are used throughout the rest of the process to compute the remaining features. As such, if the trapezoid fit fails the event is not processed further. 

The ratio of the fitted transit duration to the period (\textbf{fit\_tdur\_per}) is used as another feature for the models. We further derive the ingress duration (\textbf{ingressdur}) from the difference between the eclipse duration and the full eclipse duration. The ratio of the two is used to compute the grazing statistic (\textbf{grazestat}), which relates to the shape of the eclipse. We then split the lightcurve into two portions containing the odd and even numbered eclipses and refit each part separately with a trapezoid model to compute the ratio of their duration (\textbf{evenodd\_durratio}) and depth (\textbf{evenodd\_depthratio}). Finally, the trapezoid model for the full lightcurve is used to compute the Robust Statistic (\textbf{robstat}), which is a measurement of the significance of depth variations across all the eclipses. For a detail description of this statistic, please refer to Appendix A of \citep{Robstat} from which our implementation is based on. The statistic is computed in the \texttt{RAVEN} pipeline as follows:
\begin{equation}
    \textbf{robstat} = \frac{\textbf{m}^T \textbf{W} \textbf{f} }{\sqrt{\textbf{m}^T \textbf{W} \textbf{m} }},
\end{equation} with $\textbf{m}$ denoting the model array, $\textbf{f}$ the flux data and $\textbf{W}$ the fit weights. The weights are computed through:
\begin{equation}
    \textbf{W} = diag \frac{\mathbf{f}-\textbf{m}}{\boldsymbol{\sigma}},
\end{equation}
where $\boldsymbol{\sigma}$ represents the flux error array. For our purposes, we set the flux error as equal to the MAD of the flux in each sector of observation. Only the in-transit data points are used for the calculation of the Robust Statistic, to avoid stellar variability and other noise features of the lightcurve affecting the result.

\subsubsection{Eclipse Significance}\label{sec:significance}
The next set of features all relate to the significance of the event. The first is the \textbf{SNR}, which we derive as:
\begin{equation}
    \textbf{SNR} = \frac{D}{\sigma} \sqrt{N},
\end{equation}
with $D$ denoting the depth of the transit, $\sigma$ the MAD for the out-of-transit flux and N the number of in-transit data points. 
The rest of the significance features stem from the Single Event (SES) and Multiple Event (MES) statistics. These two features are inspired by the similarly named metrics used for the Transit Planet Search (TPS) in the \textit{TESS} SPOC pipeline \citep{SPOC}, however the implementation is different. 

\begin{figure*}
    \centering
    \includegraphics[width=\textwidth]{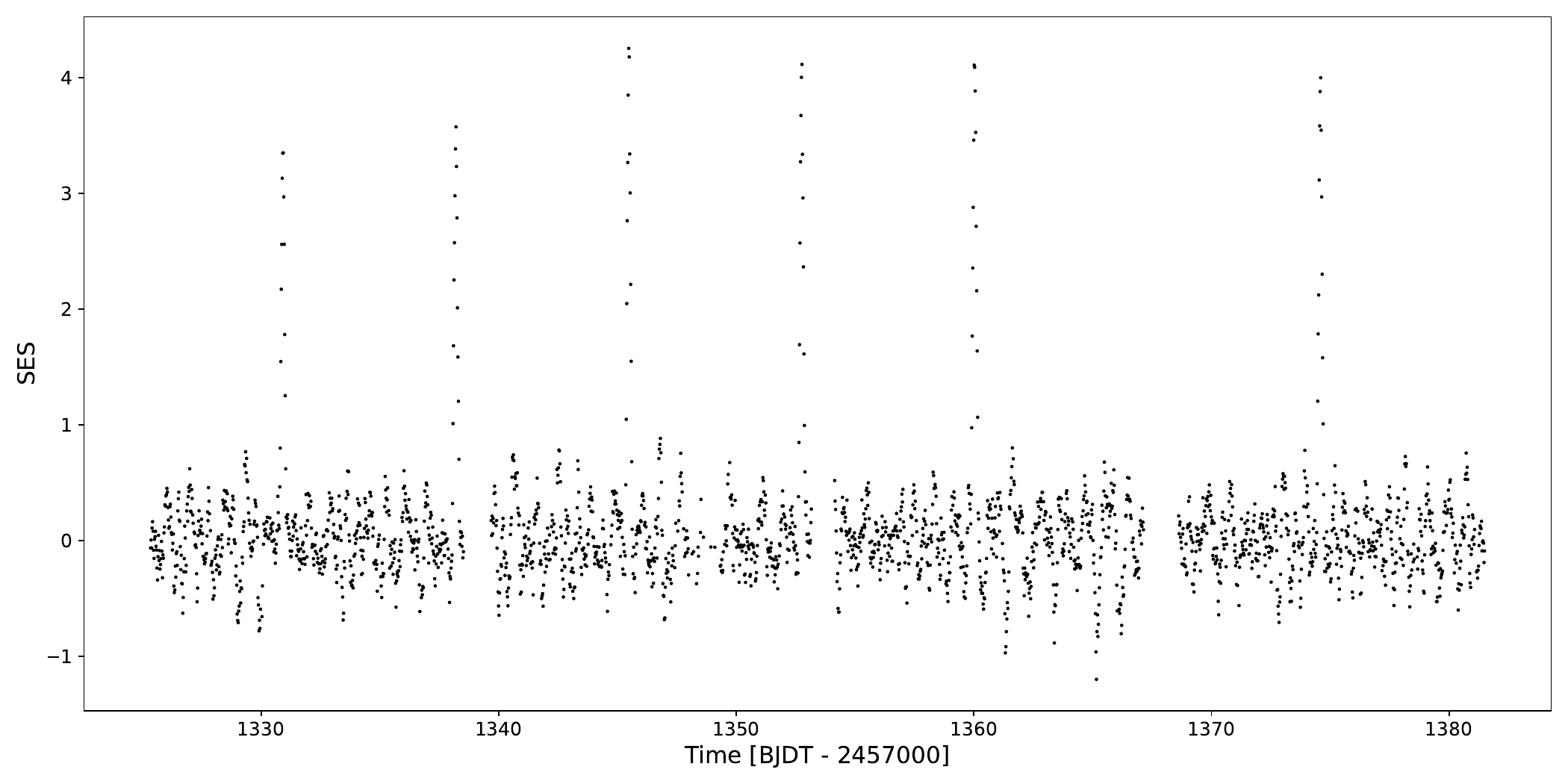}
    \caption{Single Event Statistic curve for a synthetic planet event. The peaks correspond to the injected transits.}
    \label{fig:SES}
\end{figure*}

The SES is implemented as the convolution of the flux timeseries with a square pulse that has depth and duration equal to that of the observed event. Prior to the convolution, the baseline for both the flux and the square pulse is set to 0 and the flux is further normalised by its MAD. The resulting convolution timeseries is again divided by the square root of the MAD to obtain the final SES timeseries. The overall process is expressed by the following equation:
\begin{equation}
    \textbf{SES} = \frac{\frac{\textbf{f}}{\sigma} \bullet \textbf{s}}{\sqrt{\sigma}},
\end{equation}
where $\textbf{f}$ is the flux array, $\sigma$ the MAD for the whole flux and $\textbf{s}$ the square pulse signal.
An example SES timeseries is shown in Figure \ref{fig:SES}. We derive three metrics from the timeseries, by recording the maximum SES value for each in-transit peak. These metrics are the maximum value across all peaks (\textbf{max\_SES}), along with its ratio against the median (\textbf{rSES\_med}) and minimum (\textbf{rSES\_min}) peak SES value. The two ratio metrics are used to signify variations in the depth of the event across the whole observation. 

\begin{figure}
    \centering
    \includegraphics[width=\columnwidth]{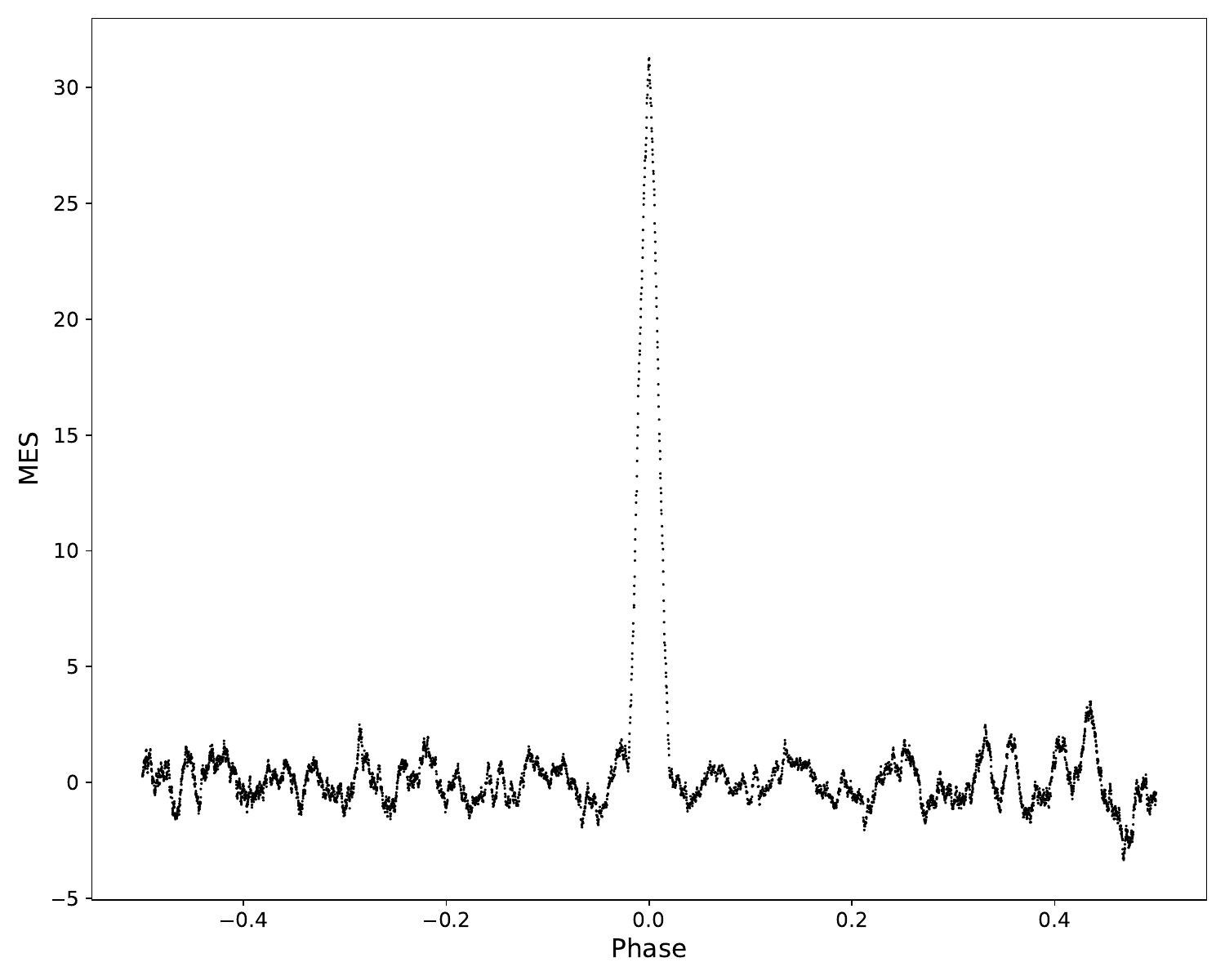}
    \caption{Multiple Event Statistic curve for a synthetic planet event. The curve is phasefolded on the injected transit.}
    \label{fig:MES}
\end{figure}

For the MES, the computation is the same as for the SES, with the only difference being that the lightcurve is folded in phase prior to the convolution. This boosts the significance of the event due to its periodicity. An example of the resulting MES curve is shown in Figure \ref{fig:MES}. The \textbf{MES} metric used in the features is computed as the maximum value of the MES curve within the transit duration in phase. The in-transit flux of the detected event is subsequently removed from the lightcurve and the MES computation is repeated, producing the secondary MES curve. From the secondary MES curve we obtain its median (\textbf{med\_MES}) and minimum (\textbf{min\_MES}) value, along with its MAD (\textbf{mad\_MES}). These 3 features are not directly used in the training, replaced instead with their ratio against the \textbf{MES} metric, denoted as \textbf{rmesmed}, \textbf{rminmes} and \textbf{rmesmad} in the feature table respectively. The 3 ratios are valuable to the training of the models as they express the significance of variability in the lightcurve, such as additional brightening and dimming events, in comparison to the detected event.

The maximum value in the secondary MES, denoted as \textbf{max\_secmes}, is treated as a potential secondary eclipse. As such we limit the maximum in the regions between phases of +0.25 and +0.5 and -0.25 to -0.5, with the primary centred at 0. We use the phase corresponding to this maximum as another metric (\textbf{max\_secmesp}), along with the ratio of the secondary maximum against the MAD of the MES as a measure of its significance (\textbf{rsecmesmad}). Furthermore, we fit a trapezoid model to determine the depth of the potential secondary. For this fit, the duration parameters are initialised based on the primary ones and allow to vary only by 5\%, while the depth is initialised as 0. In addition, we use the secondary eclipse depth to determine the geometric albedo of the candidate and compute whether it is statistically consistent with a self-luminous object (\textbf{albedo\_stat}).

\subsubsection{Nearby Metrics}\label{sec:nearby_metrics}
The \textbf{target\_fraction} and \textbf{nearby\_fraction} metrics are only relevant for the classifiers involving the Nearby FP scenarios. The two metrics are discussed in detail in Section \ref{sec: NFP}, as they are essential in the creation of the synthetic NFP events. For the 3 NFP scenarios the metrics are taken directly from their parameters. However, the metrics need to be computed for the synthetic Planet sample and the \textit{TESS} candidates. Starting with the \textbf{target\_fraction}, it as again taken as the mean of the per sector flux fraction from the target star in the aperture as determined by the SPOC pipeline. In the case of simulated Planets, as the event is already on target, the \textbf{nearby\_fraction} is sampled based on the distribution in Figure \ref{fig:NearbyFractions}. For real candidates we use their known nearby sources and estimate their flux contribution in each sector, as discussed in Section \ref{sec: NFP}. The \textbf{nearby\_fraction} is then taken as the mean fraction of their most diluting source.

\subsubsection{Self Organising Map}\label{sec:SOM}
Finally, an important feature for the pipeline is produced from an unsupervised clustering algorithm, called a Self-Organising-Map (SOM) \citep{kohonen1982SOM}, which we trained to group together events based on the shape of the eclipse. The SOM method was first used for exoplanet classification in \cite{SOM_Dave}, where it was applied on \textit{Kepler} and K2 candidates. The implementation described here is based on the method presented in \cite{SOM_Dave}, although a simpler form of the SOM-extracted statistics is used.

\begin{figure}
    \centering
    \includegraphics[width=\columnwidth]{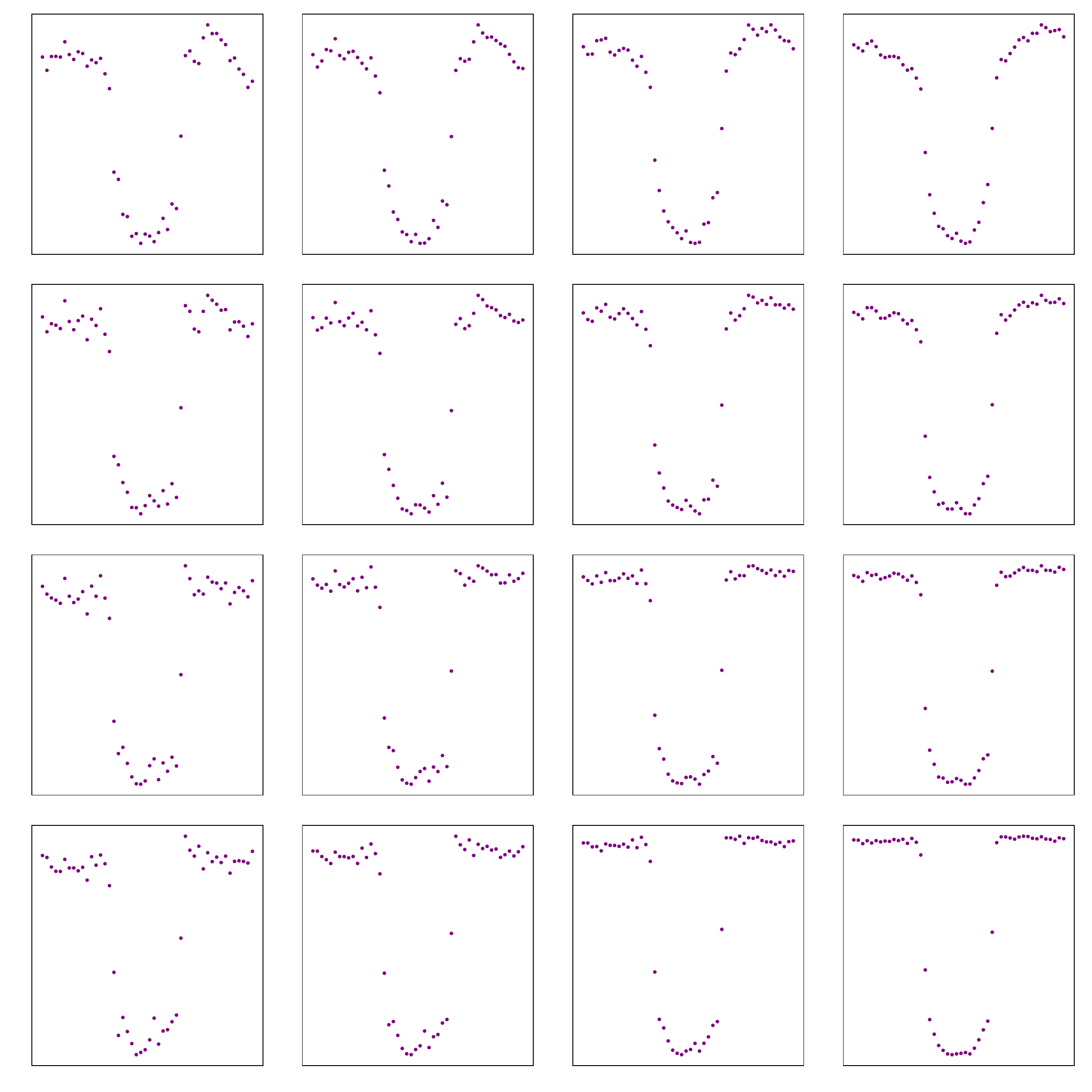}
    \caption{Cutout of the SOM pixel map trained on the synthetic lightcurves for the Planet and EB scenarios, showcasing the mean eclipse shape of the events contributing to each pixel. The transits and eclipses have been normalised, to remove any depth dependency and binned over 48 bins.}
    \label{fig:SOM_Cutout}
\end{figure}

For both the training and the application of the SOM the lightcurves are phasefolded and cut so that only the eclipse and its immediate surrounding flux, spanning 1.5 times the duration in phase on either side, remain. This cutout flux is then binned over 48 bins, with a a 5$\sigma$ outlier threshold used for the flux in each bin to exclude them from the calculation. The binned flux cutout is subsequently normalized, so that the mean of the highest quarter of data is 1 and that of the lowest quarter is 0. This normalization effectively removes any depth related information from the data, so that the only differentiating factor would be the shape of the event. We trained the SOM on the binned eclipsing events to produce a 20x20 pixel map, where each pixel represents the mean eclipse shape of similarly shaped events. One SOM is trained for each Planet-FP scenario pair, with a total of 9 maps produced. A 4x4 cutout of the SOM pixel map for the Planet-EB pair is shown in Figure \ref{fig:SOM_Cutout}, showcasing the mean eclipse shape of each pixel. For the training of each SOM we use an equal number of events for the planet and the FP scenario to ensure a balanced representation. 

\begin{figure*}
    \centering
    \includegraphics[width=\textwidth]{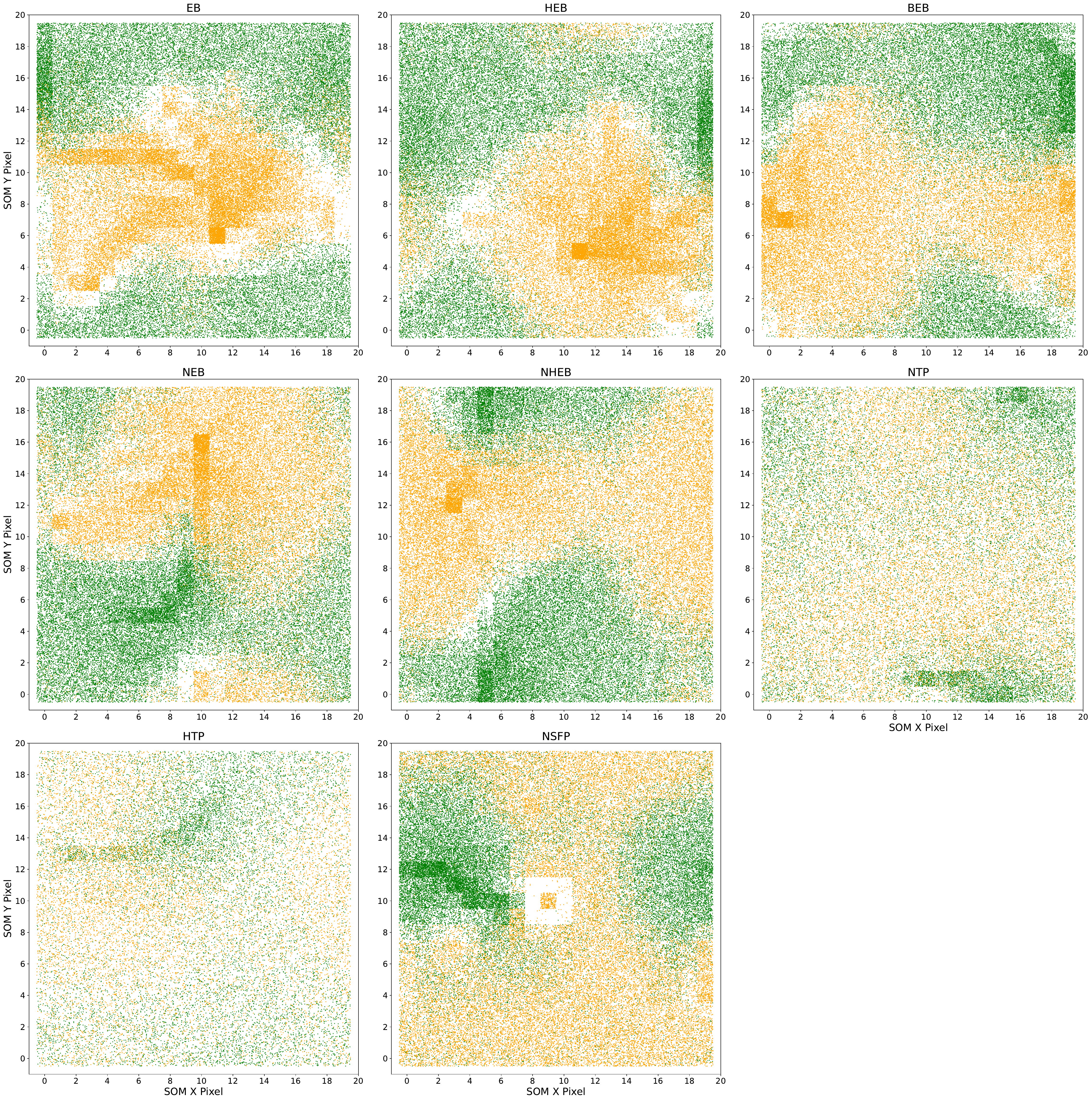}
    \caption{The result of applying the trained Planet-FP SOMs on their respective training data. The plots showcase the position of each training event on the SOM pixel map with some added random scatter up to 0.5 from the position to enhance clarity. The synthetic planet events are plotted in green and each respective FP event in orange. The plots highlight the clustering of the similarly shaped events and thus the effectiveness of each SOM in distinguishing between the Planet and FP candidates. As expected, the SOM is highly effective for EB based FP scenarios and increasingly less effective for diluted planet scenarios.}
    \label{fig:SOM_Mosaic}
\end{figure*}

The resulting SOMs are then applied on the training sets upon which they were trained. The algorithm assigns each event to the pixel that best matches the shape of their eclipse. This naturally forms clusters of similarly shaped events, as seen in the plots of Figure \ref{fig:SOM_Mosaic} where the applied SOMs for each of the 9 classification pairs are shown. The FP events are plotted with orange and the planets with green. As can be readily surmised from the plots, the ability of the algorithm to separate the planets and the FP scenarios varies per classification pair, with the planets well separated from the EBs and with less discernible separation for the HTPs and NTPs. This behaviour is not unexpected as for the latter cases we are still comparing planetary transits, even though they are diluted. 

The application of the SOMs on the training data allows us to derive the \textbf{SOM\_stat} metric that we use in the subsequent ML models. This metric expresses the ratio of planets to FPs in the pixel in which an event is assigned to. For example, an event in a pixel consisting fully of planets will get a \textbf{SOM\_stat} of 1.0, whereas an event landing in a pixel with only FPs will get a statistic of 0. Furthermore, we compare the binned eclipse array of each event to the mean eclipse array of the pixel to compute a distance statistic in the form of:
\begin{equation}
    \textbf{SOM\_dist} = \sqrt{\sum (\Bar{x}-x)^2},
\end{equation}
with $\Bar{x}$ denoting the mean eclipse array from the SOM and $x$ that of the event. It should be noted that since we train one SOM per planet-FP classification pair, the values of the SOM statistic and SOM distance metrics for the synthetic Planet events are specific and unique to each classification pair. Thus, each synthetic planet has 9 different associated SOM statistic and SOM distance metrics. As a result, the two features are computed just before the training of the classifiers and not the feature generation step. This is also the case for the \textit{TESS} candidates, as discussed later. 

\subsection{Machine Learning Training}\label{sec:ML_training}
Following the feature generation for the planets and FP scenarios, two ML classifiers were trained, namely a Gradient Boosted Decision Tree (GBDT) and a Gaussian Process (GP), per planet-FP scenario pair. As discussed in A21, the use of multiple ML methods ensures that the classification is robust and not reliant on just one method. Whilst A21 used four classifiers, we elected to use two due to the larger training sets and the need to train them for each planet-FP pair. The GP was chosen as it was the best performing for A21, while the GBDT essentially represents an upgrade over the Random Forest Classifier, the second best performing model of A21. The methodologies of these two classifiers are distinctively different, which enhances the pipeline's robustness and resilience against overfitting. The two classifiers are briefly described below.

\subsubsection{Gradient Boosted Decision Trees}\label{sec:GBDT}
\begin{figure*}
    \centering
    \includegraphics[width=\textwidth]{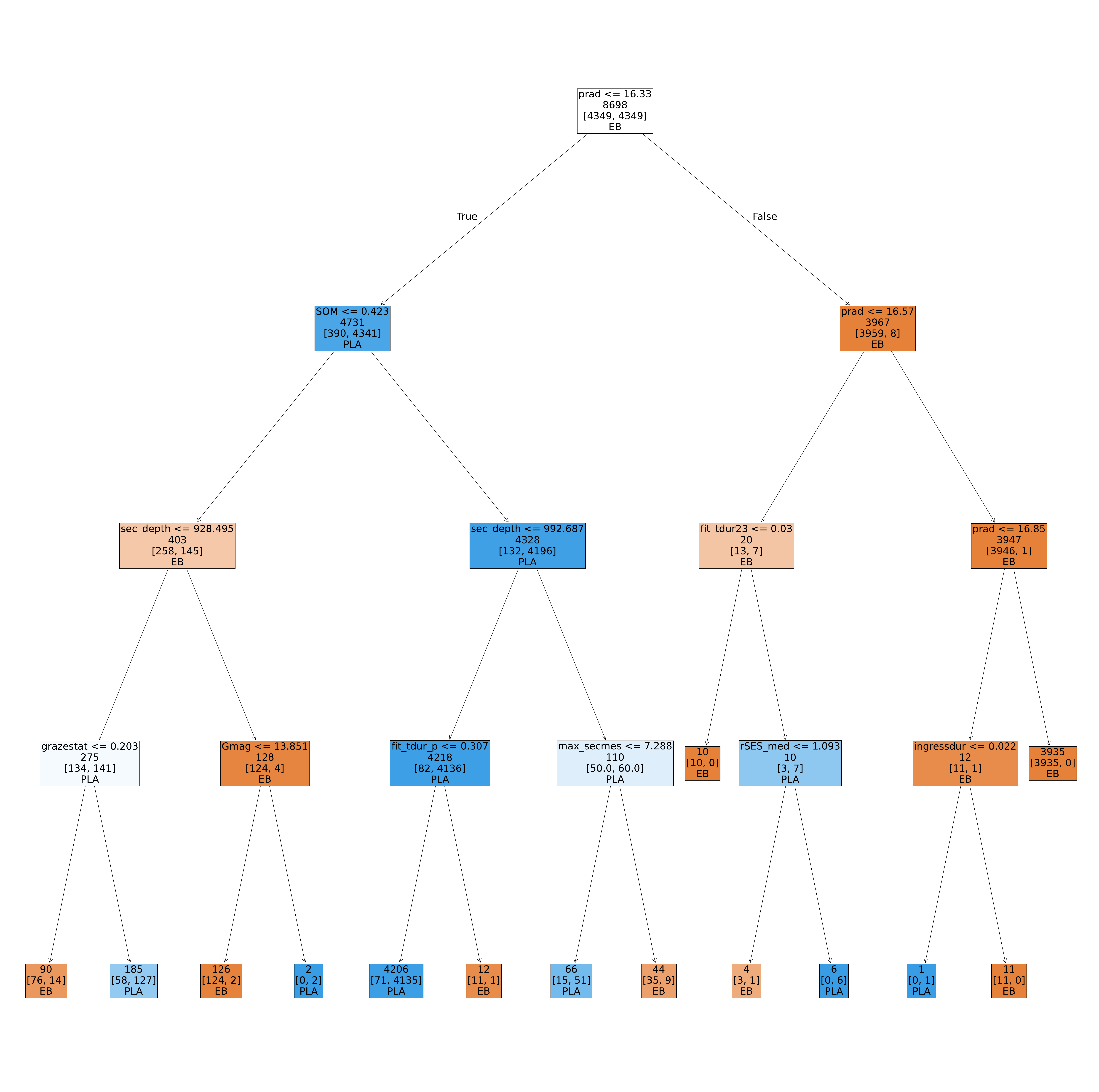}
    \caption{An example Decision Tree trained on the simulated Planet and EB training sets, limited to a depth of 4 nodes. The colours of the nodes represent the proportion of samples of each class in the node after splitting, with Planets in blue and EBs in orange. The node colour tends to white the more balanced the numbers are.}
    \label{fig:Tree}
\end{figure*}

Decision Trees are simple and yet formidable ML models, which use a combination of threshold values for the features of the datasets to split the samples into classes. The models are structured in the form of branching trees, where in each branch node a feature threshold is used to split the data, with the model branching out and growing in depth. At the end of the branch nodes are leaf nodes, which represent a class label to be assigned to the samples that fall within the branch. An advantage of Decision Trees is their interpretability, as cascading threshold values of deterministic features are an easy to follow classification method. An example Decision Tree trained on the simulated Planet and EB training sets is shown in figure \ref{fig:Tree}, to showcase the structure of the model. 

While useful, Decision Trees can be limited in their robustness and are also liable to overfitting, especially when too large in depth. These weaknesses are usually addressed by employing an ensemble of "weaker" trees. Gradient Boosted Trees \citep[for gradient boosting see][]{friedman2001gradboost} is one such ensemble method, which combines multiple, sequentially built trees to construct an enhanced final model. Its distinctive feature is that in each iteration the new tree model is trained based on the residual errors from applying the ensemble model in the previous iteration rather than the class labels themselves. As such, the goal of each new model in the ensemble is to minimise the overall loss of the model, in what is essentially a gradient descent. The results of each trained model in the ensemble are added together, scaled by a factor called the learning rate. The loss of the model is computed based on a loss function, with the residuals computed from the gradient of the function. In our pipeline, the GBDT classifier is implemented using the \texttt{XGBoost} model, introduced in \cite{xgboost}.

\subsubsection{Gaussian Process Classifier}\label{sec:GPC}

A Gaussian Process (GP) is a stochastic process that generalises the Gaussian Probability Distribution from a distribution over random variables to a distribution of functions \citep{GP}. A detailed review of the applications of GP for data analysis across a various astronomical disciplines is offered in \cite{GP_Astronomy}. Although typically used for regression, GPs can also be used in classification problems, as we do here. In the case of GP classification, the desired outcome is a discreet class label or a class probability between 0 and 1. To do this, a response function is applied on the output of the GP to force its output in the range between 0 and 1 as described in \cite{GP}. This is then combined with a probabilistic likelihood, such as the Bernoulli likelihood. However, as the likelihood is no longer Gaussian, the marginal likelihood can not be analytically determined and requires approximation. Several approximation methods exist. In this work we use a Variational approximation, as presented in \cite{hensman2015scalable}. This approximation relies on a set of inducing points, a representative subset of the data on which the GP is initially trained, to reduce the computational load and enhance its scalability.

\subsubsection{Training Set Pre-processing}\label{sec:transform}
Prior to the training, the features were scaled using the Quantile Transformer from \texttt{scikit-learn}. Scaling the data is often a beneficial process for the performance of the ML models, especially in the presence of outliers in the features. The transformer produces a normal distribution for each feature, with the number of quantiles set to 1000 for all datasets. The scaling algorithm was fit on the training data for each planet-FP pair and used for both the GBDT and the GP. While the performance of the GBDT classifier was found to be largely unaffected by the scaling, the GP was unable to be trained without the scaling applied. Only 3 features were excluded from scaling, namely the phase of the significant secondary, the grazing statistic and the SOM statistic, which are all already limited to between 0 and 1.

\subsubsection{Training and Calibration}\label{sec:train_cal}
To train and optimally tune the two classifiers an iterative process was followed, with the models trained on the synthetic training sets with different combinations of hyperparameters and their performance evaluated on the validation sets to select the best values. The tuning of the parameters was focused on 3 key FP scenarios, namely the EB, NEB and NSFP as they represented the most commonly detected FP events. While optimising, we kept parameters consistent across all scenarios to avoid overly optimising the models on each scenario and thus increasing the risk of overfitting.  

Starting with the GBDT, the classifier was implemented using the \texttt{XGBoost} model. A Logarithmic Loss (logloss) function was chosen for the model, as it is optimal for producing a better calibrated model \citep{silvafilhoClassifierCalibrationSurvey2023}. The model parameters that were tuned, and selected values, are listed in Table \ref{xgb_params}. Of particular interest were the subsampling parameters, namely \textbf{subsample}, \textbf{colsample} and \textbf{treesample}, which were used to inject randomisation into the tree boosting method. This randomisation makes the method more closely aligned to a Random Forest classifier \citep{breiman2001random} and reduces the risk of overfitting. 

Early Stopping was enabled for all models to guard against overfitting by ensuring that the training would stop as the performance of the model reached a plateau. This is achieved by monitoring the minimisation of the loss function as computed on the validation set in each iteration. If the value of the loss function did not decrease by at least 0.0001 during the last 20 iterations, the training of the model is terminated early, with the model state restored to the iteration in which the loss fuction had last been improved. 

\begin{table*}
    \centering
    \caption{Tunable model hyperparameters for the GBDT classifier. Final used values are shown in bold.}
    \begin{tabular}{l|p{11cm}|l}
    \hline
        \multicolumn{1}{l|}{\textbf{Parameter}} & 
        \multicolumn{1}{c|}{\textbf{Description}} & 
        \multicolumn{1}{c}{\textbf{Trial Values}} \\ \hline
        
        \textbf{n\_estimators} & Maximum number of iterations in which to train and add new trees to the ensemble model. & 500, 1000, 2000, 4000, \textbf{8000}, 20000 \\ 
        \textbf{ntrees} & Number of trees trained in parallel in each iteration and added to the ensebmle. & \textbf{0},10,100 \\ 
        \textbf{max\_depth} & Maximum depth level a tree can grow to. & 4,5,\textbf{6},8,10 \\ 
        \textbf{learning-rate} & Contribution of each tree in the ensemble. & \textbf{0.01}, 0.1 \\ 
        \textbf{subsample} & Number of training samples used to train each tree. & 0.5, \textbf{0.8}, 1.0 \\ 
        \textbf{colsample} & Number of randomly selected features evaluated in each tree.  & 0.25, \textbf{0.5}, 0.75, 0.8, 1.0 \\ 
        \textbf{treesample} & Number of randomly selected features evaluated in each node.  & 0.25, \textbf{0.5}, 0.75, 0.8, 1.0 \\ 
        \textbf{Early Stopping} & Controls whether the model will stop training and adding trees to the ensemble after the loss function minimisation plateaus.  & \textbf{True}/False \\ \hline
    \end{tabular}
    \label{xgb_params}
\end{table*}

For the GP, a scaled Rational Quadratic (RQ) Kernel was selected as it was found to perform better on the validation set compared to the Matern and Radial Basis Function (RBF) Kernels. The RQ Kernel is defined as \citep{GP}:
\begin{equation}
    RQ = A\left(1 + \frac{r^2}{2\alpha l^2} \right)^{-\alpha}
\end{equation}
with $A$ a scaling parameter for the whole kernel, $r=|x_{i}-x_{j}|$ the difference between two data points, $l$ the characteristic lengthscale and $\alpha$ the kernel's weighting parameter. The characteristic lengthscale was set to be independent for each training feature, allowing their significance to greatly influence the classification. Essentially, features with greater significance have a shorter lengthscale, with small variations in their value causing greater change to the classification result. On the other hand, less significant features have a larger lengthscale and thus differences in their values contribute less to the final outcome. 

To set up the GP, we used the \texttt{GPyTorch} package to create a variational GP classifier that initialises with a set of inducing points. The inducing dataset was constructed by fitting a K-Means clustering algorithm on the training data of the planet and FP scenario separately to obtain an averaged representation of the two datasets, based on the mean feature values in each cluster. An equal number of clusters were created for the two scenarios. The total size of the inducing dataset was an additional optimisable hyperparameter for the model that affected both the training time and overall performance of the model. The final size of the inducing set for each scenario is listed in Table \ref{tb:GP_Training}. 

The training of the GP is essentially the search for the optimal mean and kernel function parameters that best fit the data and allow them to be divided between the two classes. This is done by employing a gradient descent strategy, where the model is iteratively trained and evaluated, with the model parameters adjusted in each iteration in order to minimise the loss function. As the GPC implementation relies on a variational approximation of the GP, the loss function selected is the Variational Evidence Lower Bound \citep{VariationalELBO}, which defines a lower bound for the log-likelihood.

The adjustment of the model parameters is done based on a learning rate, which is by itself another hyper-parameter of the classifier that can be optimised, especially to avoid over-fitting and lengthy training times. To optimise the hyper-parameters, the Adam Optimiser \citep{Adam} was used. An essential attribute of the optimiser is its adaptive learning rate, which is adjusted based on the first and second momentum of the loss function's gradient. The learning rate was initialised as 0.01 across all scenarios and then allowed to vary by the optimiser's scheduler. 

To train the GPC, we used the framework for a binary GP classifier from the \texttt{skorch} python package, which interfaces with both \texttt{GPyTorch} and \texttt{scikit-learn}. \texttt{GPyTorch} itself is built upon the modern architecture of \texttt{PyTorch}, which allows the GP to take advantage of the tensor framework and to be trained on a GPU using \texttt{CUDA}. As a result the GP can scale well with our expansive training sets and maintain a fast training time. 

\begin{table}
    \centering
    \caption{The number of training iterations and the size of the Inducing Dataset and Training Batches for each planet-FP GP classifier.}
    \label{tb:GP_Training}
    \begin{tabular}{c|ccc}
    \hline
        \textbf{Planet -} & \textbf{Iterations} & \textbf{Inducing Size} & \textbf{Batch Size} \\ \hline
        \textbf{EB} & 150 & 512 & 1024 \\ 
        \textbf{HEB} & 200 & 512 & 1024 \\ 
        \textbf{BEB} & 200 & 512 & 1024 \\ 
        \textbf{NEB} & 200 & 512 & 1024 \\ 
        \textbf{NHEB} & 200 & 512 & 1024 \\ 
        \textbf{NTP} & 300 & 512 & 1024 \\ 
        \textbf{HTP} & 300 & 512 & 1024 \\ 
        \textbf{NSFP} & 200 & 512 & 1024 \\
        \hline
    \end{tabular}
\end{table}

To improve efficiency and memory management, the GP splits the training data into batches while training, with the model trained on each batch sequentially. Similar to the inducing dataset, the size of the batch affects both the train speed and performance of the model, thus requiring optimisation for each scenario. The selected batch size for each scenario is presented in Table \ref{tb:GP_Training}. In addition, similar to the GBDT scenario, the GP was trained with early stopping enabled to avoid potential overfitting issues. The number of iterations that each Planet-FP GP model trained for is provided in the first column of Table \ref{tb:GP_Training}.

\begin{figure*}
    \centering
    \includegraphics[width=\textwidth]{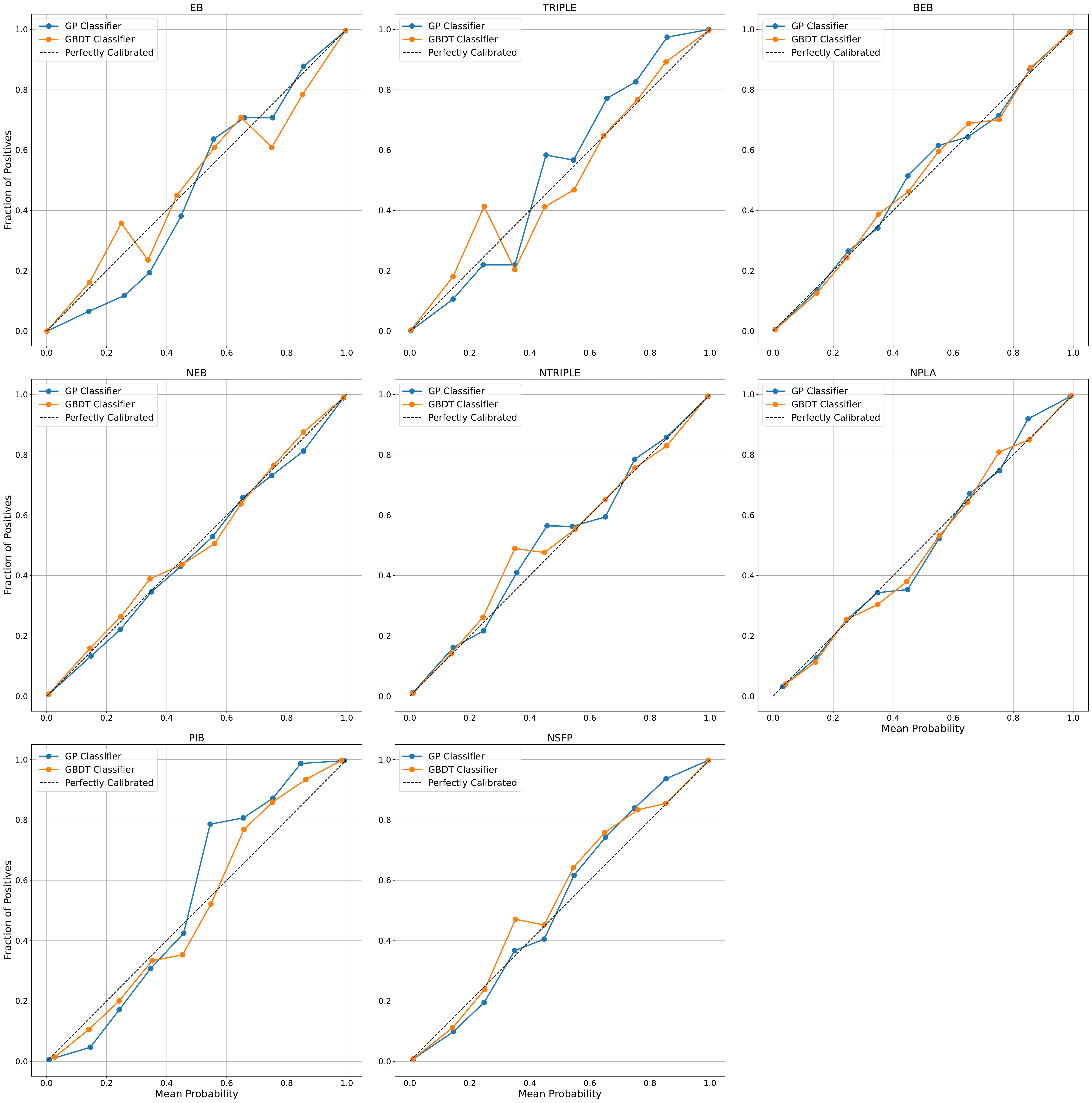}
    \caption{Calibration curves for the Gaussian Process (blue) and Gradient Boosted Decision Trees (orange) classifiers across all Planet-FP classification pairs. The calibration curves are compared to an idealistic, perfectly calibrated model plotted as a dotted line.}
    \label{fig:CalCurves}
\end{figure*}

Finally, an important attribute of our trained classifiers is how well calibrated they are, which defines their ability to provide true likelihood probabilities for our candidates. The GP output is naturally probabilistic due to the Bayesian design of the model. We tested the probability calibration of the two classifiers based on the fraction of positive samples in each probability bin from 0 to 1.0. For a perfectly calibrated classifier, the expectation is that at a probability of 0.5, only 50\% of the samples in the bin would be true positives while at 0.8 the fraction of positives would be 80\%. The calibration curves of the two classifiers are compared against the perfect calibration expectation in Figure \ref{fig:CalCurves} for all Planet-FP models. 

The plot shows that overall the two classifiers are well calibrated. However, deviations from the perfect calibration are seen in the EB, HEB, HTP and NSFP cases, especially for the GP. In the case of EBs, the deviation is mostly on the lower end of the probabilities, suggesting that the classifiers are more confident and assign low probabilities to the FP events. This results in an underestimation of the probability in the low regime. On the other hand, the deviation for the HEB, HTP and NSFP classifiers is primarily on the high probability regimen, with the classifiers being confident in assigning high probabilities to the Planet events, resulting in an overall overestimation of the probabilities across the spectrum. Considering that the GP is naturally probabilistic and that the GBDT performs similarly or better, we consider this behaviour as not significantly affecting the true posterior probabilities obtained from the classifiers.

\subsubsection{Feature Significance}\label{sec:optimisation}
The performance of ML classifiers can often benefit from the removal of redundant or minimally useful features. As the pipeline makes use of a large number of features, it was important to critically evaluate their relevance for each Planet-FP pair and decide which features should be included in the final training data. At the same time, our intention was to not overtune the included features of each model, to avoid both complications from potential overfitting on the synthetic data and the loss of potential information. As such, we decided to keep the removal of features to a minimum and keep the features near similar across all scenarios. This was possible due to the fact that both the GP and GBDT classifiers are capable of tuning the significance of each individual feature, which allows for the classifiers to adapt to each FP scenario as appropriate. 

The transit duration and depth of the event that were obtained from the simulation data were removed in favour of the ones obtained from the trapezium fit. In addition, features related to the SES and the MES, such as the minimum values, that were compared against the nominal values of the two statistics were removed as they were better represented by their ratios. Moreover, as the MES ratios were tuned for the identification of non eclipsing events, we only include them in the Planet-NSFP classifiers.

\begin{figure*}
    \centering
    \includegraphics[width=\textwidth]{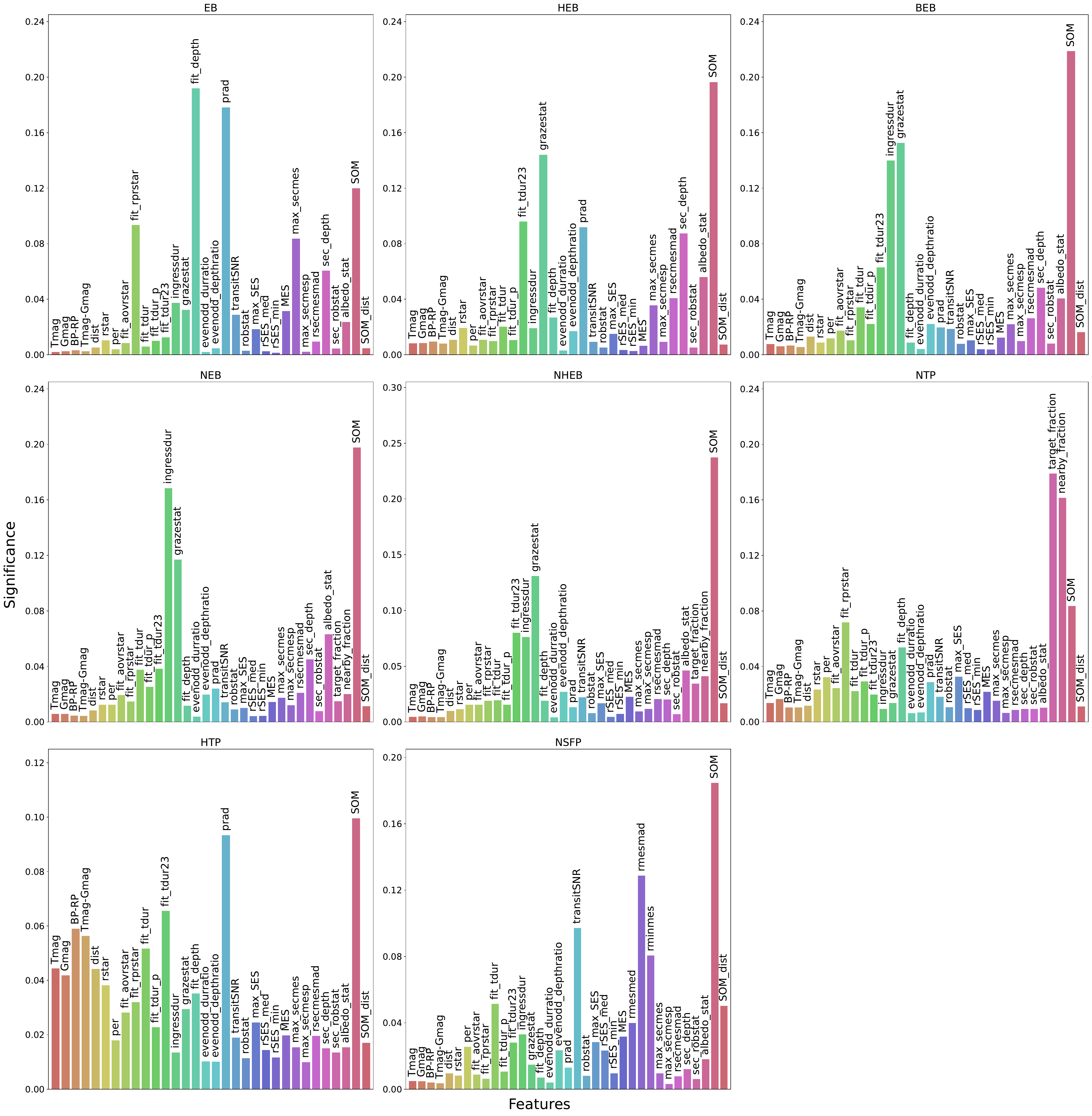}
    \caption{Significance of the features' contribution to the output of the GBDT classifier for each planet-FP classification pair.}
    \label{fig:FeaturesGBDT}
\end{figure*}

\begin{figure*}
    \centering
    \includegraphics[width=\textwidth]{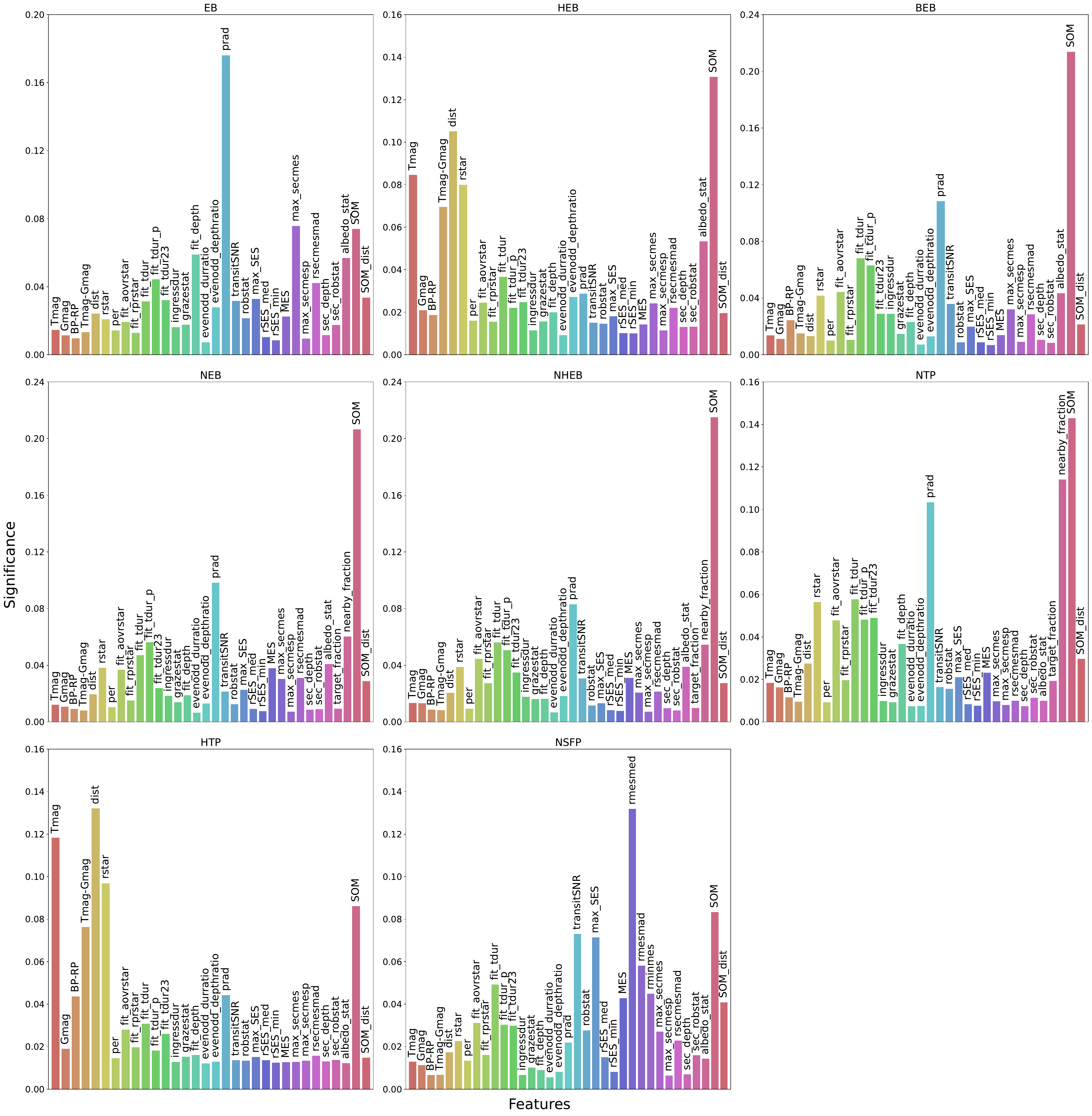}
    \caption{Significance of the features' contribution to the output of the GP classifier for each planet-FP classification pair.}
    \label{fig:FeaturesGPC}
\end{figure*}

The significance of the remaining features in each Planet-FP classifier is presented in Figures \ref{fig:FeaturesGBDT} and \ref{fig:FeaturesGPC} for the GBDT and GP respectively. The features' significance reveals the overall decision making of the classifiers and provides an insight into how the final probability is derived. There are many interesting patterns seen in the feature significance across the classifiers. One that should be highlighted is that the SOM statistic is one of the most important features across all classifiers. This is not surprising as the shape of the transit can be very powerful in revealing the nature of the event as evidenced by the proficiency of the SOM in separating the Planet and FP events shown in Figure \ref{fig:SOM_Mosaic}. Moreover, the grazing statistic and the secondary eclipse metrics become some of the most significant features for the diluted EB scenarios for the GBDT models. Interestingly, these features are not as prominent for the GP model, demonstrating the difference in the classification philosophy of the two models. This is especially true for the Planet-HEB scenario, where the GP focuses on the stellar characteristics, showcasing that it recognises and prioritises the presence of additional bound stellar companions to the target. In contrast, the GBDT model primarily focuses on the event itself. Importantly, the stellar characteristics also play a significant role for the HTP scenario, across both classifiers. This shows that the classifiers are indeed attuned to the presence of wider companions, highlighting the effectiveness of our synthetic scenarios in representing the true population. Similarly, for the other diluted planet scenario, the NTP, the focus shifts primarily onto the nearby fraction, in line with our expectations. Finally, it should be highlighted that the NSFP classifiers place significant weight on the transit significance metrics and the MES ratios, which is reassuring as they are tuned for the non eclipsing events.

\subsection{Prior Probabilities}\label{sec:priors}
The posterior probability assigned to each candidate by the ML classifiers assumes that each scenario has an equal likelihood to occur. This assumption stems from the training set disposition, where the planet and the FP events are equally represented. However, some scenarios are intrinsically more common and more likely to be detected as \textit{TESS} candidates than others. Since this information cannot be relayed to the classification models by using proportional training sets, as it would degrade their performance, we instead incorporate it externally. This is done by computing the prior probability for each candidate to belong to any of the scenarios we are considering. These prior probabilities are then used to update the ML derived posterior probability as shown in equation \ref{eq: posterior_prob} for each planet-FP pair. It should be noted that we do not compute prior probabilities for the NSFP scenario as the events are intrinsically tied to the method use to detect the \textit{TESS} candidates. Instead, a different approach is required for the NSFP, which is discussed in Section \ref{sec:nsfp_vetting}.

The priors for each synthetic scenario are a combination of 4 distinct probabilities, namely the probability that the scenario occurs, $P_\textrm{occ}$, the probability that it is detectable in the \textit{TESS} data, $P_\textrm{detection}$, and that it can be recovered, $P_\textrm{recovery}$, and the probability that the event is located on the target star or on a known nearby source, which we refer to as the Positional probability, $P_\textrm{positional}$. Each scenario prior can thus be expressed in the following form:
\begin{equation}
    P(s|I) = P_\textrm{occ} * P_\textrm{detection} * P_\textrm{recovery} * P_\textrm{positional},
\end{equation}

A detailed breakdown of the derivation of the occurrence probability for the different scenarios is presented in the sections below. It is important to highlight that we rely on empirically derived frequency rates as an estimate for the scenario's true $P_\textrm{occ}$. The frequency rates are computed across the whole region of parameter space for our data and do not take into account any specific effects of the radius or orbital period of the companions, as those are included in the distribution of the simulated events. The detection probability for each scenario is measured by the creation process of the synthetic training sets and is discussed in Section \ref{sec:training_sets}, with the probability for each scenario listed in Table \ref{tb:Detection}. Importantly, the detection probability directly incorporates the transit and eclipse probabilities for planetary and stellar companions. The recovery probability is measured by the results of our BLS survey for the synthetic sets. The recovery probabilities are presented in Table \ref{tb:Recovery}. The process for computing the positional probabilities is briefly explained in Section \ref{sec:Processing}. For a more in depth presentation and discussion on the Positional probabilities, please refer to HA24. 

\subsubsection{Planets and Eclipsing Binaries}\label{sec:planet_eb_priors}
We define the planet occurrence probability as the probability that a randomly selected star observed by \textit{TESS} has a planetary companion within our search space. This probability encompasses all planets within the parameter space of our dataset and more specifically planets with radius between 1$R_{\oplus}$ and 16$R_{\oplus}$ and orbital period of 0.5 to 16 days. To derive the probability, we use the occurrence rates for planets around F,G,K stars computed by \citet{Hsu2019}, based on planet candidates from the \textit{Kepler} mission. We retrieve the occurrence rates listed in Table 2 of the paper and then sum over the desired radius and period range to compute the cumulative occurrence probability for the planets, which we found to be $P_\textrm{occ}(PL) = 0.302$.  

Similarly, for the EBs, we compute the probability of a star having a close stellar companion. To do this, we use the results of \citet{MoeDiStefano2017} on the frequency of companions for different stellar types. We use the derived relation between companion frequency, period and the binary mass fraction, as expressed in equation 23 of the paper, to determine the frequency of companions for a solar-like star in the period range of 0.5 to 16 days. We multiply the result by a factor of 1.3 to account for binary mass fractions, q, between 0.1 and 0.3, as suggested by the authors. We determine an occurrence probability of $P_\textrm{occ}(EB) = 0.039$ for the EB prior.

\subsubsection{Hierarchical FPs}\label{sec:heb_priors}
The occurrence probability for both hierarchical FP scenarios is an extension of the probability for their respective non-hierarchical systems. More specifically, for the HTP scenario, we consider a star with a planetary companion that has an additional wider stellar companion. An important distinction is that this planetary system is on the secondary star in this configuration, as we consider a planet on the primary as a true positive. We explicitly limit the scenario only to binary systems, as we consider systems with higher stellar multiplicity unlikely to produce a detectable signal for a planet transiting one of the non-primary stars. To construct the occurrence probability for the scenario, we multiply the already computed occurrence probability for the planet scenario with the probability of the system being in a binary configuration. We adopt a binary probability of $f_\textrm{binary}=0.3$ as per the results in Table 13 of \citet{MoeDiStefano2017}. We further multiply the probability by 0.5, to account for the fact that both stars have an equal probability to host the planet and that we only consider the case of the planet being on the secondary. The HTP occurrence probability can therefore be expressed as follows:
\begin{equation}
    P_\textrm{occ}(HTP) = P_\textrm{occ}(PL) * f_\textrm{binary} * 0.5.
\end{equation}

For the HEBs, the configuration is that of a close binary system with an additional wider companion. To compute its occurrence probability, we therefore multiply the probability for close binaries found above with the fraction of solar-like stars that have one or more companions as determined by \cite{MoeDiStefano2017}. This fraction is $f_\textrm{multiple}=0.4$ and includes both binary and higher order systems. It should be noted that our simulated systems include only configurations where the target star is accompanied by a wider binary system. However, in our priors we do not limit the number of companions, nor do we distinguish on whether the close in companion orbits the primary target or the wider companion. This is because we expect those systems, even if not simulated, to still be well aligned with our synthetic training set. Thus, we account for them in the scenario prior probability. The overall occurrence probability for the HEB scenario is computed as:
\begin{equation}
    P_\textrm{occ}(HEB) = P_\textrm{occ}(EB) * f_\textrm{multiple}.
\end{equation}

The occurrence probability for the two hierarchical FP scenarios is further adjusted by incorporating the \textit{Gaia} RUWE score \citep{GaiaEDR3_astrometric_solution} for the target star of each candidate. RUWE stands for Re-normalized Unit Weight Error and is essentially a statistic that expresses how well \textit{Gaia}'s single star astrometric solution fits the observation \citep{GaiaDR2_astrometric_solution, GaiaEDR3_astrometric_solution}. It has been generally accepted that the RUWE score for well fitted stars peaks at 1.0 and extends up to 1.4, with values above that threshold suggesting that the single star solution is incompatible. Studies \citep{Gaia_Belokurov, Gaia_Binaries_Stassun} have investigated the relation of the RUWE score to stellar multiplicity finding a strong correlation with scores even below the 1.4 threshold suggestive of binary systems. In fact, the study by \cite{Gaia_Binaries_Stassun} has shown that RUWE scores above the 1.4 threshold were related to known or suspected triple star systems, while scores between 1.0 and 1.4 were found to be suggestive of multi-star systems. According to \cite{Gaia_Belokurov} the RUWE can be used to gauge stellar multiplicity for orbital periods between 0.1 and 10 AU, up to distances of 1-2kpc, with the multiplicity threshold more likely to be at about 1.1. It should be noted that \cite{Gaia_Belokurov} also found that variable stars can result in elevated RUWE scores, due to the statistic assuming a constant magnitude.  

We integrated the RUWE information into the pipeline, to constrain the prior probability for the Hierarchical FPs based on the characteristics of the target star for each candidate. To do this, we adopted a cautious approach, reducing the binary and multiple fractions for candidates whose target star has a RUWE score below 1.05 and setting them to 1.0 for stars with a score above 1.4. The fractions remain unchanged for candidates with a score between the two values. To properly compute the reduction of the binary and multiple fractions, we had to first recalculate the multiplicity frequency for solar-like stars using the relation provided in \cite{MoeDiStefano2017}, to exclude the period range that would correspond to the orbital separations that the RUWE score is sensitive to. This period range was determined to be between 8 and 8000 days, assuming a solar-like star with a solar-like companion. The reduced multiplicity frequency was found to be $f_\textrm{low RUWE multiple}=0.28$, a reduction of 0.12 from the frequency of 0.4 computed by \cite{MoeDiStefano2017}. Assuming that the rate of binaries to higher order multiples would still remain at 3:1, we then calculated the reduced binary fraction for low-RUWE targets as $f_\textrm{low RUWE binary}=0.21$.

\subsubsection{Background FPs}\label{sec:background_priors}
Similar to the hierarchical FP scenarios, the probability for BEB FPs builds upon the occurrence probabilities for the EB scenario. The differentiation comes from adjusting the occurrence to account for the probability that a background, un-resolved host for the event could exist. This probability is once again tailored to the candidate, by first determining the maximum magnitude difference between the target star and a background source that could host the event, based on the candidate's eclipse depth. We cautiously allow a full eclipse of the host star for the BEB scenario when determining the magnitude limit. We use the Galactic stellar population synthesis model TRILEGAL \citep{TRILEGAL} to determine the probability that an unbound background star within the magnitude range could exist in the target's immediate background. Recall that the background scenario considers only unresolved sources and as such our probability estimation is limited to a radius of 2'' around the target star, which is conservatively taken as the spatial resolution outside of which \textit{Gaia} would have resolved the source. The final occurrence probability for the BEB scenario is obtained by multiplying the computed background probability with the occurrence probability for EBs. For candidates for which no suitable background source could exist, the prior probability for the scenario is set to 0.

\subsubsection{Positional Probabilities and the Nearby FPs}\label{sec:positional_priors}
The positional probability of each candidate is the final component for the derivation of their prior probabilities, the process for which is described in HA24. The role of the positional probability in the priors is to incorporate the likelihood of the eclipsing event occurring on the target star or on a nearby resolved source.  
The positional probability component of the pipeline produces a probability estimate for the target and for each individual possible source of the eclipse. This allows us to not only quantify the likelihood of the event being on-target but also to potentially identify the likely host of the event. The effectiveness of this was tested in Section 3 of HA24. Although, this fine-grain probability information is not directly used in the candidate prior, the per nearby star probabilities are still provided by the pipeline and can be used alongside our validation estimate to assess \textit{TESS} candidates. To inform the priors, we utilise only the overall probability that the event is on-target or off-target.  

Due to the nature of our validation framework and the use of one-to-one planet-FP classifications, the positional probability only contributes to the result of the planet against the Nearby FP scenarios. It should be noted that for the \texttt{RAVEN} pipeline, the background scenarios are considered to be "on-target" when considering the positional probability. This is in contrast to the implementation of A21, in which the background probability was a component of the positional probability. Based on the above, the occurrence probability for the nearby scenarios is identical to their respective on-target scenarios, with the exception that no \textit{Gaia} RUWE adjustment is made for the Nearby HEB.

\subsection{Statistical Validation}\label{sec:statistical_val}
The final component of the pipeline is the derivation of the posterior probability for the planet hypothesis by combining the ML derived probability of each planet-FP with their corresponding scenario specific prior probabilities as per equation \ref{eq: posterior_prob}. As discussed in our framework presentation, this posterior probability expresses only the probability of the candidate being a planet or a specific FP scenario, a consequence of our decision to split the FP scenarios into individual classes. Therefore, our statistical validation implementation requires that a candidate must have a planet posterior probability above 0.99 for each of the 8 planet-FP classifications to be considered as validated.

\texttt{RAVEN}'s validation strategy does not directly determine the overall planet probability, nor the probability for each FP scenario. This is because the binary classifiers only provide a probability comparing the planet to a specific FP scenario. These probabilities are therefore not mutually exclusive or exhaustive, as they do not account for the remaining FP scenarios. To compute the actual probability for each scenario requires a multi-class oriented strategy. The two most prominent approaches for implementing such a strategy are to train additional binary classifiers for each combination of FP pairs or to train multi-class classification models. However, the former would be more computationally expensive and also significantly increase the complexity of the pipeline. The latter would require the training sets for each scenario to be downsized, both to satisfy the computational requirements for the training of the model and to ensure that all scenarios are equally represented. Considering that the least numerous scenario, the HTP, only has 16990 recovered events, this would result in a significant loss of information from our training sets. In comparison, our approach allows us to utilise the training sets to the maximum extent possible, while also minimising the computational resources needed. In addition, we can treat each classification as its own designated test for the potential planet candidate. Requiring the candidate to reach the statistical validation threshold for each one ensures that the potential for error is minimised. As the purpose of the \texttt{RAVEN} pipeline is ultimately to statistically validate planet candidates, we believe that our current approach is the most sensible for this initial release. However, we intend to explore a multi-class strategy in the future. 

\subsubsection{Machine Learning Classification}\label{sec:ml_class}
The first step in computing the planet posterior probability for the \textit{TESS} candidates is the application of the trained classifiers on their feature vectors. This gives the classification probability, $p(s=1|x*)$, for each classifier and for each planet-FP pair. The classifier probabilities are combined by taking their mean. This ensures that the classification probability is not reliant on just one classifier, minimising the potential for error due to overfitting, weaknesses or biases inherent to either classifier. The respective probabilities produced from each classifier are provided as an additional output by the pipeline, so that the results can be independently assessed.

\subsubsection{NSFP Vetting}\label{sec:nsfp_vetting}
The NSFP scenario represents a specialised classification case for our pipeline. This is because it is a byproduct of the survey method used to detect candidates and includes a variety of signals, both astrophysical in nature and not. In addition, the NSFP training data still includes events from the synthetic scenarios, mostly involving EBs, despite the safeguards we introduced to avoid their inclusion. As a result, an occurrence based prior probability cannot be computed as the underlying mechanisms that give rise to the detected events are not constrained. Moreover, the NSFP detections are inherently dependent on the implementations and selection strategies of each survey, which means that any prior based on the number of detected NSFP events would only be suitable for candidates that were detected from that survey. Therefore, we would not be able to use that prior probability for the TOIs, as they were detected through a variety of different methods and have undergone different forms of vetting.

Based on the above considerations, using the NSFP scenario to compute the planet posterior probability in the same manner as the rest of our FP scenarios is not appropriate. We can however, still make use of the probability derived from the ML classifications as it represents a degree of confidence in the candidate being a true planet. Therefore, we can use the NSFP probability to vet the candidates, requiring that they pass a certain threshold before being considered for validation. This allows us to identify candidates that are not likely to be planets and which are not represented from the FP events, such as stellar variability and other sources of noise. 

To select the probability threshold, we relied on the results from applying the classifier on our test set, which are presented in Section \ref{sec: Testset_results}. One simple choice for the threshold could have been the probability of 0.5. However, our desired threshold is the one that allows the vetting of as many true planets as possible while removing all events that are not represented by our FP scenarios. To this extend, we visually examined the NSFP events in the test data which received probabilities greater than 0.5. Based on this, we found that all the events with a probability above 0.87 were indeed eclipsing. As such, we decided to set the threshold to a probability of 0.9. Note that all candidates are processed by all classifiers, even if they do not pass the threshold, as the information provided can still be useful.

\subsubsection{Posterior Probabilities}\label{sec:posterior_probs}
With the classification probabilities computed, we then proceed to calculate the prior probabilities for each candidate. The final candidate priors for each scenario are then derived, taking into account both the overall scenario specific probabilities and the characteristics of the candidates themselves, as detailed in Section \ref{sec:priors}.

The planet posterior probabilities are computed by combining the classification probabilities and the prior probabilities as per equation \ref{eq: posterior_prob}. As already discussed, one posterior probability is produced for each planet-FP pair. While the results express the probability that the candidate is a true planet in the context of a specific scenario classification, we can still compare the relative probabilities for each scenario to gain an insight into the true nature of the candidate. We validate the candidate if all scenario probabilities are above 0.99, with the exception of the NSFP probability for which the threshold was set to 0.9. For general use, we urge that candidates are visually examined before declaring them validated. To rank the candidates, probabilities across the scenarios are collapsed into a single probability using the minimum. For candidates with Planet-NSFP probabilities greater than 0.9, the NSFP result is excluded from the minimum. This final probability is the \texttt{RAVEN} probability and is the pipeline's primary output.

\section{Performance on Test Sets} \label{sec: Testset_results}
Prior to applying the trained ML models on the TOIs, we first test their performance on unseen subsets of the training sets. These subsets consist of 10\% of randomly chosen events for each scenario, which were selected and isolated prior to training and as such they do not contribute to either the models or the trained SOMs.

\begin{table*}
    \centering
    \caption{Performance metrics on test sets for the GBDT and GP binary classifiers trained on Planet-FP pairs. For the accuracy, the threshold used to label candidates is 0.5, whereas for precision and recall we use a 0.9 threshold.}
    \label{tb:Testsets}
    \begin{tabular}{l|l|cccccccc}
        \hline
        \textbf{Metric} & \textbf{Classifier} & \textbf{EB} & \textbf{HEB} & \textbf{BEB} & \textbf{NEB} & \textbf{NHEB} & \textbf{NTP} & \textbf{HTP} & \textbf{NSFP} \\ \hline
        \textbf{Accuracy} & GP & 0.9907 & 0.9890 & 0.9578 & 0.9580 & 0.9616 & 0.9259 & 0.9800 & 0.9660 \\
        \textbf{} & GBDT & 0.9909 & 0.9817 & 0.9593 & 0.9609 & 0.9588 & 0.9205 & 0.9427 & 0.9730 \\
        \textbf{} & Combined & 0.9913 & 0.9879 & 0.9596 & 0.9610 & 0.9617 & 0.9249 & 0.9703 & 0.9728 \\
        \textbf{AUC} & GP & 0.9995 & 0.9995 & 0.9934 & 0.9942 & 0.9942 & 0.9728 & 0.9975 & 0.9950 \\
        \textbf{} & GBDT & 0.9994 & 0.9986 & 0.9934 & 0.9945 & 0.9941 & 0.9711 & 0.9865 & 0.9963 \\
        \textbf{} & Combined & 0.9995 & 0.9994 & 0.9939 & 0.9947 & 0.9946 & 0.9731 & 0.9959 & 0.9961 \\
        \textbf{Precision} & GP & 0.9978 & 0.9986 & 0.9911 & 0.9931 & 0.9937 & 0.9980 & 0.9986 & 0.9974 \\
        \textbf{} & GBDT & 0.9959 & 0.9968 & 0.9886 & 0.9936 & 0.9922 & 0.9954 & 0.9944 & 0.9985 \\
        \textbf{} & Combined & 0.9978 & 0.9988 & 0.9909 & 0.9944 & 0.9942 & 0.9969 & 0.9992 & 0.9982 \\
        \textbf{Recall} & GP & 0.9524 & 0.9517 & 0.8487 & 0.8655 & 0.8733 & 0.7814 & 0.8890 & 0.8820 \\
        \textbf{} & GBDT & 0.9605 & 0.9310 & 0.8549 & 0.8572 & 0.8733 & 0.7814 & 0.7609 & 0.9037 \\
        \textbf{} & Combined & 0.9542 & 0.9372 & 0.8512 & 0.8567 & 0.8705 & 0.7802 & 0.8022 & 0.8919 \\ \hline
    \end{tabular}
\end{table*}

The performance of the models is evaluated on 4 key metrics. These are the accuracy, the Area Under the ROC Curve (AUC) and the precision and recall scores. The accuracy score expresses the ratio of correct predictions to the total number of predictions. To compute it we set a threshold at 0.5 and label results over it as planets and under it as FPs. The AUC calculates the area under the curve of the True Positive Rate against the False Positive Rate and defines how often are True Positives scored higher than true FPs. The precision score is the ratio of true positives against the total number of positive predictions, with a score of 1 signifying that only true positives were classified as such. As such it expresses the classifiers' ability to recognise FPs and not label them as true planets. The recall score on the other hand is the ratio of true positive classifications to the total number of positive samples in the test set. It characterises the ability of the classifiers to label true planets correctly. For the calculation of the precision and recall scores we set the classification threshold to 0.9 and compute the scores with the planets as the positive class.

The results from the performance testing of the two classifiers are presented in Table \ref{tb:Testsets}. The two classifiers perform on a high level across all FP scenarios, especially when considering their precision. Since the primary objective of the \texttt{RAVEN} pipeline is to vet and validate true planet candidates, the precision score is the most important metric as is the one that expresses the pipelines ability to correctly identify FPs and not score them highly. The precision reached is almost 99\% for all scenarios when considering the result from combining the two classifiers. However, it is important to note that the results on the test sets do not account for the effect of applying the prior probabilities. This is because the priors rely on observed data for the true candidates, such as the centroid offset for the positional probabilities, which is not easily replicated for the synthetic lightcurves. Therefore, we need to test the classifiers on real \textit{TESS} candidates for which the disposition is known to make a definitive assessment of their performance.

In addition, the metrics reveal that the classifiers are overall better in classifying true planets against FP scenarios involving EBs rather than planets. This is apparent across all metric results and especially when considering the recall score. The result is not surprising as the photometric signature of diluted planets is similar to that of true planets, while the EBs, even diluted, remain more distinct. As such, the classifiers are more confident in assigning high probability values to the planet samples when compared against the EB scenario FPs leading to higher recall values. This suggests that it would be harder to validate planet candidates, since we require that we pass the 99\% threshold on each planet-FP pair. However, the high AUC scores achieved and to a lesser extent the accuracy scores suggest that in terms of ranking, we can expect the pipeline to perform well in providing higher probabilities for true planet candidates compared to all FPs.

\begin{figure*}
    \centering
    \includegraphics[width=\textwidth]{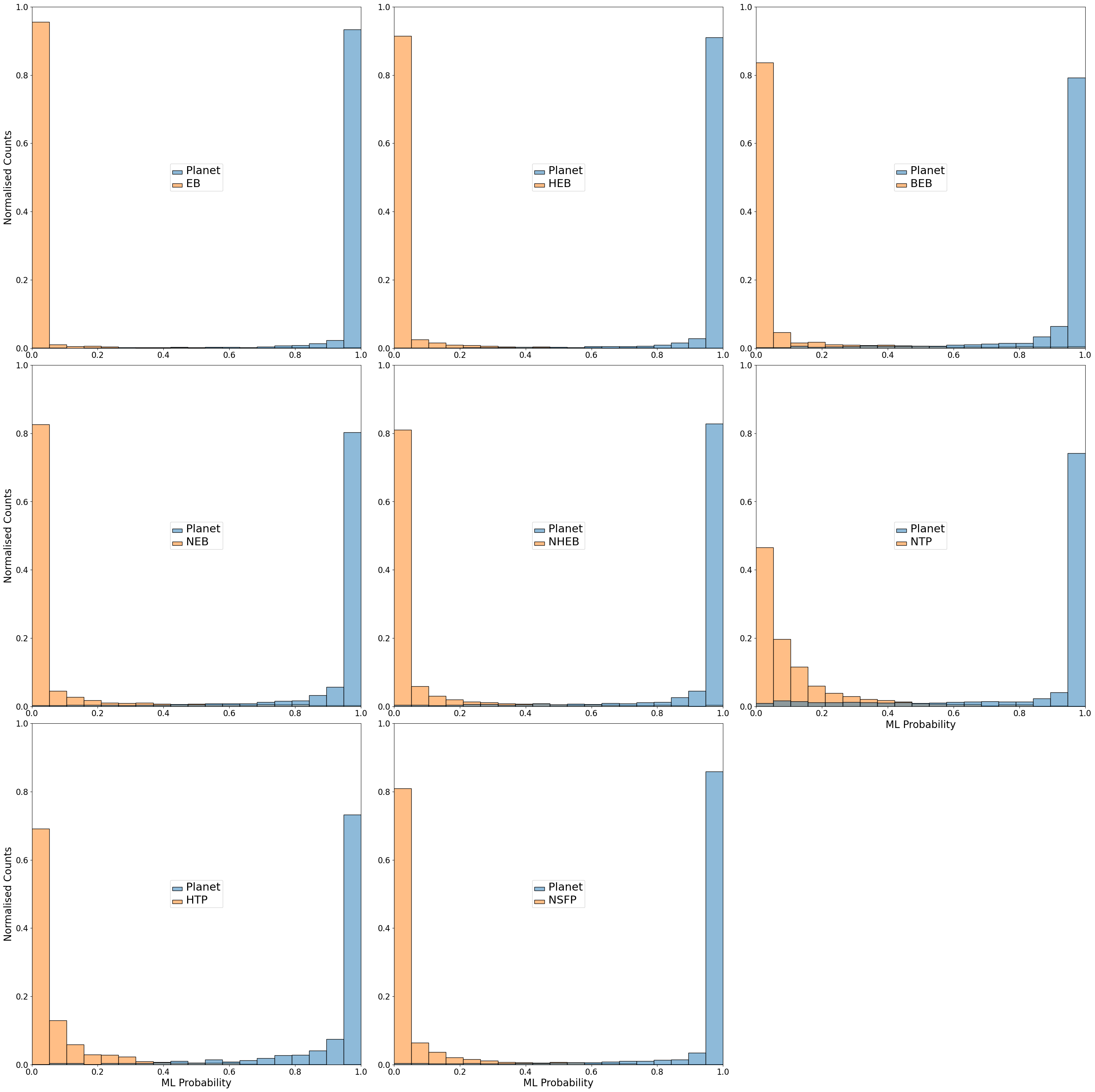}
    \caption{A collection of all scenario specific probability distributions from the combined GBDT and GP classifiers for the respective test subset of each planet-FP scenario. The resulting probability distributions showcase the varying degree of confidence of the classifiers in distinguishing between the Planet and each FP scenario.}
    \label{fig:Testset_Probs}
\end{figure*}

Finally, we plot the combined planet posterior probabilities for all sets of test events in Figure \ref{fig:Testset_Probs}, to better visualise the performance of the classifiers. Although these posterior probabilities do not account for prior probabilities, they still reveal the confidence of the classifiers when assigning probabilities for each planet-FP pair and their effectiveness in distinguishing between the synthetic true planet events and the FP events. As seen from the metrics in Table \ref{tb:Testsets}, the best performing classifiers are for the EB related FP scenarios. The classifiers confidently separate the planets from the FP events, with the vast majority of the test events occupying the extreme ends of the probability spectrum.

On the other hand, the histogram plots for the planet related FP scenarios show the increased challenge in classifying these scenarios. While the distribution of probabilities for the synthetic planets and FPs still skew towards 1 and 0 respectively, this is not as strong as for the EB FP scenarios. There are proportionally more true positive events occupying the probability spectrum between 0.05 and 0.95 and there are also more cases of events being incorrectly classified based on a 0.5 probability threshold. The classes are still well separated, and the FP planet events do not have high probabilities, as evident by the high precision values achieved for the classifiers. This shows that the pipeline has a strong capacity for vetting even for these difficult FP cases. However, validating the \textit{TESS} candidates against the planet FP scenarios is expected to be more difficult compared to their EB counterparts. The prior probabilities provide help in this regard, as the true planet scenario's larger prior, owing to its greater detection and recovery rates, will help boost the posterior probability. The candidate specific priors, specifically the positional probability for the NTP scenario and the RUWE for the HTP, further help differentiate the true planets. Finally, for the NSFP scenario, the distribution of the probabilities for the two classes showcase that the classifiers are well capable of separating the true planets and that the vast majority of them will lie above the 0.9 threshold we set for our vetting limit.

\section{Results on Pre-Classified TOIs}\label{sec: TOI_results}
\subsection{Overview}
The pipeline's performance was ultimately tested on a sample of pre-dispositioned TOIs that have already been classified as planets, FPs or FAs. These already identified TOIs have undergone extensive follow-up and classification efforts by the community, with their independent classification serving as a benchmark to test the pipeline's results. This is the most decisive test for the pipeline as it allows us to independently evaluate performance on actual \textit{TESS} detected exoplanets and FPs. It should be noted that the results and conclusions drawn in this section rely on the existing analysis of the TOIs beyond just their Planet, FP and FA label and more specifically the publicly available notes on ExoFOP\footnote{https://exofop.ipac.caltech.edu/tess/} from the \textit{TESS} Follow-Up Observation Program (TFOP) \citep{TFOP}. 

It is important to highlight that the TOIs represent a slightly different population than the one the pipeline was developed for. This is because the TOIs consist of candidates from a variety of different planet searches, from the SPOC pipeline's analysis of 2 minute cadence data to QLP's search of the FFI lightcurves and to candidates found from the Planet Hunters Initiative \citep{TOI, PlanetHuntersTess}. Thus, the candidates are the result of different processing pipelines and have been subject to different automated and manual vetting. In particular, the SPOC pipeline utilises the vetting pipeline TESS-ExoClass (TEC) to remove events not consistent with transits or stellar eclipses. Moreover, based on results from the TEC, candidates that are confidently found to be caused by EBs are not included in the TOI catalogue. As such the TOI list should be mostly free from obvious EBs and events caused by stellar variability or instrumental noise. In contrast to this, \texttt{RAVEN}'s training sets have been prepared with the expectation of being applied to candidates directly after a planet search and this is reflected in the scenario priors. As such, the pipeline is not optimised for running on the TOIs, although it is still applicable and the results should offer an indication of performance. 

The TOI list and classifications for this test were obtained from the NASA Exoplanet Archive\footnote{https://exoplanetarchive.ipac.caltech.edu/}, on the 3rd of February 2025. At that time, there was a total of 2134 pre-classified TOIs, with 548 classified as already Known Planets (KP), 485 as \textit{TESS} Confirmed Planets (CP), 1113 as FPs and 96 as FAs. However, only 1918 of the TOIs had an associated released SPOC FFI lightcurve. Finally, applying our depth and period constraints on the remaining sample left a total of 1589 TOIs to be examined. All of them were run through the full processing steps of the pipeline, except for a single FP TOI, the target star of which was flagged as a "DUPLICATE" in the TIC. From the final results, we excluded 68 TOIs due to the lack of stellar radius in the TIC for the target star. A further 87 were excluded as their \textit{TESS} magnitude was fainter than 13.5, with another 22 dropped due to having a \textit{Gaia} magnitude fainter than 14. As discussed in Section \ref{sec:training_sets}, our training sets do not contain events for which the target star was fainter than 13.5 Tmag or 14 Gmag. An additional 28 TOIs were also excluded as their MES computed during the feature generation was less than 0.8, which was the threshold we set for our training sets following our BLS survey, while 2 TOIs failed the feature generation and were thus dropped. Finally, for 21 TOIs, no positional probabilities were produced due to issues with their centroid data. As a result no posterior probabilities are provided and the TOIs are excluded from further analysis. The final number of pre-dispositioned TOIs in this test is therefore 1361 of which 705 are Known and Confirmed Planets, 630 are FPs and 26 are FAs.

In addition to the above removed cases, a further 90 events were flagged due to either the transit depth showing variations greater than 50\% of the median value or falling below 200ppm in at least one sector, as discussed in Section \ref{sec:per_sector}. This flag is intended as a fail safe, in order to catch possibly problematic variable cases which would require manual inspection. These cases are mostly FP or FA detections, either due to noise, variability or dilution from a nearby source whose contribution fluctuates as the aperture changes in each sector. However, in some cases these are also true planets that are flagged caused by an incorrectly provided period, TTVs, stellar variability that was not properly removed during de-trending or even artifacts introduced by the de-trending algorithm. In any case, the results of the pipeline for these candidates should be viewed with caution. For the analysis presented in this section, these candidates are not removed but instead pointed out when relevant as they include by design a sizable portion of the known NEB events. For reference, the flagged events include 61 FPs, 12 FAs and 17 Planets.

\subsubsection{Positional Probabilities}
\begin{figure}
    \centering
    \includegraphics[width=\columnwidth]{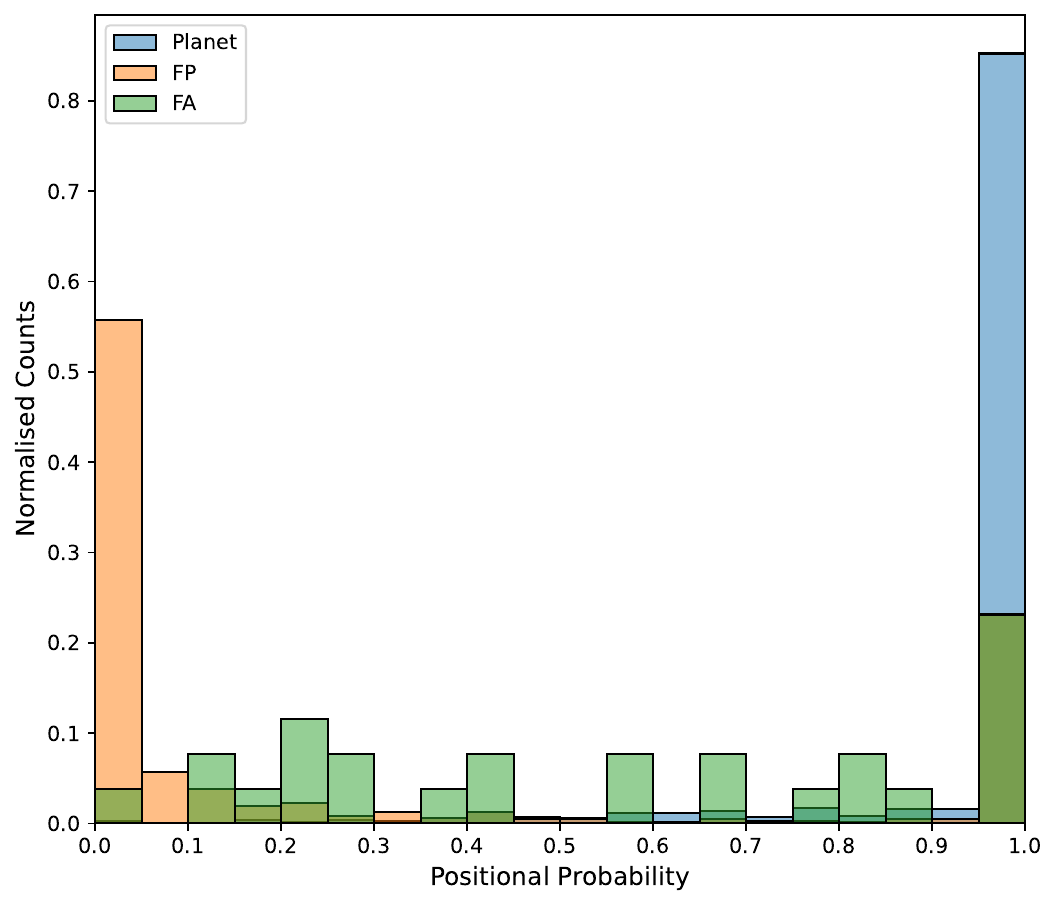}
    \caption{The distribution of the Positional Probabilities for known and confirmed Planet TOIs, False Positives and False Alarms.}
    \label{fig:TOI_PosProb}
\end{figure}

A histogram of the positional probabilities for all remaining objects, separated by disposition, is presented in Figure \ref{fig:TOI_PosProb}.
The distribution of the Positional probabilities follows the behaviour seen in the results of HA24, with the majority of the Planet TOIs having probabilities greater than 0.95. Since these should all be on target events and thus ideally have probabilities of 1.0, this result shows that pipeline has the desired capability to identify them. A more detailed discussion on the probabilities for the planet TOIs, including the reasons as to why some have low probabilities is offered in HA24. 

As evident from the histogram, most of the FP TOIs had low Positional Probabilities, suggesting that they are off-target events. A preliminary examination of the assessment notes for these TOIs available on ExoFOP supports this tendency, with 422 assessed to be NEB events, while 172 were assessed to be on target EBs.
Based on the large difference between the number of assessed NEB and EB TOIs, we can conclude that NEB events are more likely to be present in the TOI catalogue compared to EBs. This result is not surprising when considering that the automated vetting procedure of the SPOC pipeline systematically removes events that it confidently assesses to be EBs. Finally, for the few FA events in our sample, the results show that they are mostly consistent with on-target detections.

\subsubsection{\texttt{RAVEN} Probability}

\begin{figure}
    \centering
    \includegraphics[width=\columnwidth]{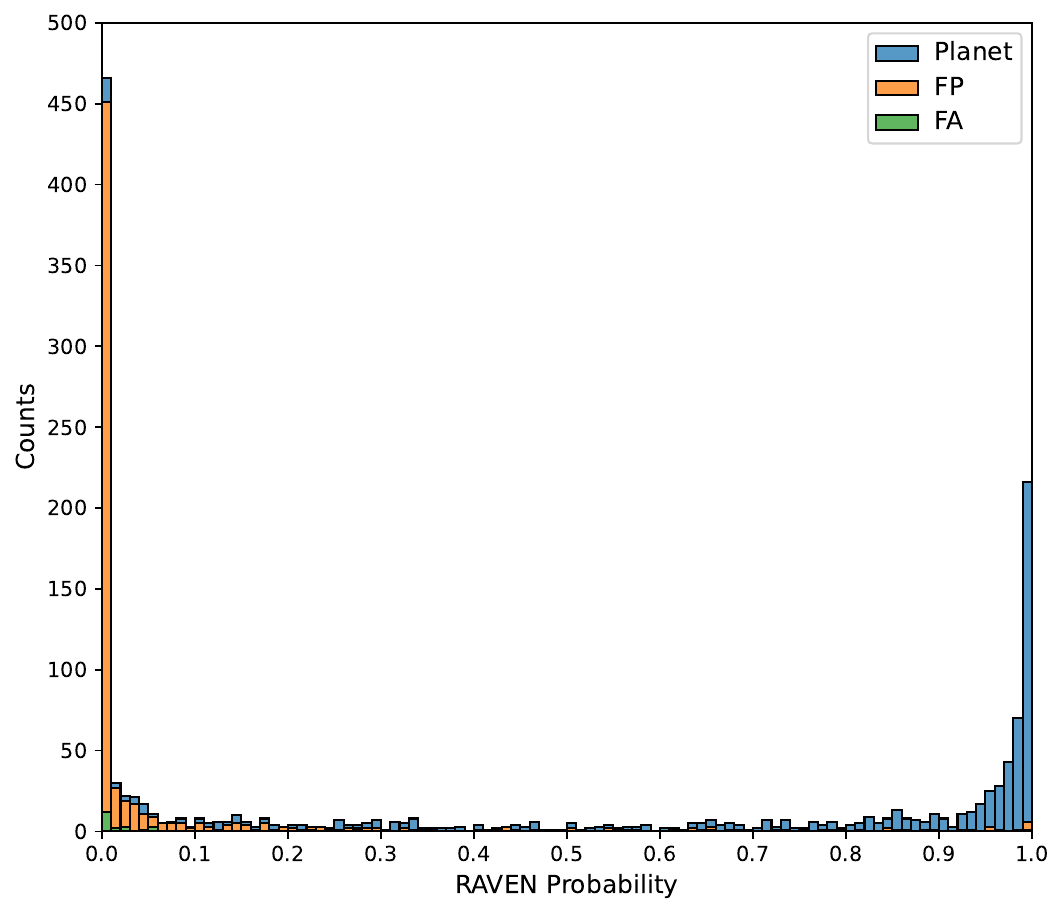}
    \caption{Stacked histogram presenting the minimum posterior probability for pre-dispositioned Planet, FP and FA TOIs. The probability tending to the extremes, especially for the FPs, highlights the ability of the \texttt{RAVEN} pipeline to perform effective vetting and validation for the \textit{TESS} candidates. }
    \label{fig:TOI_Probs}
\end{figure}

For all 1361 TOIs remaining in our sample, their \texttt{RAVEN} probability is presented in Figure \ref{fig:TOI_Probs}. The Figure shows the stark difference in probability for the 3 classes, which appear well separated and occupying the extreme ends of the probability distribution. This highlights the effectiveness of \texttt{RAVEN} in identifying FP events and assigning them appropriately low planet posterior probabilities. To be more precise, 93.8\% of the FP events had a minimum posterior probability less than 0.5 and 69.7\% below 0.01. The mean probability of FP events was 0.076, while the median probability was 0.00022. Similarly for the FA TOIs, 23 out of the 26 had probabilities lower than 0.5, with a median probability across the class of 0.016. Overall, the results for the FP and FA TOIs confirm that the pipeline is highly capable in vetting the \textit{TESS} candidates and can be used for curating candidate lists by removing the majority of FP events. 

\begin{figure}
    \centering
    \includegraphics[width=\columnwidth]{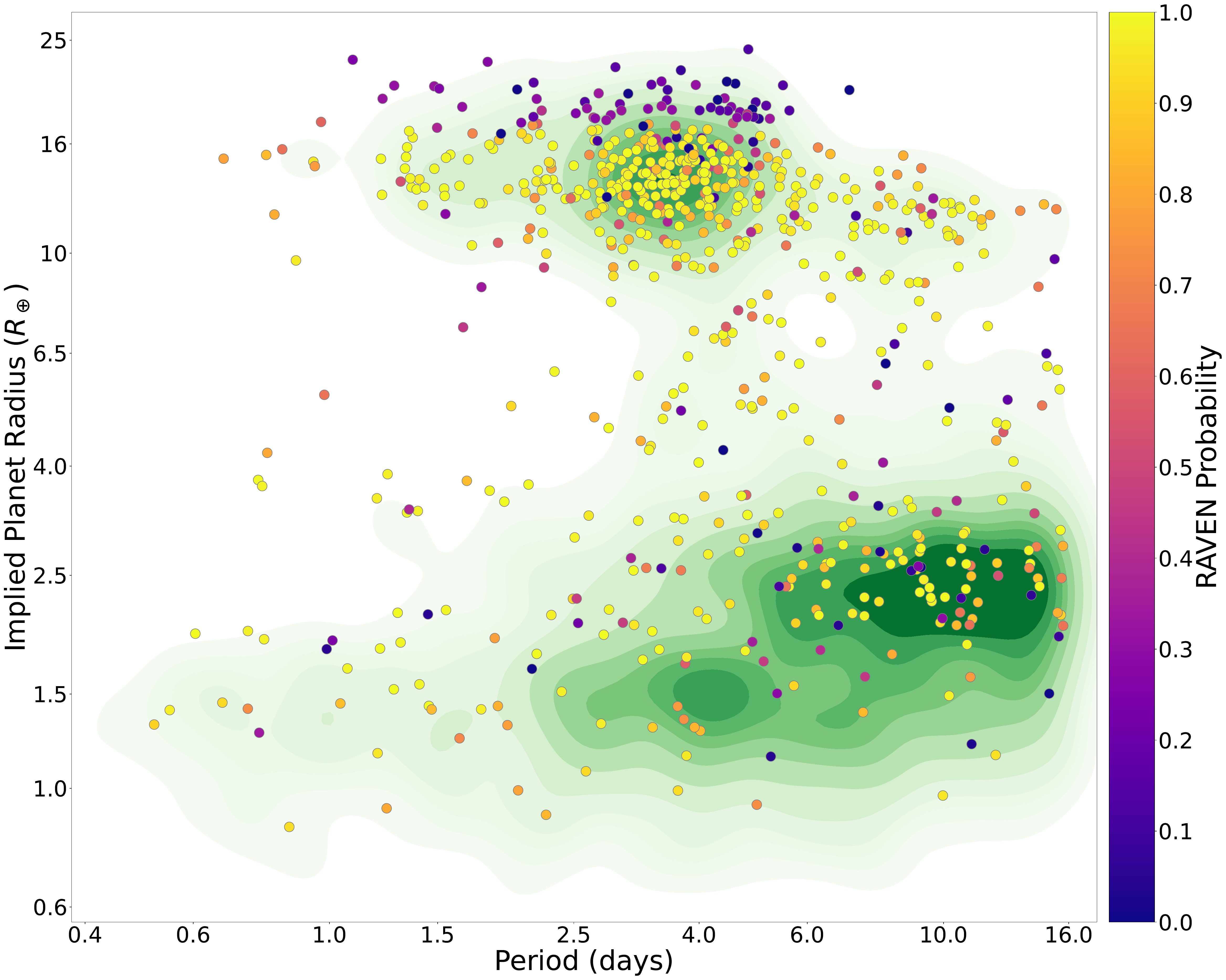}
    \caption{Radius-Period plot in log scale for the 705 Planet TOIs in our sample, overlaid over a density plot of the known planet population with period less than 16 days. The planet TOIs are shaded based on their \texttt{RAVEN} Probability to illustrate the pipeline's performance across the parameter space.}
    \label{fig:KP_per_rad}
\end{figure}

Moreover, the Figure demonstrates that the pipeline is effective in classifying true planet candidates. The distribution of probabilities for known and confirmed planets is skewed towards high probabilities, peaking at above 0.99. Overall, 81\% of the planets had probabilities higher than 0.5, with a mean and median probability of 0.79 and 0.96 respectively. It should be noted that, even though about 20\% of the Planet TOIs had probabilities lower than 0.5, almost half of those were found to have a planet radius greater than $16R_\oplus$, as shown in Figure \ref{fig:KP_per_rad}. This radius threshold is the limit we set for the pipeline's simulated Planets. As such, their low probability is not unexpected and shows the limits of the pipeline's performance when operating outside its designated parameter space.

Considering the pipeline's capability for validating candidates, 29.8\% of the known and confirmed planets lie above the standard 0.99 validation threshold. We discuss validation and further limits on when to safely validate candidates in \ref{sect:validationlimits}. Furthermore, the probability distribution across all candidates suggests that even candidates that do not pass the threshold but still have a minimum posterior probability greater than 0.9 can be considered to be likely planets and thus accounted for in population studies. This is because the region is overwhelmingly populated by true planet TOIs, with only a handful of FPs representing just 2.8\% of the total. 

Based on these results, we also compute the overall performance metrics for the pipeline on these TOIs. The pipeline achieves an accuracy of 87.1\%, using a simple probability threshold of 0.5. Setting a probability threshold of 0.9 for candidates to be likely planets, the pipeline achieves a precision of 97\% and a recall score of 60\%. In addition, for the AUC metric, which expresses the ability to rank true positives higher than false positives, \texttt{RAVEN} achieved a score of 96\%, demonstrating its potential for candidate ranking. Removing the TOIs above $16R_\oplus$, the accuracy increased to 91\%, the recall to 66\% while the precision remained at 97\%. 

We present a more detailed overview of the probabilities for all 3 classes of TOIs in Sections \ref{sec:EB&NEB_TOIs}, \ref{sec:FA_TOIs} and \ref{sec:PlanetTOIs}.

\subsubsection{Planet-NSFP Posterior Probability}\label{sec:nsfp_results}

\begin{figure}
    \centering
    \includegraphics[width=\columnwidth]{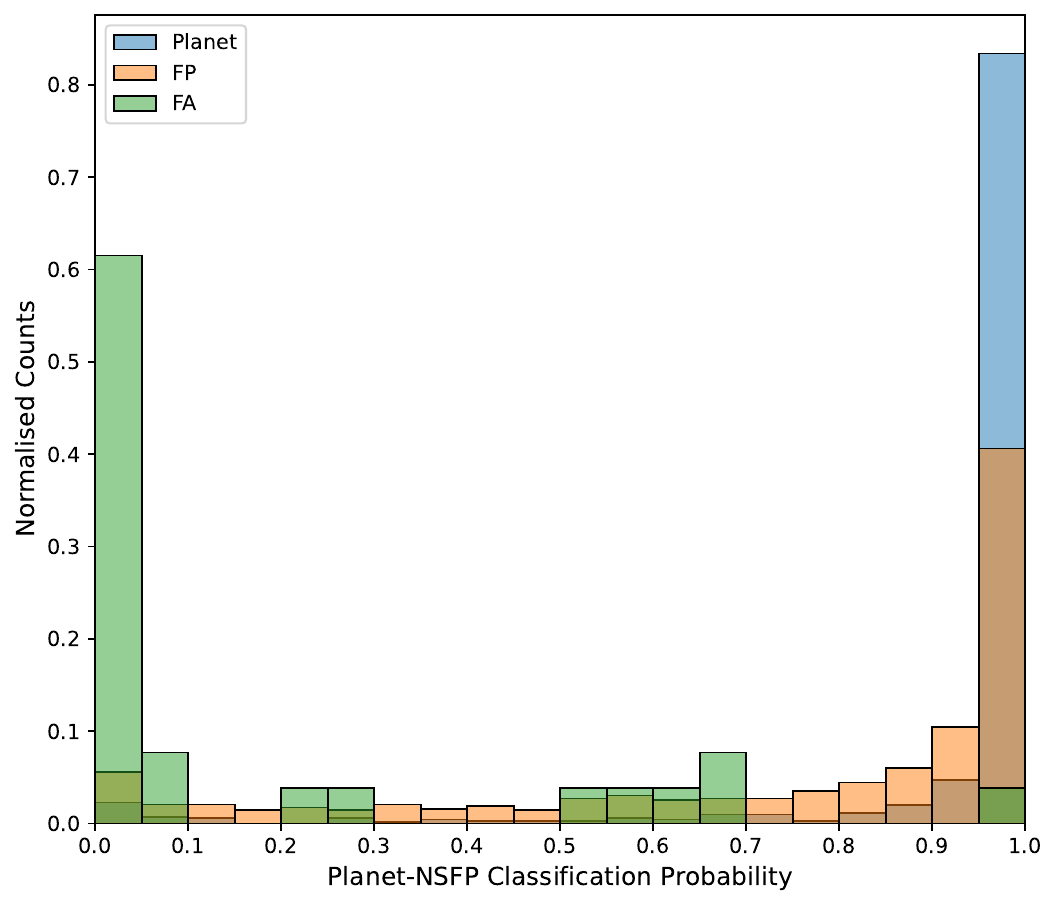}
    \caption{Histogram showcasing the distribution of the mean posterior probability from the two Planet-NSFP classifiers of the pipeline for all Planet, FP and FA TOIs in our analysis sample.}
    \label{fig:TOI_NSFP}
\end{figure}

The distribution of the planet posterior probability from the planet-NSFP classification for all events is presented in Figure \ref{fig:TOI_NSFP}. The planet TOI candidates lie in the majority above the 0.9 probability threshold. At the same time, the majority of the FA TOIs have low probabilities, most below 0.1, and thus lie well below the vetting threshold. Meanwhile, the FP TOIs are more evenly spread across the probability spectrum while still favouring higher probabilities. The Planet-NSFP classifiers are not trained to identify and classify eclipsing FP events, leaving this to the other classifiers (Section \ref{sec:NSFP}). The computed probability for these cases therefore depends on how alike they are to the NSFP sample. The expectation is that eclipsing FP events should in general align with the Planet class, as the eclipses are more similar to the transits, with events that show variability in either their eclipses or their lightcurve as a whole leaning towards the NSFP class. 

Overall, the resulting distribution demonstrates that the Planet-NSFP classifiers are effective in identifying and separating the non eclipsing events while retaining the majority of the true planet candidates. However, we do note that there are cases of FA TOIs with high NSFP probability and Planet TOIs with low probabilities. We address and investigate those cases in more detail in Sections \ref{sec:FA_TOIs} and \ref{sec:PlanetTOIs}.

\subsection{False Positives}\label{sec:EB&NEB_TOIs}
As seen in the stacked histogram of Figure \ref{fig:TOI_Probs}, only 39 FP TOIs had a minimum posterior probability greater than 0.5. Of those, 12 had probabilities greater than 0.9, pushing them into the "Likely Planet" classification. While such a result for these FPs is obviously not correct, it should not detract from the fact that more than 90\% of the 630 FPs have been correctly assigned low probabilities by the pipeline. However, it is important to investigate these problematic cases as they allow us to identify the limitations of the pipeline and potentially point towards future improvements. The 12 FP events with probabilities greater than 0.9 are presented in Table \ref{tb:fp_features}.

\begin{table*}
   \centering
   \caption{The list of FP TOIs with planet posterior probability greater than 0.9. The implied planetary radius is computed by \texttt{RAVEN}'s feature generation process.}
   \label{tb:fp_features}
    \begin{tabular}{cc|cccc}
        \hline
        \textbf{TIC ID} & \textbf{TOI} & \textbf{TFOP Disposition} & \textbf{Implied Radius ($R_\oplus$)} & \textbf{\texttt{RAVEN} Probability} & \textbf{Positional Probability} \\ \hline
            294090620 & 587.01 & EB & 15.4 & 0.952 & 1.000 \\
            157568289 & 645.01 & EB & 14.6 & 0.990 & 1.000 \\
            405904232 & 1312.01 & EB & 14.0 & 0.953 & 1.000 \\
            348673213 & 1639.01 & NEB & 3.8 & 0.961 & 0.947 \\
            82707763  & 1991.01 & EB & 15.8 & 0.939 & 1.000 \\
            99330090  & 3883.01 & EB & 12.7 & 0.996 & 1.000 \\
            18017227  & 3890.01 & EB & 7.1 & 0.928 & 1.000 \\
            139118482 & 3914.01 & EB & 13.0 & 0.993 & 0.501 \\
            237235093 & 4102.01 & NEB & 1.6 & 0.953 & 0.302 \\
            1950736   & 4610.01 & NEB & 1.5 & 0.983 & 0.070 \\
            193628179 & 5198.01 & EB & 13.5 & 0.993 & 1.000 \\
            35377903  & 6693.01 & EB & 12.2 & 0.999 & 1.000 \\ \hline
   \end{tabular}
\end{table*}

The majority of these high probability FP events are EBs, with just a total of 3 NEB events included. This is despite the fact that NEBs are the most populous type of FPs in our sample, which suggests that the pipeline is indeed effectively classifying the NEB events as FPs. In fact, for 2 of the NEB events, namely TOI-4102.01 and TOI-4610.01, the pipeline determined that they had low Positional Probabilities and correctly assigned the highest probability to the true hosts of the events as identified by follow-up observations. Moreover, TOI-4102.01 was also flagged by the pipeline, highlighting it as a problematic event. These two TOIs demonstrate that when assessing the candidates, especially for validating them, the complete output of the pipeline, and in particular the Positional Probabilities, should be taken into account as they can help identify cases where the posterior probabilities are less effective. For TOI-1639.01, neither the posterior probability nor the positional probability for the event point to its true nature. An examination of its lightcurve and centroid did not reveal any particular signs of the event being an NEB, especially since there was only one available SPOC lightcurve from Sector 18. While this is a single case and thus not representative of the pipeline's overall performance, we do urge for caution when vetting and validating candidates with a single sector of data as it increases the potential for error.

The 9 high probability EB cases arise from the known degeneracy between the photometric signal of large planets and low-mass EB companions. This degeneracy makes it practically impossible to truly separate the two cases in photometry alone, requiring the use of spectroscopic follow up observations to determine the companion's mass and thus ascertain their nature. Out of the 9 cases, 8 were identified as EBs based on their spectroscopic data, with the exception of TOI 645.01 for which the flux extracted with a different aperture revealed out-of transit variability consistent with an EB. An examination of the lightcurves for all 9 targets showed relatively shallow eclipses, with shapes consistent with planetary transits rather than the characteristic "V" shape of EBs and without any out of transit variability or secondary eclipses. As such, photometrically, the candidates appear Planet-like.

The Planet-EB photometric degeneracy has resulted in previously validated \textit{Kepler} candidates to be later re-classified as FPs following further follow-up observations, for example in \citet{Avi2017_vespa_issue}. As per the study, validation methods should be limited to candidates with radius less than $8R_\oplus$, as above it the degeneracy between planets, brown dwarfs and low-mass EBs can lead to cases of missclassification. This prompted the \texttt{TRICERATOPS} validation framework to adopt this limit \citep{Triceratops}. Moreover, the ExoMiner+++ vetting framework has reported reduced precision for their classifications in the $10$ to $16R_\oplus$ regime \citep{ExoMinerTESS}. In \texttt{RAVEN}'s case, we also adopt the $8R_\oplus$ validation limit. This ensures that none of the problematic cases would be validated, with the only remaining high probability case being TOI 3890.01 at 0.928. It should be noted that results above this limit can still be used for ranking purposes, as they greatly reduce the number of EBs that would need to be followed up, with True planets having much higher probabilities in general.

\begin{table*}
\centering
\caption{A breakdown of the number of TOI events in each TFOP disposition group and their minimum probability scenarios. Candidates with probabilities above 0.9 were assigned to the Planet scenario.}
\label{tb:FPs}
\begin{tabular}{l|ccccccccc|c}
\hline
\textbf{TFOP Disposition} & \textbf{Planet} & \textbf{EB} & \textbf{HEB} & \textbf{BEB} & \textbf{NEB} & \textbf{NHEB} & \textbf{NTP} & \textbf{HTP} & \textbf{NSFP} & \textbf{Total} \\
\hline
EB    & 9  & 98 & 37 & 1 & 17  & 1  & 0  & 4 & 5  & 172 \\
NEB   & 3  & 2  & 4  & 0 & 296 & 52 & 45 & 5 & 15 & 422 \\
NPC   & 0  & 0  & 0  & 0 & 12  & 3  & 12 & 1 & 3  & 31  \\
BEB   & 0  & 2  & 0  & 0 & 0   & 0  & 0  & 0 & 0  & 2   \\
FP    & 0  & 0  & 0  & 0 & 1   & 0  & 2  & 0 & 0  & 3   \\
\hline
\textbf{Total} & 12 & 102 & 41 & 1 & 326 & 56 & 59 & 10 & 23 & 630 \\
\hline
\end{tabular}
\end{table*}

Beyond overall FP identification, we test whether \texttt{RAVEN} can be used to infer the true scenario of the FP events. Specifically, we assess the pipeline's propensity to assign the lowest Planet-FP probability to the true scenario of the FP TOI events. For this test, we rely exclusively on the dispositions assigned to the TOI events by the TESS Working Group as published on ExoFOP. It should be noted that the EB label includes TOIs for which their TFOP disposition is either EB, SEB, SEB1 or SEB2. The later 3 refer specifically to spectroscopically confirmed EBs, with lines from single (SEB1) or two stars (SEB2) detected. The EB label also includes hierarchical binary systems, as they are not differentiated in their classification.

We examine the planet-FP posterior probabilities for all 630 FP TOI events in our sample. Of these, 172 were assessed to be EBs and 422 to be NEBs, representing the two most populous cases. Moreover, 31 FP TOIs were classified as NPCs, which are candidates where the event occurs on a known nearby star and for which the depth of the event on the true host suggests that it could be a planet. Finally, the sample included 2 BEB TOIs and 3 events for which their exact classification was not determined and as such are simply labelled as FPs. For all events, the FP scenario with the lowest planet posterior probability was identified to investigate whether it was consistent with their TFOP classification. TOIs with probabilities greater than 0.9 across all scenarios were assigned to the Planet scenario. For the NPC TOIs, determining the `correct' FP scenario is not straightforward as the true nature of the companions is not specified. The number of TOIs from each TFOP classification that had each minimum \texttt{RAVEN} FP scenario as their lowest is presented in Table \ref{tb:FPs}. 

In the majority of cases, the minimum scenario is aligned with the TFOP disposition. This is particularly true for the NEB events, where the NEB scenario was the lowest for 70\% of them, while any nearby FP scenario was the lowest in about 93\% of the cases. Considering that differentiating between the 3 nearby FP scenarios would require in practice dedicated observations of the true host to ascertain the nature of the companion, the distribution of the NEB TOIs across them is not of paramount importance. This is also true for the NPC events, where again the nearby FP events were in the majority the lowest scenario. For this class of TFOP FPs, it is interesting that the events are split between the NEB and NTP scenarios, in contrast to the NEB TOIs. This highlights that a significant portion of them could indeed be planetary companions on nearby stars. 

For the EB TOIs, the EB scenario was the lowest for more than 50\% of the cases. This result is particularly noteworthy when considering that \texttt{RAVEN}'s EB scenario is primarily geared towards EBs with deep eclipses, which are removed by the SPOC pipeline's vetting process. For reference, the median depth across the EB TOIs was only 7000 ppm, while for the synthetic EBs it was 95000 ppm. The EB TOIs in our sample were therefore more likely to be aligned with the diluted EB scenarios. The fact that they were not, showcases the strength of the pipeline in assessing the scenarios beyond just the shape and depth of the event. Examining the distribution of the EB TOIs across all EB scenario configurations, we find that they accounted for 154 out of the 172 EB TOIs, representing 89.5\% of the total. 

It is also noteworthy that the HEB was the second most populous selected scenario for the EB events. This is important because the TFOP disposition does not differentiate between the EB and HEB events. One notable example is TOI-4838.01, which was a previously validated planet, designated as K2-399 b, and was recently found to be a HEB \citep{VESPA_HEB}. \texttt{RAVEN}'s Planet-HEB probability correctly had the lowest planet posterior probability for this TOI, demonstrating that the pipeline is capable of identifying such events. As such, we consider that many of the TFOP EB TOIs where the lowest scenario was the HEB might indeed be true HEB events. This is further supported by the fact that the median \textit{Gaia} RUWE score across these 37 TOIs was 1.88, well above the threshold of 1.4 used to signify the presence of a wider stellar companion. As noted in \cite{Gaia_Belokurov}, stars above this limit were mostly in hierarchical systems. In comparison, the median RUWE score for the remaining EB TOIs was only 1.08, clearly showcasing the difference between the two groups.

Overall, the conclusion from the test on the scenario specific probabilities is that the pipeline is highly capable of offering an insight into the true nature of the \textit{TESS} candidates, even if the actual posterior probability for the scenarios is not calculated. As shown by our analysis, the scenario with the lowest probability is in most of cases a good proxy for the true FP classification. However, it is important to clarify that often the difference in the probabilities between some of the scenarios could be marginal and that this should be considered if one attempts to use the \texttt{RAVEN} probabilities in practice to identify the FP event. The probabilities provided by the pipeline do include the planet-FP probabilities across all scenarios, offering flexibility and depth when assessing the candidate scenario.

\subsection{False Alarms}\label{sec:FA_TOIs}
Next we investigate the pipeline's performance on the False Alarm TOIs.  
In Table \ref{tb:FA} we present the minimum posterior probability for the 26 FA TOIs in our sample, along with the FP scenario that had the lowest probability. As discussed in Section \ref{sec:nsfp_results}, the expectation is that most FA TOIs should have the lowest planet posterior probability against the NSFP scenario as it is the one that encompasses variability and systematic noise. Additionally, we note whether or not the TOI was flagged by the pipeline as having high variability in its transit depth across sectors. 

\begin{table*}
    \centering
    \caption{The list of FA TOIs, with their minimum planet posterior probability and the FP scenario associated with it. The last column provides a brief reasoning for their classification based on the assessments on ExoFOP and from our own findings.}
    \label{tb:FA}
    \begin{tabular}{cc|cccl}
    \hline
        \textbf{TIC ID} & \textbf{TOI} & \textbf{\texttt{RAVEN} Probability} & \textbf{Minimum Scenario} & \textbf{Flagged} & \textbf{Notes} \\ \hline
        326453034 & 223.01  & 0.0003 & NSFP   & 0 & Systematic FA \\
        160074939 & 230.01  & 0.0036 & NSFP   & 0 & No significant transiting event detected. \\
        93125144  & 523.01  & 0.0274 & NSFP   & 1 & Stellar Variability \\
        365690646 & 524.01  & 0.0005 & NSFP   & 0 & No significant transiting event detected. \\
        150247134 & 718.01  & 0.0001 & NSFP   & 1 & Stellar variability, which evolved over sectors. \\
        125405602 & 821.01  & 0.2740 & NSFP   & 0 & Potentially systematic FA \\
        219380235 & 878.01  & 0.6589 & NTP    & 1 & No significant transiting event detected. \\
        102283403 & 896.01  & 0.0220 & NSFP   & 0 & Stellar variability \\
        245947683 & 950.01  & 0.0561 & NSFP   & 1 & No significant transiting event detected. \\
        47384844  & 1022.01 & 0.0491 & NSFP   & 0 & Stellar variability \\
        13499636  & 1275.01 & 0.0002 & NSFP   & 0 & No significant transiting event detected. \\
        43064903  & 1276.01 & 0.0014 & NEB    & 0 & Appears to be a NEB. \\
        16920150  & 1459.01 & 0.5776 & NSFP   & 0 & Systematic FA. Event confirmed as FA based on sector 57 data. \\
        129991079 & 1602.01 & 0.0023 & NSFP   & 0 & Stellar variability. \\
        229951289 & 1609.01 & 0.0035 & NSFP   & 1 & No significant transiting event detected. \\
        165551882 & 1633.01 & 0.0149 & NSFP   & 1 & No significant transiting event detected. \\
        459903429 & 1671.01 & 0.0014 & NSFP   & 1 & No significant transiting event detected. \\
        310981412 & 1790.01 & 0.0175 & NSFP   & 1 & No significant transiting event detected. \\
        159244568 & 2078.01 & 0.0518 & NSFP   & 1 & No significant transiting event detected. \\
        75656058  & 2171.01 & 0.0018 & NSFP   & 0 & No significant transiting event detected. \\
        76197937  & 2172.01 & 0.0256 & NTP    & 1 & No significant transiting event detected. \\
        441798995 & 2269.01 & 0.0522 & NHEB    & 1 & Shallow eclipses in some sectors. \\
        321068176 & 2409.01 & 0.0014 & NEB    & 1 & Eclipses appear in only one sector. \\
        416539317 & 3794.01 & 0.2256 & NEB    & 0 & Eclipses appear in only one sector. \\
        143022742 & 4323.01 & 0.0083 & NSFP   & 0 & No significant transiting event detected. \\
        376981340 & 4528.01 & 0.9990 & Planet & 0 & Duplicate of TOI 4545.01 (TIC 610976842). \\
        \hline
    \end{tabular}
\end{table*}

Almost all of the FA TOIs had probabilities inconsistent with the Planet scenario, showcasing that the pipeline is capable of identifying them. There were only 3 FA TOIs for which the minimum posterior probability was above 0.5, with only one above 0.9. This was TOI 4528.01, which represents a unique case of a TOI issued for a duplicated entry of a star in the TIC catalogue. TOI 4528.01 is essentially a duplicate of TOI-4545.01, which is the known planet K2-295 b. As such, its high probability is justified. 
As for the remaining 2 FA TOIs with probabilities above 0.5, their probabilities are well below the 0.9 vetting threshold, with 1 of the 2 also flagged by the pipeline as problematic. As such, these TOIs would not be considered as likely planets.

Finally, the scenario with lowest planet posterior probability for all FA TOIs was for the majority the NSFP scenario, in line with our initial expectations. This was especially true for the events assessed to be due to stellar variability, which are among the main targets that the NSFP scenario aims to identify. There were only a few exceptions to the above, which had a nearby FP scenario as the lowest. The most notable of these is TOI-1276.01, for which we have visually confirmed that eclipses were present in all 5 sectors of TESS observations. The event also has a low positional probability and based on its planet-NEB probability, we consider this likely to be an NEB. Overall, \texttt{RAVEN}'s results for the FA TOIs showcase again the pipeline's potential in providing a useful insight into the true nature of the events.

\subsection{Confirmed and Known Planets}\label{sec:PlanetTOIs}

Following the reduction of our initial TOI sample, a total of 397 Known Planets and 308 Confirmed Planets were left, for a total of 705 planets. As a reminder, Confirmed Planets refer exclusively to candidates discovered through the \textit{TESS} mission and which have subsequently been confirmed as true exoplanets. The Known Planets on the other hand include planets that have been detected and confirmed from a variety of different missions prior to TESS, such as \textit{Kepler} \citep{Kepler} and its K2 Mission extension \citep{K2}, WASP \citep{WASP} and others. For the purpose of this analysis we do not differentiate between the two groups and refer to them simply as Planets. This collection of Planets includes both those confirmed through follow-up radial velocity to determine their mass and validated planets. For the latter, their validation was achieved using mostly \texttt{VESPA} and \texttt{TRICERATOPS}. 

The Planet TOIs also include Brown Dwarfs (BD), which are not differentiated either in ExoFOP or the Exoplanet Archive. \texttt{RAVEN} does not currently incorporate BDs in either the simulated Planet scenario or as a stand-alone simulated FP scenario. Given BDs are typically similar in radius to Jupiter-like planets, they are likely to be classified as planets by the pipeline.

As already seen in Figure \ref{fig:TOI_Probs}, most Planet TOIs have high planet posterior probability, with 81\% above the 0.5 threshold. The majority of these, specifically 420, lay above the 0.9 probability threshold, crossing into the "Likely Planet" regime. Moreover, a total of 210 Planet TOIs crossed the 0.99 threshold required for statistical validation, representing about 30\% of the total Planet sample. These results, in light of the pipeline's performance on the FP and FA TOIs, show that \texttt{RAVEN} succeeds in vetting, ranking and validating true Planet candidates. However, the results also show that there is room for improvement, with a total of 134 Planet TOIs having probabilities lower than 0.5. A true planet having a probability lower than 0.5 is not by itself a wrong outcome as it simply conveys the pipeline's uncertainty about the planetary nature, and some known planets may not be confidently identifiable from \textit{TESS} photometry alone. We investigate these cases here to gain further insight into the pipeline.

We plot the Planet TOIs as a function of their Implied Planet Radius, calculated through our feature generation, and orbital period in Figure \ref{fig:KP_per_rad}. The Figure clearly shows that the probability drops sharply above a planetary radius of $16R_{\oplus}$. As addressed before, this behaviour is explainable as the simulated Planet scenario was limited to a radius range between $1R_{\oplus}$ and $16R_{\oplus}$. The pipeline therefore cannot consider candidates above this radius as true planets as they are not included in its training. While these candidates could have been removed prior to testing the pipeline's performance as they fall outside its parameter space, we decided to include them to demonstrate the strong dependence of the performance on the training sets and the hard limits they place to the pipeline's effective operating space. 

Excluding the TOIs with radius above $16R_{\oplus}$, the number of Planets with probabilities less than 0.5 falls from 134 to 67. Examining their distribution in Figure \ref{fig:KP_per_rad} does not reveal any clear patterns in terms of the location of these Planet TOIs, particularly in relation to the known Exoplanet population. This is also true for the high probability planets, which appear in both lightly and densely populated regions. As such, we can surmise that the final probability is not strongly dependent on the planetary occurrence rates.

\begin{table*}
\centering
\caption{Known and Confirmed planets with \texttt{RAVEN} planet probability below 0.01.
\label{tb:low_prob_planets}}
\begin{tabular}{cc|ccl}
\hline
\textbf{TIC ID} & \textbf{TOI} & \textbf{\texttt{RAVEN} Probability} & \textbf{Minimum Scenario} & \textbf{Notes} \\
\hline
410214986  & 200.01   & $7.7 \times 10^{-5}$  & NSFP & Young active host star, Variability \\
224225541  & 251.01   & $2.8 \times 10^{-5}$  & NSFP & Young active host star, Variability \\
257605131  & 451.03  & $4.8 \times 10^{-3}$  & NSFP & Confirmed multi-planet system \\
146520535  & 942.01   & $5.3 \times 10^{-5}$  & NSFP & Young active host star, Variability \\
383390264  & 1098.01  & $4.0 \times 10^{-3}$  & NSFP & Young active host star, Variability \\
267572272  & 4489.01  & $1.4 \times 10^{-20}$ & NEB  & Variability \\
374180079  & 5520.01  & $7.4 \times 10^{-5}$  & NSFP & Confirmed multi-planet system \\
456945304  & 5559.01  & $2.7 \times 10^{-4}$  & NSFP & Variability (non-stellar) \\
\hline
\end{tabular}
\end{table*}

To gain a deeper insight into the Planet TOIs with low probabilities, we examined all 8 cases which had a probability below 0.01. We list all 8 TOIs in Table \ref{tb:low_prob_planets}, along with their \texttt{RAVEN} probability, their corresponding lowest FP scenario and an observation note. Based on the latter, a clear picture emerges on the reasons for their low probability. The 8 cases can be split into two distinct groups. The first are planets orbiting stars with high photometric variability, most of them young, for which the pipeline was unable to detrend the stellar signal. As a result, the variability in the lightcurve results in low NSFP probabilities. This outcome is not only understandable but also desirable as the pipeline essentially attempts to classify variability. To properly assess these cases, more specialised and versatile detrending methods are required to remove the stellar signal before applying the pipeline. This should be considered an essential pre-processing step for such cases and should be performed prior to using the \texttt{RAVEN} pipeline.

The second group involves Planet TOIs in systems with multiple candidates, most of which are also confirmed. This group accounts for 2 of the 8 TOIs below 0.01. Extending this up to the probability threshold of 0.1, we found that multi-candidate systems represented 9 of the 25 Planet TOIs below it. A naive interpretation of this would be that the pipeline struggles with correctly assessing multi-candidate systems, especially since they are not represented in the simulated Planet events. However, examining the probabilities of all 91 Planet TOIs in our sample with at least one additional detected TOI in their system, we found that they occupied the whole of the probability spectrum, with a total of 38 having probabilities greater than 0.9. The mean and median probabilities across the whole group were 0.70 and 0.86 respectively. Overall, this shows that the pipeline's performance on candidates with additional candidates in the system is complicated and depends on more factors than just the mere presence of additional transits in the lightcurve. 

A more in depth examination of the low probability Planet TOIs in multiplanet systems revealed that the final probability was also dependent on various other features including the depth of the event, its SNR, whether it was grazing, on its ephemeris and on the level of dilution. Regardless of the causes that led to the low probability, it is important to acknowledge that the presence of additional candidates can cause the pipeline's performance to degrade. As such, we suggest that when evaluating a candidate in a multi-candidate system, the transits of any additional candidates should be masked. While the pipeline is able to perform this function, we opted to not enable it for this paper, as we wanted to evaluate the pipeline's performance as it would perform when run on candidates retrieved directly from a survey. In such a scenario, removing the additional candidates would be problematic due to the potentially recovered aliases of the event. This option therefore should only be reserved for targeted assessment of \textit{TESS} candidates. However, we can confirm that masking the other candidates in the multi-candidate systems when applying the pipeline resulted in the improvement of their probabilities, with the number of Planets above 0.9 increasing to 51 and the median and mean probability for the group increasing to 0.78 and 0.94 respectively.

Finally, examining the rest of the low probability Planet TOIs, again showcased that different reasons were responsible for the assigned probability. Most common among them were extreme grazing transits and low SNR. For the former, we found that the median probability for candidates with a grazing statistic less than 0.2 was 0.39, suggesting that pipeline is more likely to rank these candidates lower as a result of their shape. As for the latter, the median SNR and MES for those with probability less than 0.5 was 33 and 41 respectively, while for those above the threshold their median was 101 and 240. The difference between the two groups shows that the pipeline considers less significant events as more unlikely to be true planets. It should be noted though that the final probability assigned to a candidate is not dependent on just a single feature but a combination of them. In some cases with low transit SNR, we found that the period of the event was not correct, with the resulting probability well justified.

Overall, the results demonstrate the pipeline's effectiveness in assessing true Planet candidates. They also showcase the importance of examining the results on each individual candidate based on not just the final \texttt{RAVEN} probability, but also the scenario specific and the positional probabilities, to obtain a more in-depth understanding of the pipeline's reasoning. This fine-grained analysis provided by the pipeline is one of its greatest strengths and should not be understated.

\subsection{Validation, Likely Planets and the 8$R_{\oplus}$ Threshold}
\label{sect:validationlimits}
A total of 210 Planet TOIs results were above the 0.99 probability threshold and would have been considered validated if they were new discoveries. However, as discussed in Section \ref{sec:EB&NEB_TOIs},5 FP TOI also crossed the validation threshold, all with implied radii above $8R_\oplus$. We intend to explore in the future if this problem can be overcome by adding to the pipeline separate training sets focusing on low mass EBs and BDs. For this paper we adopt the upper limit of 8$R_{\oplus}$ for planets to be considered validated by \texttt{RAVEN}, consistent with other validation works \citep{Triceratops}. The probabilities for candidates with radius above this limit should be used for vetting and ranking purposes.

For general use, we also recommend avoiding validation of candidates flagged by the pipeline for transit depth variations, and candidates with Positional Probability less than 0.5. Applying these cuts to our TOI test sample leaves 249 TOIs, of which 226 are true Planets, 16 FPs and 7 FAs. Of these, 141 Planets and 2 FPs crossed the pipeline's 0.9 `likely planet' vetting threshold, with the FPs representing just 1.4\% of the total. Moreover, none of the FPs had probabilities above the 0.99 validation threshold, which was successfully crossed by 68 Planet TOIs. They represent 26\% of the Planet TOIs in our test sample with radius below the 8$R_{\oplus}$ limit, demonstrating the pipeline's validation potential.

\section{Conclusions}
This work presents a new pipeline for vetting, ranking and statistically validating \textit{TESS} candidates, called \texttt{RAVEN}. The pipeline uses two different Machine Learning models, namely a Gradient Boosted Decision Tree and a Gaussian Process, to derive the planet posterior probability against 8 FP scenarios.
The collection of probabilities per scenario is reduced to a single `\texttt{RAVEN}' probability by taking the minimum, expressing the lowest confidence of the pipeline of the candidate being a true planet, to provide a single vetting output.

A key improvement on previous work is the substantially larger training set that is split by FP scenario, enabled by extensive synthetic simulations. This allows for detailed vetting analysis of specific false positive types and the ability to test performance and individual candidates as a function of FP scenario, a first for an ML-based validation pipeline. New developments also include the incorporation of the \textit{Gaia} RUWE score and source information to inform FP scenario priors, and the ability to automate \textit{TESS} planet discovery and validation at scale. Fast vetting and validation without a human component is a critical element for future bias-controlled studies of planet demographics. The \texttt{RAVEN} pipeline is almost fully self contained, relying only on \textit{TESS} SPOC lightcurve data, and stellar characteristics for the host stars of the candidates from the TIC and \textit{Gaia}.

Currently the pipeline can be run on candidates with an orbital period between 0.5 and 16 days, a \textit{TESS} SPOC FFI lightcurve and a transit depth greater than 300ppm. Candidates with apparent radii above $8R_\oplus$ can be vetted but should not be validated, due to the photometric degeneracy with brown dwarfs and very low-mass stars at these radii. Subject to these constraints, we consider candidates with a \texttt{RAVEN} probability greater than 0.9 to be "Likely Planets" and those above 0.99 to be statistically validated.

The pipeline was tested on subsets of the training dataset, showcasing its effectiveness in identifying FPs and FAs and in ranking candidates, with AUC scores of 97--99\% depending on FP scenario. To robustly test the performance of the pipeline in practice on a completely independent sample, we also ran the pipeline on 1361 pre-classified TOIs, showing remarkably strong performance despite these TOIs often being classified using further follow-up observations we did not use, with an accuracy of 90\% for TOIs in our pipeline's parameter space. The results from the Planet TOIs also established the pipeline's validation potential with 26\% of TOI planets below 8$R_{\oplus}$ crossing the 0.99 validation threshold.

Future development may be able to extend the pipeline's orbital period and radius limits, as well as supporting \textit{TESS} QLP lightcurves. We also hope that our extensive simulation infrastructure will be independently useful, for testing other algorithms and supporting the development of future missions such as PLATO or NASA Roman. Code to run the \texttt{RAVEN} pipeline on new \textit{TESS} candidates and generate new simulations has been made available to the community. The pipeline is also deployed online as a cloud-hosted app, for ease of use without any additional setup requirements.

\section*{Data Availability}

Code to generate our simulated planets and false positives is publicly available at \url{https://github.com/ckm3/pastis-dev}. The \texttt{RAVEN} pipeline code itself is available at \url{https://github.com/ahadjigeorghiou/RAVEN-Pipeline}, including scripts to run the pipeline on new candidates. 

\section*{Acknowledgments}
This paper includes data collected by the \TESS\ mission. Funding for the \TESS\ mission is provided by the NASA Explorer Program. Resources supporting this work were provided by the NASA High-End Computing (HEC) Program through the NASA Advanced Supercomputing (NAS) Division at Ames Research Center for the production of the \textit{SPOC} data products.

We acknowledge the use of public \TESS\ Alert data from pipelines at the \TESS\ Science Office and at the \TESS\ Science Processing Operations Center.

This research has made use of the NASA Exoplanet Archive, which is operated by the California Institute of Technology, under contract with the National Aeronautics and Space Administration under the Exoplanet Exploration Program.

This research has made use of the Exoplanet Follow-up Observation Program (ExoFOP; DOI: 10.26134/ExoFOP5) website, which is operated by the California Institute of Technology, under contract with the National Aeronautics and Space Administration under the Exoplanet Exploration Program.

AH was supported by an STFC studentship. 
This research was funded by the UKRI (Grants ST/X001121/1, EP/X027562/1).





\bibliographystyle{mnras}
\bibliography{ref}



\bsp	
\label{lastpage}
\end{document}

%% file: sim_table.tex
\begin{center}
\onecolumn
\begin{longtable}{c|c}
\caption{The parameters used for the creation of the simulated astrophysical scenarios along with the corresponding distribution, method or catalogue from which they were drawn. In the table, the parameters of the target sources, which correspond to an observed TIC star, are listed first as they are common to all scenarios. The parameters specific to the planetary or eclipsing binary companions are listed after, which were kept the same for each scenario involving a transiting planet or eclipsing binary. Last are the parameters of the host star for the bound hierarchical scenarios and the unresolved background scenarios.}

\label{tb:astro_param_distributions}\\
\hline
\textbf{Parameter} & \textbf{Source} \\
\hline
\multicolumn{2}{c}{\textbf{Target Star Parameters}} \\
\hline
Effective Temperature ($T_{\text{eff}}$) & SPOC-Gaia \citep{LaurenSample}\\
Surface Gravity (log $g$) & SPOC-Gaia \citep{LaurenSample} \\
Metallicity ([Fe/H]) & SPOC-Gaia \citep{LaurenSample} \\
Radius ($R_{\odot}$) & SPOC-Gaia \citep{LaurenSample} \\
Mass ($M_{\odot}$) & Derived from log $g$ and radius from SPOC-Gaia \citep{LaurenSample} \\
TESS magnitude ($T_{\text{mag}}$) & SPOC-Gaia \citep{LaurenSample} \\
Gaia magnitude ($G_{\text{mag}}$) & SPOC-Gaia \citep{LaurenSample} \\
BP-RP color & SPOC-Gaia \citep{LaurenSample} \\
Extinction ($A_v$) & 3.1 $\times$ E(B$-$V) from SPOC-Gaia \citep{LaurenSample}\\
Age & Isochrone fitting \\
\hline
\multicolumn{2}{c}{\textbf{Planetary Companions}} \\
\hline
Planet radius & $ 1R_{\oplus} \leq R_{P} \leq 16R_{\oplus} $, \citep{Hsu2019} \\
Planet mass & Statistical mass-radius relation \citep{mullerMassradiusRelationExoplanets2024} \\
Albedo & $\mathcal{U}(0, 1)$ \\
Orbital period (P) & $0.5d \leq P \leq 16d$, \citep{Hsu2019} \\
Eccentricity & Statistical relation with orbital period \citep{MoeDiStefano2017} \\
Argument of periastron ($\omega$) & $\mathcal{U}(0, 2\pi)$ \\
Inclination angle ($i$) & $\sin(i)$, $0 \leq i \leq \pi/2$ \\
\hline
\multicolumn{2}{c}{\textbf{Eclipsing Binary Companions}} \\
\hline
\textit{Secondary Star Parameters} & \\
Mass ratio & \citep{MoeDiStefano2017}, constrained based on primary's mass and the binary orbital period \\
Albedo & $\mathcal{U}(0.4, 1)$ \\
Mass & Derived from mass ratio and target mass \\
Age & Same as primary star \\
Distance & Primary star distance \\
$T_{\text{eff}}$ & Isochrone fitting \\
log $g$ & Isochrone fitting \\
$T_{\text{mag}}$ & Isochrone fitting \\
$G_{\text{mag}}$ & Isochrone fitting \\
BP-RP & Isochrone fitting \\
Radius & Isochrone fitting \\
\textit{Orbital Parameters} & \\
Orbital period & $\log_{10} P \sim \mathcal{N}(\mu=5.03, \sigma=2.28)$ truncated at $(-\infty, \log_{10} 16)$ \citep{Raghavan2010}\\
Eccentricity & Statistical relation with orbital period \citep{MoeDiStefano2017} \\
Argument of periastron ($\omega$) & $\mathcal{U}(0, 2\pi)$ \\
Inclination angle ($i$) & $\sin(i)$, $0 \leq i \leq \pi/2$ \\
\hline
\multicolumn{2}{c}{\textbf{Hierarchical Systems (HTP and HEB)}} \\
\hline
\textit{Secondary Star Parameters} & \\
Mass & IMF \citep{robinSyntheticViewStructure2003} \\
Age (yr) & Target Star age \\
Metallicity ([Fe/H]) & Target Star [Fe/H] \\
Distance & Target Star distance \\
Albedo & $\mathcal{U}(0.4, 1)$ \\
$T_{\text{eff}}$ & Isochrone fitting \\
log $g$ & Isochrone fitting \\
$T_{\text{mag}}$ & Isochrone fitting \\
$G_{\text{mag}}$ & Isochrone fitting \\
BP-RP & Isochrone fitting \\
Radius & Isochrone fitting \\
Orbital period & 10000 days \\
Eccentricity & Statistical relation with orbital period \citep{MoeDiStefano2017} \\
Argument of periastron ($\omega$) & $\mathcal{U}(0, 2\pi)$ \\
Inclination angle ($i$) & $\sin(i)$, $0 \leq i \leq \pi/2$ \\
\textit{Tertiary Parameters} &  \\
HTP & As per Planet Companion above \\
HEB & As per Eclipsing Binary Companion above \\
\hline
\multicolumn{2}{c}{\textbf{Background Blended Systems (BEB and BTP)}} \\
\hline
\textit{Primary Star Parameters} & \\
Mass & IMF \citep{robinSyntheticViewStructure2003} \\
Age (Gyr) & $\mathcal{U}(6, 10)$ \\
Metallicity ([Fe/H]) & $\mathcal{U}(-2.5, 0.5)$ \\
Distance & Milky Way exponential decreasing space density \citep{bailer-jonesEstimatingDistanceParallaxes2018} \\
Albedo & $\mathcal{U}(0.4, 1)$ \\
$T_{\text{eff}}$ & Isochrone fitting \\
log $g$ & Isochrone fitting \\
$T_{\text{mag}}$ & Isochrone fitting \\
$G_{\text{mag}}$ & Isochrone fitting \\
BP-RP & Isochrone fitting \\
Radius & Isochrone fitting \\
\textit{Secondary Parameters} & \\
BTP & As per Planet Companion above \\
BEB & As per Eclipsing Binary Companion above \\
\hline
\end{longtable}
\twocolumn
\end{center}

%% file: main.bbl
\begin{thebibliography}{}
\makeatletter
\relax
\def\mn@urlcharsother{\let\do\@makeother \do\$\do\&\do\#\do\^\do\_\do\%\do\~}
\def\mn@doi{\begingroup\mn@urlcharsother \@ifnextchar [ {\mn@doi@} {\mn@doi@[]}}
\def\mn@doi@[#1]#2{\def\@tempa{#1}\ifx\@tempa\@empty \href {http://dx.doi.org/#2} {doi:#2}\else \href {http://dx.doi.org/#2} {#1}\fi \endgroup}
\def\mn@eprint#1#2{\mn@eprint@#1:#2::\@nil}
\def\mn@eprint@arXiv#1{\href {http://arxiv.org/abs/#1} {{\tt arXiv:#1}}}
\def\mn@eprint@dblp#1{\href {http://dblp.uni-trier.de/rec/bibtex/#1.xml} {dblp:#1}}
\def\mn@eprint@#1:#2:#3:#4\@nil{\def\@tempa {#1}\def\@tempb {#2}\def\@tempc {#3}\ifx \@tempc \@empty \let \@tempc \@tempb \let \@tempb \@tempa \fi \ifx \@tempb \@empty \def\@tempb {arXiv}\fi \@ifundefined {mn@eprint@\@tempb}{\@tempb:\@tempc}{\expandafter \expandafter \csname mn@eprint@\@tempb\endcsname \expandafter{\@tempc}}}

\bibitem[\protect\citeauthoryear{{Aigrain} \& {Foreman-Mackey}}{{Aigrain} \& {Foreman-Mackey}}{2023}]{GP_Astronomy}
{Aigrain} S.,  {Foreman-Mackey} D.,  2023, \mn@doi [\araa] {10.1146/annurev-astro-052920-103508}, \href {https://ui.adsabs.harvard.edu/abs/2023ARA&A..61..329A} {61, 329}

\bibitem[\protect\citeauthoryear{{Armstrong}, {Pollacco}  \& {Santerne}}{{Armstrong} et~al.}{2017}]{SOM_Dave}
{Armstrong} D.~J.,  {Pollacco} D.,   {Santerne} A.,  2017, \mn@doi [\mnras] {10.1093/mnras/stw2881}, \href {https://0-ui-adsabs-harvard-edu.pugwash.lib.warwick.ac.uk/abs/2017MNRAS.465.2634A} {465, 2634}

\bibitem[\protect\citeauthoryear{{Armstrong}, {Gamper}  \& {Damoulas}}{{Armstrong} et~al.}{2021}]{KeplerPipeline}
{Armstrong} D.~J.,  {Gamper} J.,   {Damoulas} T.,  2021, \mn@doi [\mnras] {10.1093/mnras/staa2498}, \href {https://ui.adsabs.harvard.edu/abs/2021MNRAS.504.5327A} {504, 5327}

\bibitem[\protect\citeauthoryear{{Bailer-Jones}, {Smith}, {Tiede}, {Sordo}  \& {Vallenari}}{{Bailer-Jones} et~al.}{2008}]{QuasarClassification}
{Bailer-Jones} C.~A.~L.,  {Smith} K.~W.,  {Tiede} C.,  {Sordo} R.,   {Vallenari} A.,  2008, \mn@doi [\mnras] {10.1111/j.1365-2966.2008.13983.x}, \href {https://ui.adsabs.harvard.edu/abs/2008MNRAS.391.1838B} {391, 1838}

\bibitem[\protect\citeauthoryear{{Bailer-Jones}, Rybizki, Fouesneau, Mantelet  \& Andrae}{{Bailer-Jones} et~al.}{2018}]{bailer-jonesEstimatingDistanceParallaxes2018}
{Bailer-Jones} C. A.~L.,  Rybizki J.,  Fouesneau M.,  Mantelet G.,   Andrae R.,  2018, \mn@doi [AJ] {10.3847/1538-3881/aacb21}, 156, 58

\bibitem[\protect\citeauthoryear{{Bell} \& {Higgins}}{{Bell} \& {Higgins}}{2022}]{TESS_PRF_software}
{Bell} K.~J.,  {Higgins} M.~E.,  2022, {TESS\_PRF: Display the TESS pixel response function}, Astrophysics Source Code Library, record ascl:2207.008

\bibitem[\protect\citeauthoryear{{Belokurov} et~al.,}{{Belokurov} et~al.}{2020}]{Gaia_Belokurov}
{Belokurov} V.,  et~al., 2020, \mn@doi [\mnras] {10.1093/mnras/staa1522}, \href {https://ui.adsabs.harvard.edu/abs/2020MNRAS.496.1922B} {496, 1922}

\bibitem[\protect\citeauthoryear{{Blei}, {Kucukelbir}  \& {McAuliffe}}{{Blei} et~al.}{2016}]{VariationalELBO}
{Blei} D.~M.,  {Kucukelbir} A.,   {McAuliffe} J.~D.,  2016, \mn@doi [arXiv e-prints] {10.48550/arXiv.1601.00670}, \href {https://ui.adsabs.harvard.edu/abs/2016arXiv160100670B} {p. arXiv:1601.00670}

\bibitem[\protect\citeauthoryear{{Borucki} et~al.,}{{Borucki} et~al.}{2010}]{Kepler}
{Borucki} W.~J.,  et~al., 2010, \mn@doi [Science] {10.1126/science.1185402}, \href {https://ui.adsabs.harvard.edu/abs/2010Sci...327..977B} {327, 977}

\bibitem[\protect\citeauthoryear{{Borucki} et~al.,}{{Borucki} et~al.}{2013}]{BLENDER_5_Hz}
{Borucki} W.~J.,  et~al., 2013, \mn@doi [Science] {10.1126/science.1234702}, \href {https://ui.adsabs.harvard.edu/abs/2013Sci...340..587B} {340, 587}

\bibitem[\protect\citeauthoryear{Breiman}{Breiman}{2001}]{breiman2001random}
Breiman L.,  2001, Machine learning, 45, 5

\bibitem[\protect\citeauthoryear{{Caldwell} et~al.,}{{Caldwell} et~al.}{2020}]{FFI}
{Caldwell} D.~A.,  et~al., 2020, \mn@doi [Research Notes of the American Astronomical Society] {10.3847/2515-5172/abc9b3}, \href {https://ui.adsabs.harvard.edu/abs/2020RNAAS...4..201C} {4, 201}

\bibitem[\protect\citeauthoryear{Catanzarite}{Catanzarite}{2015}]{AutovetterDR24}
Catanzarite J.~H.,  2015, \mn@doi [NASA Ames Research Center] {10.26133/NEA23}

\bibitem[\protect\citeauthoryear{{Chen} \& {Guestrin}}{{Chen} \& {Guestrin}}{2016}]{xgboost}
{Chen} T.,  {Guestrin} C.,  2016, \mn@doi [arXiv e-prints] {10.48550/arXiv.1603.02754}, \href {https://ui.adsabs.harvard.edu/abs/2016arXiv160302754C} {p. arXiv:1603.02754}

\bibitem[\protect\citeauthoryear{Choi, Dotter, Conroy, Cantiello, Paxton  \& Johnson}{Choi et~al.}{2016}]{choiMesaIsochronesStellar2016}
Choi J.,  Dotter A.,  Conroy C.,  Cantiello M.,  Paxton B.,   Johnson B.~D.,  2016, \mn@doi [The Astrophysical Journal] {10.3847/0004-637X/823/2/102}, 823, 102

\bibitem[\protect\citeauthoryear{Claret \& Bloemen}{Claret \& Bloemen}{2011}]{claretGravityLimbdarkeningCoefficients2011}
Claret A.,  Bloemen S.,  2011, \mn@doi [Astronomy and Astrophysics] {10.1051/0004-6361/201116451}, 529, A75

\bibitem[\protect\citeauthoryear{{Collins}, {Quinn}, {Latham}, {Christiansen}, {Ciardi}, {Dragomir}, {Crossfield}  \& {Seager}}{{Collins} et~al.}{2018}]{TFOP}
{Collins} K.,  {Quinn} S.~N.,  {Latham} D.~W.,  {Christiansen} J.,  {Ciardi} D.,  {Dragomir} D.,  {Crossfield} I.,   {Seager} S.,  2018, in American Astronomical Society Meeting Abstracts \#231. p. 439.08

\bibitem[\protect\citeauthoryear{{Coughlin} et~al.,}{{Coughlin} et~al.}{2016}]{Robovetter24}
{Coughlin} J.~L.,  et~al., 2016, \mn@doi [\apjs] {10.3847/0067-0049/224/1/12}, \href {https://ui.adsabs.harvard.edu/abs/2016ApJS..224...12C} {224, 12}

\bibitem[\protect\citeauthoryear{{Delrez} et~al.,}{{Delrez} et~al.}{2022}]{triceratops_validated1}
{Delrez} L.,  et~al., 2022, \mn@doi [\aap] {10.1051/0004-6361/202244041}, \href {https://ui.adsabs.harvard.edu/abs/2022A&A...667A..59D} {667, A59}

\bibitem[\protect\citeauthoryear{{D{\'\i}az}, {Almenara}, {Santerne}, {Moutou}, {Lethuillier}  \& {Deleuil}}{{D{\'\i}az} et~al.}{2014}]{pastis}
{D{\'\i}az} R.~F.,  {Almenara} J.~M.,  {Santerne} A.,  {Moutou} C.,  {Lethuillier} A.,   {Deleuil} M.,  2014, \mn@doi [\mnras] {10.1093/mnras/stu601}, \href {https://ui.adsabs.harvard.edu/abs/2014MNRAS.441..983D} {441, 983}

\bibitem[\protect\citeauthoryear{Dotter}{Dotter}{2016}]{dotterMESAIsochronesStellar2016}
Dotter A.,  2016, \mn@doi [The Astrophysical Journal Supplement Series] {10.3847/0067-0049/222/1/8}, 222, 8

\bibitem[\protect\citeauthoryear{{Doyle}, {Armstrong}, {Bayliss}, {Rodel}  \& {Kunovac}}{{Doyle} et~al.}{2024}]{LaurenSample}
{Doyle} L.,  {Armstrong} D.~J.,  {Bayliss} D.,  {Rodel} T.,   {Kunovac} V.,  2024, \mn@doi [\mnras] {10.1093/mnras/stae616}, \href {https://ui.adsabs.harvard.edu/abs/2024MNRAS.529.1802D} {529, 1802}

\bibitem[\protect\citeauthoryear{{Eisner} et~al.,}{{Eisner} et~al.}{2021}]{PlanetHuntersTess}
{Eisner} N.~L.,  et~al., 2021, \mn@doi [\mnras] {10.1093/mnras/staa3739}, \href {https://ui.adsabs.harvard.edu/abs/2021MNRAS.501.4669E} {501, 4669}

\bibitem[\protect\citeauthoryear{Friedman}{Friedman}{2001}]{friedman2001gradboost}
Friedman J.~H.,  2001, \mn@doi [The Annals of Statistics] {10.1214/aos/1013203451}, 29, 1189

\bibitem[\protect\citeauthoryear{{Gaia Collaboration} et~al.,}{{Gaia Collaboration} et~al.}{2018}]{GaiaDR2}
{Gaia Collaboration} et~al., 2018, \mn@doi [\aap] {10.1051/0004-6361/201833051}, \href {https://ui.adsabs.harvard.edu/abs/2018A&A...616A...1G} {616, A1}

\bibitem[\protect\citeauthoryear{{Gaia Collaboration} et~al.,}{{Gaia Collaboration} et~al.}{2023}]{GaiaDR3}
{Gaia Collaboration} et~al., 2023, \mn@doi [\aap] {10.1051/0004-6361/202243940}, \href {https://ui.adsabs.harvard.edu/abs/2023A&A...674A...1G} {674, A1}

\bibitem[\protect\citeauthoryear{{Giacalone} et~al.,}{{Giacalone} et~al.}{2021}]{Triceratops}
{Giacalone} S.,  et~al., 2021, \mn@doi [\aj] {10.3847/1538-3881/abc6af}, \href {https://ui.adsabs.harvard.edu/abs/2021AJ....161...24G} {161, 24}

\bibitem[\protect\citeauthoryear{{Giacalone} et~al.,}{{Giacalone} et~al.}{2022}]{TRICERATOPS_Validated_large1}
{Giacalone} S.,  et~al., 2022, \mn@doi [\aj] {10.3847/1538-3881/ac4334}, \href {https://ui.adsabs.harvard.edu/abs/2022AJ....163...99G} {163, 99}

\bibitem[\protect\citeauthoryear{{Girardi}, {Groenewegen}, {Hatziminaoglou}  \& {da Costa}}{{Girardi} et~al.}{2005}]{TRILEGAL}
{Girardi} L.,  {Groenewegen} M.~A.~T.,  {Hatziminaoglou} E.,   {da Costa} L.,  2005, \mn@doi [\aap] {10.1051/0004-6361:20042352}, \href {https://0-ui-adsabs-harvard-edu.pugwash.lib.warwick.ac.uk/abs/2005A&A...436..895G} {436, 895}

\bibitem[\protect\citeauthoryear{{Gomez Barrientos} et~al.,}{{Gomez Barrientos} et~al.}{2025}]{triceratops_plus}
{Gomez Barrientos} J.,  et~al., 2025, \mn@doi [\aj] {10.3847/1538-3881/ade68b}, \href {https://ui.adsabs.harvard.edu/abs/2025AJ....170..148G} {170, 148}

\bibitem[\protect\citeauthoryear{{Guerrero} et~al.,}{{Guerrero} et~al.}{2021}]{TOI}
{Guerrero} N.~M.,  et~al., 2021, \mn@doi [\apjs] {10.3847/1538-4365/abefe1}, \href {https://ui.adsabs.harvard.edu/abs/2021ApJS..254...39G} {254, 39}

\bibitem[\protect\citeauthoryear{{Hadjigeorghiou} \& {Armstrong}}{{Hadjigeorghiou} \& {Armstrong}}{2024}]{PosProbs}
{Hadjigeorghiou} A.,  {Armstrong} D.~J.,  2024, \mn@doi [\mnras] {10.1093/mnras/stad3286}, \href {https://ui.adsabs.harvard.edu/abs/2024MNRAS.527.4018H} {527, 4018}

\bibitem[\protect\citeauthoryear{Hensman, Matthews  \& Ghahramani}{Hensman et~al.}{2015}]{hensman2015scalable}
Hensman J.,  Matthews A.,   Ghahramani Z.,  2015, in Artificial Intelligence and Statistics. pp 351--360

\bibitem[\protect\citeauthoryear{{Hippke} \& {Heller}}{{Hippke} \& {Heller}}{2019}]{TLS}
{Hippke} M.,  {Heller} R.,  2019, \mn@doi [\aap] {10.1051/0004-6361/201834672}, \href {https://ui.adsabs.harvard.edu/abs/2019A&A...623A..39H} {623, A39}

\bibitem[\protect\citeauthoryear{{Hoffman}}{{Hoffman}}{2022}]{cuvarbase}
{Hoffman} J.,  2022, {cuvarbase: fast period finding utilities for GPUs}, Astrophysics Source Code Library, record ascl:2210.030

\bibitem[\protect\citeauthoryear{{Howell} et~al.,}{{Howell} et~al.}{2014}]{K2}
{Howell} S.~B.,  et~al., 2014, \mn@doi [\pasp] {10.1086/676406}, \href {https://ui.adsabs.harvard.edu/abs/2014PASP..126..398H} {126, 398}

\bibitem[\protect\citeauthoryear{{Hsu}, {Ford}, {Ragozzine}  \& {Ashby}}{{Hsu} et~al.}{2019}]{Hsu2019}
{Hsu} D.~C.,  {Ford} E.~B.,  {Ragozzine} D.,   {Ashby} K.,  2019, \mn@doi [\aj] {10.3847/1538-3881/ab31ab}, \href {https://ui.adsabs.harvard.edu/abs/2019AJ....158..109H} {158, 109}

\bibitem[\protect\citeauthoryear{{Huang} et~al.,}{{Huang} et~al.}{2020a}]{QLP1}
{Huang} C.~X.,  et~al., 2020a, \mn@doi [Research Notes of the American Astronomical Society] {10.3847/2515-5172/abca2e}, \href {https://ui.adsabs.harvard.edu/abs/2020RNAAS...4..204H} {4, 204}

\bibitem[\protect\citeauthoryear{{Huang} et~al.,}{{Huang} et~al.}{2020b}]{QLP2}
{Huang} C.~X.,  et~al., 2020b, \mn@doi [Research Notes of the American Astronomical Society] {10.3847/2515-5172/abca2d}, \href {https://ui.adsabs.harvard.edu/abs/2020RNAAS...4..206H} {4, 206}

\bibitem[\protect\citeauthoryear{{Jenkins} et~al.,}{{Jenkins} et~al.}{2016}]{SPOC}
{Jenkins} J.~M.,  et~al., 2016, in {Chiozzi} G.,  {Guzman} J.~C.,  eds,  Society of Photo-Optical Instrumentation Engineers (SPIE) Conference Series Vol. 9913, Software and Cyberinfrastructure for Astronomy IV. p. 99133E, \mn@doi{10.1117/12.2233418}

\bibitem[\protect\citeauthoryear{{Kingma} \& {Ba}}{{Kingma} \& {Ba}}{2014}]{Adam}
{Kingma} D.~P.,  {Ba} J.,  2014, \mn@doi [arXiv e-prints] {10.48550/arXiv.1412.6980}, \href {https://ui.adsabs.harvard.edu/abs/2014arXiv1412.6980K} {p. arXiv:1412.6980}

\bibitem[\protect\citeauthoryear{Kohonen}{Kohonen}{1982}]{kohonen1982SOM}
Kohonen T.,  1982, Biological cybernetics, 43, 59

\bibitem[\protect\citeauthoryear{{Kov{\'a}cs}, {Zucker}  \& {Mazeh}}{{Kov{\'a}cs} et~al.}{2002}]{BLS}
{Kov{\'a}cs} G.,  {Zucker} S.,   {Mazeh} T.,  2002, \mn@doi [\aap] {10.1051/0004-6361:20020802}, \href {https://ui.adsabs.harvard.edu/abs/2002A&A...391..369K} {391, 369}

\bibitem[\protect\citeauthoryear{{Kreidberg}}{{Kreidberg}}{2015}]{batman}
{Kreidberg} L.,  2015, \mn@doi [\pasp] {10.1086/683602}, \href {https://ui.adsabs.harvard.edu/abs/2015PASP..127.1161K} {127, 1161}

\bibitem[\protect\citeauthoryear{{Kunimoto} et~al.,}{{Kunimoto} et~al.}{2022}]{TESS-FaintStar}
{Kunimoto} M.,  et~al., 2022, \mn@doi [\apjs] {10.3847/1538-4365/ac5688}, \href {https://ui.adsabs.harvard.edu/abs/2022ApJS..259...33K} {259, 33}

\bibitem[\protect\citeauthoryear{{Kunimoto} et~al.,}{{Kunimoto} et~al.}{2025}]{LEO-Vetter}
{Kunimoto} M.,  et~al., 2025, \mn@doi [arXiv e-prints] {10.48550/arXiv.2509.10619}, \href {https://ui.adsabs.harvard.edu/abs/2025arXiv250910619K} {p. arXiv:2509.10619}

\bibitem[\protect\citeauthoryear{{Lillo-Box} et~al.,}{{Lillo-Box} et~al.}{2024}]{VESPA_HEB}
{Lillo-Box} J.,  et~al., 2024, \mn@doi [\aap] {10.1051/0004-6361/202451398}, \href {https://ui.adsabs.harvard.edu/abs/2024A&A...689L...8L} {689, L8}

\bibitem[\protect\citeauthoryear{{Lindegren} et~al.,}{{Lindegren} et~al.}{2018}]{GaiaDR2_astrometric_solution}
{Lindegren} L.,  et~al., 2018, \mn@doi [\aap] {10.1051/0004-6361/201832727}, \href {https://ui.adsabs.harvard.edu/abs/2018A&A...616A...2L} {616, A2}

\bibitem[\protect\citeauthoryear{{Lindegren} et~al.,}{{Lindegren} et~al.}{2021}]{GaiaEDR3_astrometric_solution}
{Lindegren} L.,  et~al., 2021, \mn@doi [\aap] {10.1051/0004-6361/202039709}, \href {https://ui.adsabs.harvard.edu/abs/2021A&A...649A...2L} {649, A2}

\bibitem[\protect\citeauthoryear{{Mayo} et~al.,}{{Mayo} et~al.}{2018}]{vespa_k2_validated}
{Mayo} A.~W.,  et~al., 2018, \mn@doi [\aj] {10.3847/1538-3881/aaadff}, \href {https://ui.adsabs.harvard.edu/abs/2018AJ....155..136M} {155, 136}

\bibitem[\protect\citeauthoryear{{McCauliff} et~al.,}{{McCauliff} et~al.}{2015}]{Autovetter}
{McCauliff} S.~D.,  et~al., 2015, \mn@doi [\apj] {10.1088/0004-637X/806/1/6}, \href {https://0-ui-adsabs-harvard-edu.pugwash.lib.warwick.ac.uk/abs/2015ApJ...806....6M} {806, 6}

\bibitem[\protect\citeauthoryear{{Mistry} et~al.,}{{Mistry} et~al.}{2023}]{VaTEST2}
{Mistry} P.,  et~al., 2023, \mn@doi [\aj] {10.3847/1538-3881/acd548}, \href {https://ui.adsabs.harvard.edu/abs/2023AJ....166....9M} {166, 9}

\bibitem[\protect\citeauthoryear{{Mistry} et~al.,}{{Mistry} et~al.}{2024}]{VaTEST3}
{Mistry} P.,  et~al., 2024, \mn@doi [\pasa] {10.1017/pasa.2024.29}, \href {https://ui.adsabs.harvard.edu/abs/2024PASA...41...30M} {41, e030}

\bibitem[\protect\citeauthoryear{{Moe} \& {Di Stefano}}{{Moe} \& {Di Stefano}}{2017}]{MoeDiStefano2017}
{Moe} M.,  {Di Stefano} R.,  2017, \mn@doi [\apjs] {10.3847/1538-4365/aa6fb6}, \href {https://ui.adsabs.harvard.edu/abs/2017ApJS..230...15M} {230, 15}

\bibitem[\protect\citeauthoryear{{Morton}}{{Morton}}{2012}]{vespa}
{Morton} T.~D.,  2012, \mn@doi [\apj] {10.1088/0004-637X/761/1/6}, \href {https://ui.adsabs.harvard.edu/abs/2012ApJ...761....6M} {761, 6}

\bibitem[\protect\citeauthoryear{Morton}{Morton}{2015}]{mortonIsochronesStellarModel2015}
Morton T.~D.,  2015, Astrophysics Source Code Library, p. ascl:1503.010

\bibitem[\protect\citeauthoryear{{Morton}, {Bryson}, {Coughlin}, {Rowe}, {Ravichandran}, {Petigura}, {Haas}  \& {Batalha}}{{Morton} et~al.}{2016}]{vespa_validated}
{Morton} T.~D.,  {Bryson} S.~T.,  {Coughlin} J.~L.,  {Rowe} J.~F.,  {Ravichandran} G.,  {Petigura} E.~A.,  {Haas} M.~R.,   {Batalha} N.~M.,  2016, \mn@doi [\apj] {10.3847/0004-637X/822/2/86}, \href {https://ui.adsabs.harvard.edu/abs/2016ApJ...822...86M} {822, 86}

\bibitem[\protect\citeauthoryear{M{\"u}ller, Baron, Helled, Bouchy  \& Parc}{M{\"u}ller et~al.}{2024}]{mullerMassradiusRelationExoplanets2024}
M{\"u}ller S.,  Baron J.,  Helled R.,  Bouchy F.,   Parc L.,  2024, \mn@doi [A\&A] {10.1051/0004-6361/202348690}, 686, A296

\bibitem[\protect\citeauthoryear{Paxton, Bildsten, Dotter, Herwig, Lesaffre  \& Timmes}{Paxton et~al.}{2011}]{paxtonModulesExperimentsStellar2011}
Paxton B.,  Bildsten L.,  Dotter A.,  Herwig F.,  Lesaffre P.,   Timmes F.,  2011, \mn@doi [The Astrophysical Journal Supplement Series] {10.1088/0067-0049/192/1/3}, 192, 3

\bibitem[\protect\citeauthoryear{Paxton et~al.,}{Paxton et~al.}{2013}]{paxtonModulesExperimentsStellar2013}
Paxton B.,  et~al., 2013, \mn@doi [The Astrophysical Journal Supplement Series] {10.1088/0067-0049/208/1/4}, 208, 4

\bibitem[\protect\citeauthoryear{Paxton et~al.,}{Paxton et~al.}{2015}]{paxtonModulesExperimentsStellar2015}
Paxton B.,  et~al., 2015, \mn@doi [The Astrophysical Journal Supplement Series] {10.1088/0067-0049/220/1/15}, 220, 15

\bibitem[\protect\citeauthoryear{{Pollacco} et~al.,}{{Pollacco} et~al.}{2006}]{WASP}
{Pollacco} D.~L.,  et~al., 2006, \mn@doi [\pasp] {10.1086/508556}, \href {https://ui.adsabs.harvard.edu/abs/2006PASP..118.1407P} {118, 1407}

\bibitem[\protect\citeauthoryear{{Raghavan} et~al.,}{{Raghavan} et~al.}{2010}]{Raghavan2010}
{Raghavan} D.,  et~al., 2010, \mn@doi [\apjs] {10.1088/0067-0049/190/1/1}, \href {https://ui.adsabs.harvard.edu/abs/2010ApJS..190....1R} {190, 1}

\bibitem[\protect\citeauthoryear{{Rasmussen} \& {Williams}}{{Rasmussen} \& {Williams}}{2006}]{GP}
{Rasmussen} C.~E.,  {Williams} C. K.~I.,  2006, {Gaussian Processes for Machine Learning}

\bibitem[\protect\citeauthoryear{{Ricker} et~al.,}{{Ricker} et~al.}{2015}]{TESS}
{Ricker} G.~R.,  et~al., 2015, \mn@doi [Journal of Astronomical Telescopes, Instruments, and Systems] {10.1117/1.JATIS.1.1.014003}, \href {https://ui.adsabs.harvard.edu/abs/2015JATIS...1a4003R} {1, 014003}

\bibitem[\protect\citeauthoryear{Robin, Reyl{\'e}, Derri{\`e}re  \& Picaud}{Robin et~al.}{2003}]{robinSyntheticViewStructure2003}
Robin A.~C.,  Reyl{\'e} C.,  Derri{\`e}re S.,   Picaud S.,  2003, \mn@doi [A\&A] {10.1051/0004-6361:20031117}, 409, 523

\bibitem[\protect\citeauthoryear{{Santerne} et~al.,}{{Santerne} et~al.}{2015}]{pastis2}
{Santerne} A.,  et~al., 2015, \mn@doi [\mnras] {10.1093/mnras/stv1080}, \href {https://ui.adsabs.harvard.edu/abs/2015MNRAS.451.2337S} {451, 2337}

\bibitem[\protect\citeauthoryear{{Savitzky} \& {Golay}}{{Savitzky} \& {Golay}}{1964}]{SavitzkyGolay}
{Savitzky} A.,  {Golay} M.~J.~E.,  1964, \mn@doi [Analytical Chemistry] {10.1021/ac60214a047}, \href {https://ui.adsabs.harvard.edu/abs/1964AnaCh..36.1627S} {36, 1627}

\bibitem[\protect\citeauthoryear{{Shallue} \& {Vanderburg}}{{Shallue} \& {Vanderburg}}{2018}]{AstroNET-Kepler}
{Shallue} C.~J.,  {Vanderburg} A.,  2018, \mn@doi [\aj] {10.3847/1538-3881/aa9e09}, \href {https://ui.adsabs.harvard.edu/abs/2018AJ....155...94S} {155, 94}

\bibitem[\protect\citeauthoryear{{Shporer} et~al.,}{{Shporer} et~al.}{2017}]{Avi2017_vespa_issue}
{Shporer} A.,  et~al., 2017, \mn@doi [\apjl] {10.3847/2041-8213/aa8bff}, \href {https://ui.adsabs.harvard.edu/abs/2017ApJ...847L..18S} {847, L18}

\bibitem[\protect\citeauthoryear{Silva~Filho, Song, {Perello-Nieto}, {Santos-Rodriguez}, Kull  \& Flach}{Silva~Filho et~al.}{2023}]{silvafilhoClassifierCalibrationSurvey2023}
Silva~Filho T.,  Song H.,  {Perello-Nieto} M.,  {Santos-Rodriguez} R.,  Kull M.,   Flach P.,  2023, \mn@doi [Mach Learn] {10.1007/s10994-023-06336-7}, 112, 3211

\bibitem[\protect\citeauthoryear{{Stassun} \& {Torres}}{{Stassun} \& {Torres}}{2021}]{Gaia_Binaries_Stassun}
{Stassun} K.~G.,  {Torres} G.,  2021, \mn@doi [\apjl] {10.3847/2041-8213/abdaad}, \href {https://ui.adsabs.harvard.edu/abs/2021ApJ...907L..33S} {907, L33}

\bibitem[\protect\citeauthoryear{{Stumpe} et~al.,}{{Stumpe} et~al.}{2012a}]{Correction}
{Stumpe} M.~C.,  et~al., 2012a, \mn@doi [\pasp] {10.1086/667698}, \href {https://ui.adsabs.harvard.edu/abs/2012PASP..124..985S} {124, 985}

\bibitem[\protect\citeauthoryear{{Stumpe} et~al.,}{{Stumpe} et~al.}{2012b}]{KeplerPDC}
{Stumpe} M.~C.,  et~al., 2012b, \mn@doi [\pasp] {10.1086/667698}, \href {https://ui.adsabs.harvard.edu/abs/2012PASP..124..985S} {124, 985}

\bibitem[\protect\citeauthoryear{{Tenenbaum} et~al.,}{{Tenenbaum} et~al.}{2013}]{Robstat}
{Tenenbaum} P.,  et~al., 2013, \mn@doi [\apjs] {10.1088/0067-0049/206/1/5}, \href {https://ui.adsabs.harvard.edu/abs/2013ApJS..206....5T} {206, 5}

\bibitem[\protect\citeauthoryear{{Tey} et~al.,}{{Tey} et~al.}{2023}]{AstroNET-TriageV2}
{Tey} E.,  et~al., 2023, \mn@doi [\aj] {10.3847/1538-3881/acad85}, \href {https://ui.adsabs.harvard.edu/abs/2023AJ....165...95T} {165, 95}

\bibitem[\protect\citeauthoryear{{Thompson} et~al.,}{{Thompson} et~al.}{2018}]{Robovveter25}
{Thompson} S.~E.,  et~al., 2018, \mn@doi [\apjs] {10.3847/1538-4365/aab4f9}, \href {https://ui.adsabs.harvard.edu/abs/2018ApJS..235...38T} {235, 38}

\bibitem[\protect\citeauthoryear{{Torres} et~al.,}{{Torres} et~al.}{2011}]{BLENDER}
{Torres} G.,  et~al., 2011, \mn@doi [\apj] {10.1088/0004-637X/727/1/24}, \href {https://ui.adsabs.harvard.edu/abs/2011ApJ...727...24T} {727, 24}

\bibitem[\protect\citeauthoryear{{Torres} et~al.,}{{Torres} et~al.}{2015}]{BLENDER_12HZ}
{Torres} G.,  et~al., 2015, \mn@doi [\apj] {10.1088/0004-637X/800/2/99}, \href {https://ui.adsabs.harvard.edu/abs/2015ApJ...800...99T} {800, 99}

\bibitem[\protect\citeauthoryear{{Twicken} et~al.,}{{Twicken} et~al.}{2018}]{KEPLER_DVP}
{Twicken} J.~D.,  et~al., 2018, \mn@doi [\pasp] {10.1088/1538-3873/aab694}, \href {https://ui.adsabs.harvard.edu/abs/2018PASP..130f4502T} {130, 064502}

\bibitem[\protect\citeauthoryear{{Valizadegan} et~al.,}{{Valizadegan} et~al.}{2022}]{ExoMiner}
{Valizadegan} H.,  et~al., 2022, \mn@doi [\apj] {10.3847/1538-4357/ac4399}, \href {https://ui.adsabs.harvard.edu/abs/2022ApJ...926..120V} {926, 120}

\bibitem[\protect\citeauthoryear{{Valizadegan} et~al.,}{{Valizadegan} et~al.}{2025}]{ExoMinerTESS}
{Valizadegan} H.,  et~al., 2025, \mn@doi [arXiv e-prints] {10.48550/arXiv.2502.09790}, \href {https://ui.adsabs.harvard.edu/abs/2025arXiv250209790V} {p. arXiv:2502.09790}

\bibitem[\protect\citeauthoryear{{Yu} et~al.,}{{Yu} et~al.}{2019}]{AstroNET-Triage}
{Yu} L.,  et~al., 2019, \mn@doi [\aj] {10.3847/1538-3881/ab21d6}, \href {https://ui.adsabs.harvard.edu/abs/2019AJ....158...25Y} {158, 25}

\makeatother
\end{thebibliography}
